\documentclass{emulateapj}

\tabletypesize{\scriptsize}
\usepackage{url}
\usepackage{multirow}

\newcommand{\DataRunningDays}{0.008}
\newcommand{\DataRunningMinutes}{11.5}
\newcommand{\INSERTNIGHTS}{22}	
\newcommand{\INSERTNIGHTSSIMULTANEOUS}{7}	
\newcommand{\INSERTSPREADDAYS}{120}		
\newcommand{\TOTALHOURS}{180}			
\newcommand{\TOTALNONOVERLAPPINGHOURS}{150}			
\newcommand{\TSpotKelvinValue}{1650}
\newcommand{\TSpotKelvinValueError}{50}
\newcommand{\TSpotHotKelvinValue}{1900}
\newcommand{\TSpotHotKelvinValueError}{200}
\newcommand{\TSpotKelvinSimultaneousValue}{1600}
\newcommand{\TSpotKelvinSimultaneousValueError}{100}
\newcommand{\TSpotHotKelvinSimultaneousValue}{1750}
\newcommand{\TSpotHotKelvinSimultaneousValueError}{50}

\newcommand{\InsertRadius}{11}
\newcommand{\InsertPeriod}{11}
\newcommand{\InsertPhase}{21}
\newcommand{\fsedUseSpectra}{2}
\newcommand{\loggUseSpectra}{5.5}
\newcommand{\TemperatureUCDUseSpectra}{1700}
\newcommand{\TemperatureMinModel}{1800}	
\newcommand{\fsedMinModel}{1}		
\newcommand{\PeriodHoursTwoMassZeroZeroThreeSixApprox}{3.08}
\newcommand{\PeriodHoursTwoMassZeroZeroThirtySixAllbandsQuote}{3.080}
\newcommand{\PeriodHoursTwoMassZeroZeroThirtySixAllbands}{3.080}
\newcommand{\PeriodHoursErrorTwoMassZeroZeroThirtySixAllbands}{0.001}
\newcommand{\PeriodHoursTwoMassZeroZeroThirtySixJband}{3.080}
\newcommand{\PeriodHoursErrorTwoMassZeroZeroThirtySixJband}{0.005}
\newcommand{\PeriodHoursTwoMassZeroZeroThirtySixRband}{3.081}
\newcommand{\PeriodHoursErrorTwoMassZeroZeroThirtySixRband}{0.123}
\newcommand{\PeriodHoursTwoMassZeroZeroThirtySixIband}{3.071}
\newcommand{\PeriodHoursErrorTwoMassZeroZeroThirtySixIband}{0.125}
\newcommand{\PeriodHoursTwoMassZeroZeroThirtySixzband}{3.079}
\newcommand{\PeriodHoursErrorTwoMassZeroZeroThirtySixzband}{0.054}
\newcommand{\SinusoidPeaktoPeakAmplitudePercentageTwoMassZeroZeroThirtySixAllData}{1.36}
\newcommand{\SinusoidPeaktoPeakAmplitudePercentageErrorTwoMassZeroZeroThirtySixAllData}{0.03}

\newcommand{\SinusoidPhaseTwoMassZeroZeroThirtySixAllData}{0.896}
\newcommand{\SinusoidPhaseErrorTwoMassZeroZeroThirtySixAllData}{0.004}

\newcommand{\SinusoidPeaktoPeakAmplitudePercentageTwoMassZeroZeroThirtySixKsband}{1.07}
\newcommand{\SinusoidPeaktoPeakAmplitudePercentageErrorTwoMassZeroZeroThirtySixKsband}{0.08}

\newcommand{\SinusoidPhaseTwoMassZeroZeroThirtySixKsband}{0.628}
\newcommand{\SinusoidPhaseErrorTwoMassZeroZeroThirtySixKsband}{0.013}

\newcommand{\SinusoidPeaktoPeakAmplitudePercentageTwoMassZeroZeroThirtySixHband}{0.45}
\newcommand{\SinusoidPeaktoPeakAmplitudePercentageErrorTwoMassZeroZeroThirtySixHband}{0.05}

\newcommand{\SinusoidPhaseTwoMassZeroZeroThirtySixHband}{0.869}
\newcommand{\SinusoidPhaseErrorTwoMassZeroZeroThirtySixHband}{0.016}

\newcommand{\SinusoidPeaktoPeakAmplitudePercentageTwoMassZeroZeroThirtySixJband}{1.22}
\newcommand{\SinusoidPeaktoPeakAmplitudePercentageErrorTwoMassZeroZeroThirtySixJband}{0.04}

\newcommand{\SinusoidPhaseTwoMassZeroZeroThirtySixJband}{0.913}
\newcommand{\SinusoidPhaseErrorTwoMassZeroZeroThirtySixJband}{0.005}

\newcommand{\SinusoidPeaktoPeakAmplitudePercentageTwoMassZeroZeroThirtySixzband}{2.74}
\newcommand{\SinusoidPeaktoPeakAmplitudePercentageErrorTwoMassZeroZeroThirtySixzband}{0.08}

\newcommand{\SinusoidPhaseTwoMassZeroZeroThirtySixzband}{0.878}
\newcommand{\SinusoidPhaseErrorTwoMassZeroZeroThirtySixzband}{0.005}

\newcommand{\SinusoidPeaktoPeakAmplitudePercentageTwoMassZeroZeroThirtySixIband}{2.11}
\newcommand{\SinusoidPeaktoPeakAmplitudePercentageErrorTwoMassZeroZeroThirtySixIband}{0.09}

\newcommand{\SinusoidPhaseTwoMassZeroZeroThirtySixIband}{0.879}
\newcommand{\SinusoidPhaseErrorTwoMassZeroZeroThirtySixIband}{0.006}

\newcommand{\SinusoidPeaktoPeakAmplitudePercentageTwoMassZeroZeroThirtySixRband}{3.40}
\newcommand{\SinusoidPeaktoPeakAmplitudePercentageErrorTwoMassZeroZeroThirtySixRband}{0.11}

\newcommand{\SinusoidPhaseTwoMassZeroZeroThirtySixRband}{0.884}
\newcommand{\SinusoidPhaseErrorTwoMassZeroZeroThirtySixRband}{0.005}

\newcommand{\SinusoidPeaktoPeakAmplitudePercentageTwoMassZeroZeroThirtySixFifteenSeptemberTwentyThreePerkinsIband}{1.98}
\newcommand{\SinusoidPeaktoPeakAmplitudePercentageErrorTwoMassZeroZeroThirtySixFifteenSeptemberTwentyThreePerkinsIband}{0.07}
\newcommand{\SinusoidPhaseTwoMassZeroZeroThirtySixFifteenSeptemberTwentyThreePerkinsIband}{0.950}
\newcommand{\SinusoidPhaseErrorTwoMassZeroZeroThirtySixFifteenSeptemberTwentyThreePerkinsIband}{0.005}

\newcommand{\SinusoidPeaktoPeakAmplitudePercentageTwoMassZeroZeroThirtySixFifteenSeptemberTwentyThreeHallIband}{2.19}
\newcommand{\SinusoidPeaktoPeakAmplitudePercentageErrorTwoMassZeroZeroThirtySixFifteenSeptemberTwentyThreeHallIband}{0.15}
\newcommand{\SinusoidPhaseTwoMassZeroZeroThirtySixFifteenSeptemberTwentyThreeHallIband}{0.929}
\newcommand{\SinusoidPhaseErrorTwoMassZeroZeroThirtySixFifteenSeptemberTwentyThreeHallIband}{0.011}

\newcommand{\SinusoidPeaktoPeakAmplitudePercentageTwoMassZeroZeroThirtySixFifteenSeptemberTwentyFourzband}{2.51}
\newcommand{\SinusoidPeaktoPeakAmplitudePercentageErrorTwoMassZeroZeroThirtySixFifteenSeptemberTwentyFourzband}{0.24}
\newcommand{\SinusoidPhaseTwoMassZeroZeroThirtySixFifteenSeptemberTwentyFourzband}{0.951}
\newcommand{\SinusoidPhaseErrorTwoMassZeroZeroThirtySixFifteenSeptemberTwentyFourzband}{0.013}

\newcommand{\SinusoidPeaktoPeakAmplitudePercentageTwoMassZeroZeroThirtySixFifteenSeptemberTwentyFourIband}{3.21}
\newcommand{\SinusoidPeaktoPeakAmplitudePercentageErrorTwoMassZeroZeroThirtySixFifteenSeptemberTwentyFourIband}{0.17}
\newcommand{\SinusoidPhaseTwoMassZeroZeroThirtySixFifteenSeptemberTwentyFourIband}{0.921}
\newcommand{\SinusoidPhaseErrorTwoMassZeroZeroThirtySixFifteenSeptemberTwentyFourIband}{0.009}

\newcommand{\SinusoidPeaktoPeakAmplitudePercentageTwoMassZeroZeroThirtySixFifteenSeptemberTwentyFivezband}{2.80}
\newcommand{\SinusoidPeaktoPeakAmplitudePercentageErrorTwoMassZeroZeroThirtySixFifteenSeptemberTwentyFivezband}{0.26}
\newcommand{\SinusoidPhaseTwoMassZeroZeroThirtySixFifteenSeptemberTwentyFivezband}{0.918}
\newcommand{\SinusoidPhaseErrorTwoMassZeroZeroThirtySixFifteenSeptemberTwentyFivezband}{0.016}

\newcommand{\SinusoidPeaktoPeakAmplitudePercentageTwoMassZeroZeroThirtySixFifteenSeptemberTwentyFiveRband}{2.68}
\newcommand{\SinusoidPeaktoPeakAmplitudePercentageErrorTwoMassZeroZeroThirtySixFifteenSeptemberTwentyFiveRband}{0.16}
\newcommand{\SinusoidPhaseTwoMassZeroZeroThirtySixFifteenSeptemberTwentyFiveRband}{0.922}
\newcommand{\SinusoidPhaseErrorTwoMassZeroZeroThirtySixFifteenSeptemberTwentyFiveRband}{0.009}

\newcommand{\SinusoidPeaktoPeakAmplitudePercentageTwoMassZeroZeroThirtySixFifteenSeptemberTwentySixzband}{2.83}
\newcommand{\SinusoidPeaktoPeakAmplitudePercentageErrorTwoMassZeroZeroThirtySixFifteenSeptemberTwentySixzband}{0.09}
\newcommand{\SinusoidPhaseTwoMassZeroZeroThirtySixFifteenSeptemberTwentySixzband}{0.914}
\newcommand{\SinusoidPhaseErrorTwoMassZeroZeroThirtySixFifteenSeptemberTwentySixzband}{0.005}

\newcommand{\SinusoidPeaktoPeakAmplitudePercentageTwoMassZeroZeroThirtySixFifteenSeptemberTwentySixRband}{4.45}
\newcommand{\SinusoidPeaktoPeakAmplitudePercentageErrorTwoMassZeroZeroThirtySixFifteenSeptemberTwentySixRband}{0.10}
\newcommand{\SinusoidPhaseTwoMassZeroZeroThirtySixFifteenSeptemberTwentySixRband}{0.926}
\newcommand{\SinusoidPhaseErrorTwoMassZeroZeroThirtySixFifteenSeptemberTwentySixRband}{0.004}

\newcommand{\SinusoidPeaktoPeakAmplitudePercentageTwoMassZeroZeroThirtySixFifteenOctoberFourteenzband}{3.22}
\newcommand{\SinusoidPeaktoPeakAmplitudePercentageErrorTwoMassZeroZeroThirtySixFifteenOctoberFourteenzband}{0.03}
\newcommand{\SinusoidPhaseTwoMassZeroZeroThirtySixFifteenOctoberFourteenzband}{0.960}
\newcommand{\SinusoidPhaseErrorTwoMassZeroZeroThirtySixFifteenOctoberFourteenzband}{0.002}

\newcommand{\SinusoidPeaktoPeakAmplitudePercentageTwoMassZeroZeroThirtySixFifteenOctoberFourteenRband}{2.78}
\newcommand{\SinusoidPeaktoPeakAmplitudePercentageErrorTwoMassZeroZeroThirtySixFifteenOctoberFourteenRband}{0.05}
\newcommand{\SinusoidPhaseTwoMassZeroZeroThirtySixFifteenOctoberFourteenRband}{0.989}
\newcommand{\SinusoidPhaseErrorTwoMassZeroZeroThirtySixFifteenOctoberFourteenRband}{0.003}

\newcommand{\SinusoidPeaktoPeakAmplitudePercentageTwoMassZeroZeroThirtySixFifteenNovemberEightIband}{2.20}
\newcommand{\SinusoidPeaktoPeakAmplitudePercentageErrorTwoMassZeroZeroThirtySixFifteenNovemberEightIband}{0.30}
\newcommand{\SinusoidPhaseTwoMassZeroZeroThirtySixFifteenNovemberEightIband}{0.951}
\newcommand{\SinusoidPhaseErrorTwoMassZeroZeroThirtySixFifteenNovemberEightIband}{0.019}

\newcommand{\SinusoidPeaktoPeakAmplitudePercentageTwoMassZeroZeroThirtySixFifteenNovemberEightJband}{1.22}
\newcommand{\SinusoidPeaktoPeakAmplitudePercentageErrorTwoMassZeroZeroThirtySixFifteenNovemberEightJband}{0.09}
\newcommand{\SinusoidPhaseTwoMassZeroZeroThirtySixFifteenNovemberEightJband}{0.003}
\newcommand{\SinusoidPhaseErrorTwoMassZeroZeroThirtySixFifteenNovemberEightJband}{0.010}

\newcommand{\SinusoidPeaktoPeakAmplitudePercentageTwoMassZeroZeroThirtySixFifteenNovemberNineHband}{0.47}
\newcommand{\SinusoidPeaktoPeakAmplitudePercentageErrorTwoMassZeroZeroThirtySixFifteenNovemberNineHband}{0.06}
\newcommand{\SinusoidPhaseTwoMassZeroZeroThirtySixFifteenNovemberNineHband}{0.989}
\newcommand{\SinusoidPhaseErrorTwoMassZeroZeroThirtySixFifteenNovemberNineHband}{0.016}

\newcommand{\SinusoidPeaktoPeakAmplitudePercentageTwoMassZeroZeroThirtySixFifteenNovemberNineIband}{2.08}
\newcommand{\SinusoidPeaktoPeakAmplitudePercentageErrorTwoMassZeroZeroThirtySixFifteenNovemberNineIband}{0.34}
\newcommand{\SinusoidPhaseTwoMassZeroZeroThirtySixFifteenNovemberNineIband}{0.968}
\newcommand{\SinusoidPhaseErrorTwoMassZeroZeroThirtySixFifteenNovemberNineIband}{0.028}

\newcommand{\PeriodHoursTwoMassZeroZeroThirtySixJbandFifteenNovemberTwentyThird}{7.5}
\newcommand{\PeriodHoursErrorTwoMassZeroZeroThirtySixJbandFifteenNovemberTwentyThird}{0.9}
\newcommand{\SinusoidPeaktoPeakAmplitudePercentageTwoMassZeroZeroThirtySixFifteenOctoberFourteenADJzband}{2.87}
\newcommand{\SinusoidPeaktoPeakAmplitudePercentageErrorTwoMassZeroZeroThirtySixFifteenOctoberFourteenADJzband}{0.03}

\newcommand{\SinusoidPeaktoPeakAmplitudePercentageTwoMassZeroZeroThirtySixFifteenOctoberFourteenADJRband}{3.36}
\newcommand{\SinusoidPeaktoPeakAmplitudePercentageErrorTwoMassZeroZeroThirtySixFifteenOctoberFourteenADJRband}{0.06}

\shorttitle{Multiwavelength Photometry of the L3.5 dwarf 2MASS 0036+18} 
\shortauthors{Croll et al.}

\begin{document}

\title{Long-term, Multiwavelength Light Curves of Ultra-cool Dwarfs: I. An Interplay of Starspots \& Clouds Likely Drive the Variability of the L3.5 dwarf 2MASS 0036+18}

\author{
	Bryce Croll\altaffilmark{1},
	Philip S. Muirhead\altaffilmark{1} \altaffilmark{2},	
 	Eunkyu Han\altaffilmark{2},
	Paul A. Dalba\altaffilmark{2},
	Jacqueline Radigan\altaffilmark{3},
	Caroline V. Morley\altaffilmark{4},
	Marko Lazarevic\altaffilmark{5},
	Brian Taylor\altaffilmark{6}
}

\altaffiltext{1}{Institute for Astrophysical Research, Boston University, 725 Commonwealth Ave. Room 506, Boston, MA 02215; croll@bu.edu}

\altaffiltext{2}{Department of Astronomy, Boston University, 725 Commonwealth Ave., Boston, MA 02215, USA}

\altaffiltext{3}{Utah Valley University, Orem, UT 84058, USA}

\altaffiltext{4}{Department of Astronomy and Astrophysics, University of California, Santa Cruz, CA 95064, USA}

\altaffiltext{5}{Department of Physics, Northeastern University, 100 Forsyth St, Boston, MA 02115}

\altaffiltext{6}{TI Research, 2915 Kletha Trail, Flagstaff Az. 86005}

\begin{abstract}

We present
multi-telescope, ground-based, multiwavelength optical and near-infrared photometry of the variable L3.5 ultra-cool
dwarf 2MASSW J0036159+182110.
We present \INSERTNIGHTS \ nights of photometry of 
2MASSW J0036159+182110,
including \INSERTNIGHTSSIMULTANEOUS \ nights of simultaneous, multiwavelength photometry, spread over $\sim$\INSERTSPREADDAYS \ days allowing 
us to determine the rotation period of this ultra-cool dwarf to be 
\PeriodHoursTwoMassZeroZeroThirtySixAllbands \ $\pm$ \PeriodHoursErrorTwoMassZeroZeroThirtySixAllbands \ hr. Our many nights
of multiwavelength photometry allow us to observe the evolution, or more specifically the lack thereof,
of the light curve over a great many rotation periods.
The lack of discernible phase shifts in our multiwavelength photometry, and that the amplitude 
of variability generally decreases
as one moves to longer wavelengths for 2MASSW J0036159+182110,
is generally consistent with starspots driving the variability on this ultra-cool dwarf,
with starspots that are $\sim$100 degrees $K$ hotter or cooler than the $\sim$1700 $K$ photosphere.
Also, reasonably thick clouds are required to fit the spectra of 
2MASSW J0036159+182110,
suggesting there likely exists some complex interplay between the starspots driving the
variability of this ultra-cool dwarf and the clouds that appear to envelope this ultra-cool dwarf.


\end{abstract}


\section{Introduction}






 Detecting and characterizing the variability of ultra-cool dwarfs is an area that has attracted growing attention in recent years,
with a wealth of variability detections from the late-M, L \& T spectral classes (e.g. \citealt{Gelino02,Irwin11,Radigan14,Buenzli14,Metchev15}).
For L-dwarfs, ground and space-based optical, near-infrared and infrared 
observations have recently shown that low level variability of L-dwarfs is common \citep{BailerJonesMundt01,Gelino02,Lane07,Harding13,Koen13,Metchev15};
intriguingly, once viewing geometry is taken into account, all L-dwarfs might be variable \citep{Metchev15}.
The questions this prompts 
are: what is the astrophysical cause of the observed variability, is it consistent across the L-spectral class, and if not where does the transition
region lie between various astrophysical causes of the observed variability.




On the stellar side of the hydrogen-fusing limit, starspots are the usual explanation for the observed variability.
M-dwarfs are notoriously active, with detections of H$\alpha$, a common marker of activity,
rising throughout the M-spectral class, including that nearly all very late M-dwarfs are
active \citep{West04,Schmidt15}. 
Rotation periods revealed by photometry \citep{Rockenfeller06,Irwin11},
and Doppler imaging techniques for M-dwarfs \citep{Barnes01,Barnes04},
have indicated that starspots
are ubiquitous on M-dwarfs.

Until recently there was reason to doubt that this 
starspot driven variability extended into the L-spectral class.
The neutral atmospheres of L dwarfs were believed to be too electrically resistive,
with too small of magnetic Reynolds numbers, for magnetic starspots to form 
\citep{Mohanty02,Gelino02}.
This conclusion was previously supported by studies that found 
the frequency of H$\alpha$ detections fell sharply at the M/L transition, reaching
negligible levels by the $\sim$L3 spectral class \citep{West04}.
This did not stop speculation that the variability displayed by L-dwarfs 
might arise from magnetic starspots \citep{Clarke02,Lane07}.
Such speculation might prove to be prescient, as a 
recent study has cast doubts on 
previous 
H$\alpha$ null-detections
for L-dwarfs and therefore indicated that
starspots might be present throughout 
the L spectral class:
\citet{Schmidt15}
analyzed higher signal-to-noise L-dwarf spectra and were able to detect
H$\alpha$ for approximately 90\% of L0 dwarfs, and
more than half of L-dwarfs as late as L5.
Therefore, starspots might be driving the variability
for cloudy early and even-late L dwarfs.



Once ultra-cool dwarf effective temperatures drop 
and silicate clouds begin to clear at the L/T transition,
cloud condensate
variability
has become the accepted explanation
for the large amplitude variability that has been observed for these objects
\citep{Artigau09,Radigan12,Gillon13,Radigan14,Buenzli14,Crossfield14Nature}. 
Multiwavelength photometry of these apparent cloudy ultra-cool dwarfs has returned light curves with different amplitudes
at different wavelengths \citep{Radigan12}, 
and with significant temporal phase shifts in the observed variability at different wavelengths \citep{Buenzli12,Yang16}, including
multiwavelength light curves that are anti-correlated in phase \citep{Biller13}.
However, multiwavelength light curves of some of these brown dwarfs that display variability that is believed to be driven by heterogeneous cloud cover
have not displayed significant phase offsets \citep{Apai13,Buenzli15}.
The likely explanation is that multiwavelength phase shifts will not be caused by cloud variability if the clouds span multiple pressure layers,
and specifically, if the clouds span the range of pressures that are probed by the various wavelengths of observation.

Intriguingly, cloud condensate variability
might not be constrained to the L/T transition.
Spitzer/IRAC observations have indicated
that nearly all L-dwarfs are likely variable \citep{Metchev15},
and anti-correlated light curves have been observed on a
late-M dwarf \citep{Littlefair08} -- in both cases clouds are
a likely explanation for the observed variability.





Another possibility that has recently arose, is that the observed variability of ultra-cool dwarfs results from
auroral activity, similar to the aurorae observed on planets in our own solar system (e.g. \citealt{Clarke80}) 
including the Earth.
Such auroral activity has been observed at the end of the main sequence on an M8.5 dwarf \citep{Hallinan15}.
Multiwavelength photometry of this dwarf
displays light curves that are anti-correlated in phase, and \citet{Hallinan15} speculate that auroral activity might explain
the anti-correlated light curves of another late M-dwarf
\citep{Littlefair08}. 
Similar auroral
activity may be indicated from radio detections
of polarized, pulsed emissions from a
T2.5 dwarf \citep{Kao16} 
and a T6.5 dwarf
\citep{RouteWolszczan12,WilliamsBerger15}.
Therefore, another possibility
is that auroral activity may be responsible for 
some
of the variability at the 
L/T transition \citep{Artigau09,Radigan12}
that has 
previously been believed to be due to holes in condensate clouds.



 Finally, another explanation for the observed variability is atmospheric temperature variations, 
arising either deep in the atmosphere of these ultra-cool dwarfs, or at other pressure layers,
that are communicated via radiative heating to the 
altitude regions that are probed by optical to near- and mid-infrared observations \citep{ShowmanKaspi13,RobinsonMarley14,Morley14}.

 It is also possible that ultra-cool dwarfs are variable due to more than one of the aforementioned astrophysical reasons. 
Time evolving clouds 
may periodically 
obscure magnetically driven cool or hot starspots
\citep{Lane07,Heinze13,Metchev15}, or
aurorae may play an occasional role on predominantly 
cloudy brown dwarfs \citep{Hallinan15}. 
Many variability studies of ultra-cool dwarfs 
to date have examined targets for only tens of minutes \citep{Buenzli14} to hours at a time \citep{Radigan14}.
This method is likely sufficient for ultra-cool dwarfs that display constant variability \citep{Gizis13,Gizis15}, but
will inadequately address the variability of objects with light curves that
evolve rapidly, such as have been observed for several ultra-cool dwarfs \citep{Artigau09,Gillon13,Metchev15}. 
The rapid evolution of these brown dwarfs might be due to the properties of a single variability
mechanism changing --- such as the size of starspots growing, or the thickness of clouds decreasing ---
but could also be due to the growing 
or waning strength of a secondary variability mechanism compared 
to the primary mechanism.
It is therefore imperative to monitor ultra-cool dwarfs for multiple rotation cycles
spread over days, weeks, months and even years.

Precise long-term monitoring of ultra-cool dwarfs is also a perfect data-set to search for transiting exoplanets
that are as small or smaller than Earth-sized planets in the ``habitable zones'' of these dwarfs.
Ultra-cool dwarfs typically have stellar radii similar to that of Jupiter \citep{Charbrier00}, meaning that even 
Earth-sized planets produce $\sim$1\% transit depths.
For late M to early T dwarf spectral types the habitable zones stretch from periods of a day, up to a
few days (depending on the influence of various atmospheric
compositions and albedos, and the exact effects of tidal heating; \citealt{Bolmont11}; \citealt{BarnesHeller13}; \citealt{Zsom13}).
The number of small planets appears to increase with decreasing stellar effective temperature \citep{Howard12,DressingCharbonneau15}
-- 
trends that may continue into the substellar regime.
There are already examples of M-dwarfs 
orbited by multiple rocky planets in short orbits:
there are three rocky planets with periods less than two days orbiting an M4 dwarf \citep{Muirhead12},
and two rocky planets have been found to orbit an M8 dwarf with an additional rocky planet in a longer period orbit \citep{Gillon16}.
Perhaps most importantly, a habitable rocky planet around an ultra-cool dwarf would be more favourable
for follow-up than 
hosts with earlier spectral types; 
atmospheric spectral features of a planet around a brown dwarf should prove detectable with
a feasible two weeks of total {\it James Webb Space Telescope} observing time \citep{Belu13}.


 Here we attempt to characterize and determine the astrophysical cause of the variability for
2MASSW J0036159+182110; 
in Section \ref{SecIntroTwoMassZeroZeroThreeSix} we provide an overview of this intriguing L3.5 ultra-cool dwarf.
To determine the physical mechanism causing the variability of 
2MASSW J0036159+182110
in Section \ref{SecObs} we 
present \INSERTNIGHTS \ nights of photometry of 2MASSW J0036159+182110 spread out over $\sim$\INSERTSPREADDAYS \ days, 
from optical to near-infrared wavelengths (the 
Ks, H, J, I, z' and R-bands), including \INSERTNIGHTSSIMULTANEOUS \ nights of simultaneous multiwavelength photometry
in the optical and the near-infrared.
In Section \ref{SecAnalysis} we analyze our multiwavelength photometry and determine that the rotation period 
of this ultra-cool dwarf is approximately
\PeriodHoursTwoMassZeroZeroThirtySixAllbands \ $\pm$ \PeriodHoursErrorTwoMassZeroZeroThirtySixAllbands \ hr,
and this is consistent from wavelength to wavelength for our optical and near-infrared photometry and the previously detected
radio period,
allowing us to place a limit on differential rotation for 2MASSW J0036159+182110.
We also demonstrate a lack of significant phase offsets in our simultaneous multiwavelength variability,
and therefore starspots rotating in and out of view, or gaps in clouds that span multiple pressure layers probed by our observations,
are the most likely explanations for the observed variability
on 2MASSW J0036159+182110.
Our photometry does not reveal any significant flares, such as have been observed at radio wavelengths, and 
we are therefore
able to place a limit on the frequency of flaring for this ultra-cool dwarf (Section \ref{SecFlares}).
We also rule out the presence of transiting super-Earth sized planets in the habitable zone of this 
ultra-cool dwarf (Section \ref{SecPlanets}).
In Section \ref{SecDiscussion} we demonstrate that the decreasing amplitudes of variability that we observe
with increasing wavelength in our photometry are most consistent with starspots or gaps in clouds that are 
that are approximately $\sim$100 $K$ cooler 
or hotter 
than the 
photosphere of this ultra-cool dwarf.
In Section \ref{SecConclusions} we conclude that the variability we observe from 2MASS 0036+18
is likely driven by starspots, but there probably exists some complex interplay between starspots and the clouds
that appear to envelope this ultra-cool dwarf.


\section{The L3.5 ultra-cool dwarf 2MASS 0036+18}
\label{SecIntroTwoMassZeroZeroThreeSix}


The L3.5 ultra-cool dwarf 2MASSW J0036159+182110 (hereafter 2MASS 0036+18) has been
subject to intensive multiwavelength photometric observations: from the X-ray and radio \citep{Berger02,Berger05},
to the optical and near-infrared \citep{Gelino02,Lane07,Maiti07,Koen13,Harding13}, and also the infrared \citep{Metchev15}.
At several of these wavelengths 2MASS 0036+18 has been reported to be variable.
\citet{Berger05} detected periodic radio emission from 
2MASS 0036+18 with a period of $\sim$3 $hr$,
but failed to detect X-ray or H$\alpha$ emission from this dwarf.
\citet{Hallinan08} observed 2MASS 0036+18 in the radio for 12 hours and observed a periodicity of 
3.08 $\pm$ 0.05 hours that was attributed to rotation \& the electron cyclotron maser instability.
2MASS 0036+18 was announced as an L-dwarf by \citet{Kirkpatrick00},
and is believed to be at or below the hydrogen burning limit \citep{Hallinan08}.
\citet{Vrba04} inferred an effective temperature of $\sim$1900 K,
although a model fit to spectra of this ultra-cool dwarf suggest a slightly lower effective temperature
of $\sim$1700 K \citep{Cushing08}.

Optical and very near-infrared photometric monitoring has led to a confusing mixture of detections and upper-limits
on variability, possibly due to the impact of systematic errors in the optical and near-infrared: (i) \citet{Gelino02} found 2MASS 0036+18 to not 
be obviously variable in
the I-band above their precision of $\sim$1\%, 
(ii) \citet{Maiti07} reported that 2MASS 0036+18 was likely variable in several nights of quasi-simultaneous R and I-band observations,
and variable in R-band but not in I-band 
on one night; inspection of the \citet{Maiti07} light curves indicate that statistical and systematic errors in their photometry limit
the wider applicability of their conclusions,
(iii) \citet{Lane07} reported up to 5\% peak-to-peak I-band variability of 2MASS 0036+18 from a single night of photometric monitoring,
(iv) \citet{Koen13} reported 2MASS 0036+18 to be variable in I-band on two nights, and not variable in R-band on the 
second night that was contemporaneous with their I-band variability detection, 
(v) \citet{Harding13} found 2MASS 0036+18 to be variable at the 4\% level from two nights of I-band photometry, although these data were affected by 
heavy clouds.
Most recently, \citet{Metchev15} used Spitzer/IRAC \citep{Fazio04} to obtain 8 hours of
3.6 $\mu m$ photometry immediately followed by 6 hours of 4.5 $\mu m$ photometry 
of 2MASS 0036+18; the ultra-cool dwarf
displayed variability with a period of $\sim$2.7 $h$ that appeared to evolve from rotation period to rotation period
with peak-to-peak amplitudes of
0.47 $\pm$ 0.05\% at 3.6 $\mu m$ and
0.19 $\pm$ 0.04\% at 4.5 $\mu m$. 
Although, the smaller amplitude variability displayed
in the 4.5 $\mu m$ channel than the 3.6 $\mu m$ channel is consistent with what one
expects if the variability is due to cool starspots on 2MASS 0036+18, as the 
two channel photometry is not
simultaneous, and this ultra-cool dwarf appears to display irregular variability that rapidly evolves,
this 
conclusion is not definitive.


Furthermore, after a number of searches for H$\alpha$ that resulted in 
null-results (\citealt{Kirkpatrick00}; \citealt{Berger05}; \citealt{ReinersBasri08}), recently \citet{Pineda16}
detected H$\alpha$ emission from 2MASS 0036+18, suggesting the possibility
of intermittent H$\alpha$ emission from this ultra-cool dwarf.

Therefore the detections of variability of 2MASS 0036+18 to
date may be the result of a single physical process, or the result of more than one physical process occuring simultaneously or at different epochs in time.
The likely physical mechanisms leading to the observed variability include: 
starspots, aurorae, cloud variability, temperature fluctuations, etc. 
The radio period, apparent intermittent H$\alpha$ emission, and various optical and near-infrared variability detections to date,
suggest that 
2MASS 0036+18 is active and starspots are a likely physical mechanism driving the variability of this ultra-cool dwarf.
Here we present considerable additional ground-based observations of
2MASS 0036+18 to determine whether starspots, or another physical mechanism, is driving the observed variability of 
this ultra-cool dwarf.

\section{Observations}
\label{SecObs}

\begin{deluxetable*}{ccccccccc}
\tablecaption{Observing Log}
\tablehead{
\colhead{Date} 		& \colhead{Telescope}		& \colhead{Band}	& \colhead{Duration\tablenotemark{a}}	& \colhead{Exposure} 	& \colhead{Overhead\tablenotemark{b}}	& \colhead{Airmass}				& \colhead{Conditions}	& \colhead{Aperture \tablenotemark{c}}\\
\colhead{(UTC)}		& \colhead{\& Instrument}	& \colhead{}		& \colhead{(hours)}			& \colhead{Time (sec)}	& \colhead{(sec)}			& \colhead{}					& \colhead{}		& \colhead{(pixels)}\\
}
\centering
\startdata
2015/08/20			& Perkins/MIMIR			& J			& 6.32 					& 15			& 3	& 2.04 $\rightarrow$ 1.04 $\rightarrow$ 1.15	& Clear				& 5.5, 15, 25  \\	%
2015/08/22			& Perkins/MIMIR			& J			& 4.05 					& 15			& 3	& 1.20 $\rightarrow$ 1.04 $\rightarrow$ 1.15	& Clear				& 6, 15, 25  \\	%
2015/08/23			& Perkins/MIMIR			& J			& 4.63 					& 15			& 3	& 1.36 $\rightarrow$ 1.04 $\rightarrow$ 1.13	& Clear				& 6, 15, 25  \\	%
2015/08/29			& Perkins/MIMIR			& J			& 2.94 					& 15			& 3	& 1.04 $\rightarrow$ 1.33 			& Thin clouds at the start	& 6, 15, 25  \\	%
2015/08/30			& Perkins/MIMIR			& J			& 6.93 					& 10, 15		& 3	& 1.66 $\rightarrow$ 1.04 $\rightarrow$ 1.41	& Occasional clouds		& 5, 15, 25  \\	%
2015/09/01			& Perkins/MIMIR			& J			& 6.08 					& 15, 20		& 3	& 1.80 $\rightarrow$ 1.04 $\rightarrow$ 1.10	& Occasional thick clouds	& 5, 15, 25  \\	%
2015/09/02			& Perkins/MIMIR			& J			& 4.61 					& 15, 20		& 3	& 1.12 $\rightarrow$ 1.04 $\rightarrow$ 1.38	& Occasional clouds		& 6, 15, 25  \\	%
2015/09/03			& Perkins/MIMIR			& J			& 5.68 					& 20			& 3	& 1.24 $\rightarrow$ 1.04 $\rightarrow$ 1.45	& Occasional clouds		& 5, 15, 25 \\	%
2015/09/23			& Perkins/PRISM			& I			& 5.91 					& 20			& 18	& 1.12 $\rightarrow$ 1.04 $\rightarrow$ 1.99	& Clear				& 7, 15, 27  \\	%
2015/09/23			& Hall/NASA42			& I			& 4.71 					& 180			& 9	& 1.05 $\rightarrow$ 1.04 $\rightarrow$ 2.06	& Clear				& 4.5, 14, 26  \\	%
2015/09/24			& Perkins/PRISM			& z'			& 5.38 					& 40			& 18	& 1.11 $\rightarrow$ 1.04 $\rightarrow$ 1.68	& Clear				& 7, 15, 27  \\	%
2015/09/24			& Hall/NASA42			& I			& 5.78 					& 180			& 9	& 1.09 $\rightarrow$ 1.04 $\rightarrow$ 2.10	& Clear				& 4.5, 14, 26  \\	%
2015/09/25			& Perkins/PRISM			& R			& 8.91 					& 150			& 18	& 1.89 $\rightarrow$ 1.04 $\rightarrow$ 2.27	& Clear				& 6, 15, 27  \\	%
2015/09/25			& Hall/NASA42			& z'			& 8.64 					& 100			& 9	& 1.67 $\rightarrow$ 1.04 $\rightarrow$ 2.38	& Clear				& 4.5, 14, 26  \\	%
2015/09/26			& Perkins/PRISM			& R			& 9.01 					& 150			& 18	& 1.95 $\rightarrow$ 1.04 $\rightarrow$ 2.27	& Clear				& 6, 15, 27  \\	%
2015/09/26\tablenotemark{d}	& Hall/NASA42			& z'			& 8.56 					& 100			& 9	& 1.64 $\rightarrow$ 1.04 $\rightarrow$ 2.38	& Clear				& 4.5, 14, 26  \\	%
2015/09/29			& Perkins/PRISM			& z'			& 9.41 					& 50			& 18	& 2.25 $\rightarrow$ 1.04 $\rightarrow$ 2.30	& Thin clouds			& 7, 15, 27  \\	
2015/10/14			& DCT/LMI			& z' \& R		& 8.49					& 20, 60		& 8.5	& 1.63 $\rightarrow$ 1.04 $\rightarrow$	2.31 	& Clear				& 8, 20, 30  \\
2015/11/07			& Perkins/MIMIR			& J			& 3.52 					& 20			& 3	& 1.06 $\rightarrow$ 1.90 			& Clear				& 5, 15, 25  \\	%
2015/11/08			& Perkins/MIMIR			& J			& 7.97 					& 15, 20		& 3	& 1.46 $\rightarrow$ 1.04 $\rightarrow$ 2.26 	& Clear				& 7, 15, 25  \\	%
2015/11/08			& Hall/NASA42			& I			& 7.09 					& 180			& 9	& 1.29 $\rightarrow$ 1.04 $\rightarrow$ 2.03	& Clear				& 4.5, 14, 26  \\	%
2015/11/09			& Perkins/MIMIR			& H			& 7.73 					& 5,6,7			& 3	& 1.40 $\rightarrow$ 1.04 $\rightarrow$ 2.20 	& Clear				& 5.5, 15, 25  \\	%
2015/11/09			& Hall/NASA42			& I			& 8.23 					& 180			& 9	& 1.55 $\rightarrow$ 1.04 $\rightarrow$ 2.26	& Clear				& 4.5, 14, 26  \\	%
2015/11/11			& Perkins/MIMIR			& Ks			& 4.31 					& 20 			& 3	& 1.16 $\rightarrow$ 1.04 $\rightarrow$ 1.24	& Clear				& 10, 15, 25  \\	%
2015/11/23			& Perkins/MIMIR			& J			& 7.50 					& 10 			& 3	& 1.37 $\rightarrow$ 1.04 $\rightarrow$ 2.09	& Clear				& 4.5, 15, 25  \\	%
2015/11/24			& Perkins/MIMIR			& J			& 7.71 					& 10 			& 3	& 1.36 $\rightarrow$ 1.04 $\rightarrow$ 2.33	& Clear				& 6, 15, 25  \\	%
2015/11/26			& Perkins/MIMIR			& J			& 2.75 					& 20 			& 3	& 1.05 $\rightarrow$ 1.04 $\rightarrow$ 1.21	& Steady thin clouds		& 6, 15, 25  \\	%
2015/12/16			& Hall/NASA42			& z'			& 4.96					& 180			& 9	& 1.09 $\rightarrow$ 1.04 $\rightarrow$	1.64	& Clear				& 4.5, 14, 26	\\
2015/12/18			& Perkins/MIMIR			& J			& 6.50					& 15 			& 3	& 1.09 $\rightarrow$ 1.04 $\rightarrow$ 2.96	& Clear				& 6, 15, 25  \\	%
\enddata
\tablenotetext{a}{The duration indicates the time between the first and last observation of the evening, and does not take into account gaps in the data due to clouds, humidity or poor weather.}
\tablenotetext{b}{The overhead includes time for read-out, and any other applicable overheads.}
\tablenotetext{c}{We give the radius of the aperture, the radius of the inner annulus and the radius of the outer annulus that we use for sky subtraction in pixels.}
\tablenotetext{d}{For this data fringes were removed using the technique discussed in Section \ref{SecObs}.}
\label{TableObs}
\end{deluxetable*}

\begin{figure*}
\includegraphics[scale=0.50, angle = 270]{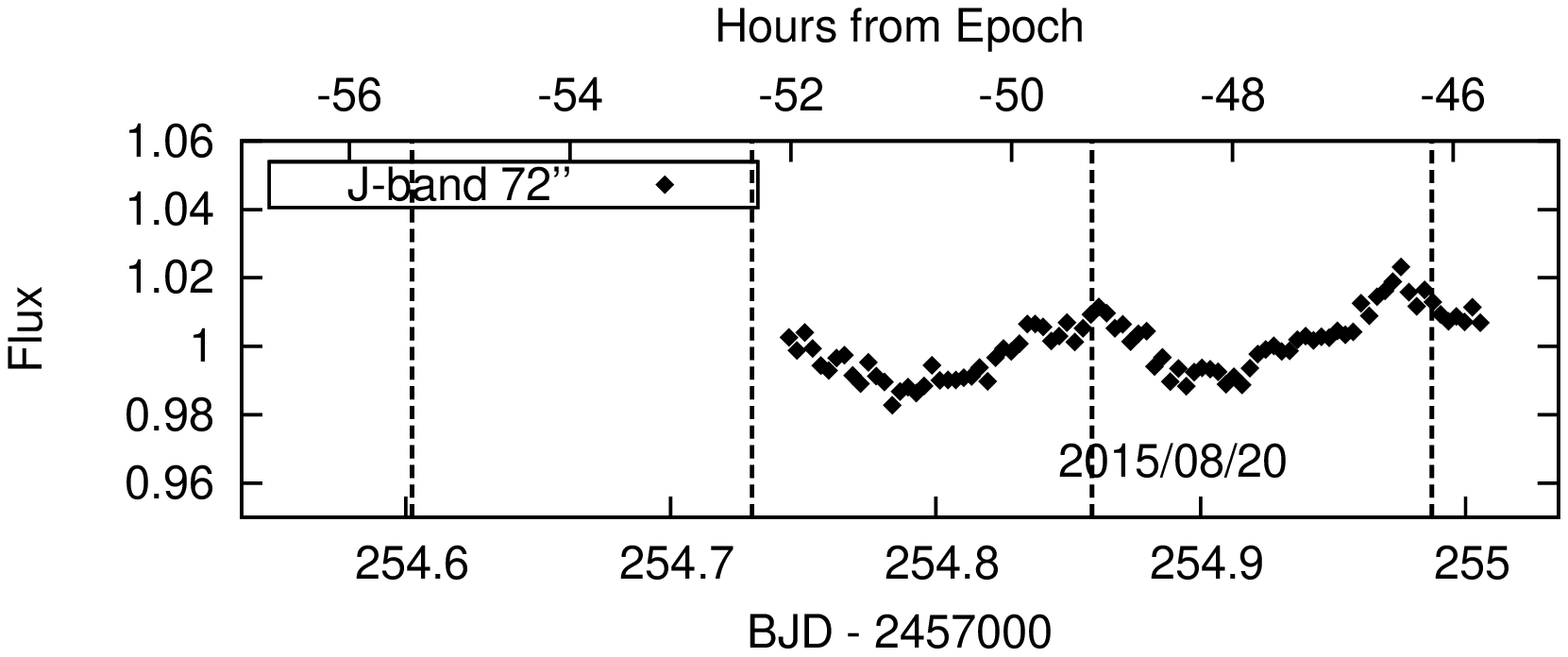}
\includegraphics[scale=0.50, angle = 270]{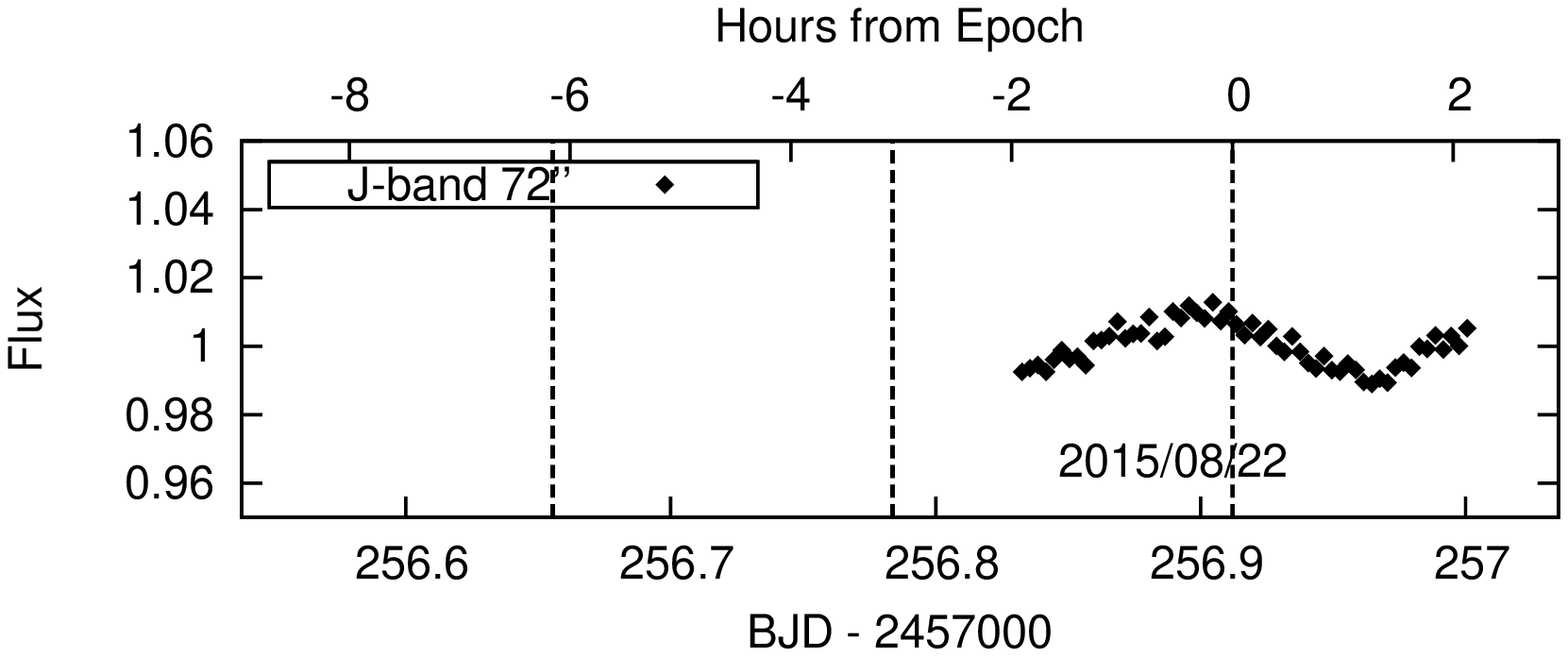}
\includegraphics[scale=0.50, angle = 270]{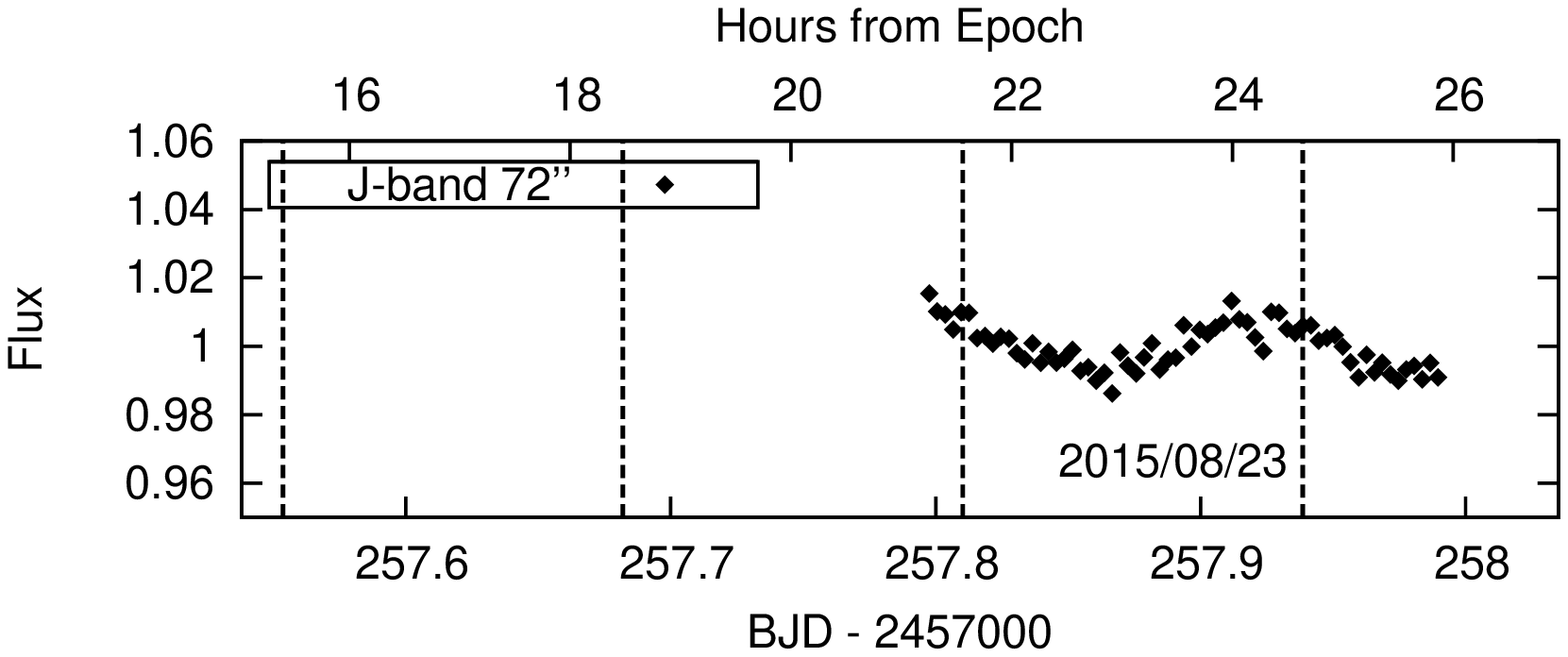}
\includegraphics[scale=0.50, angle = 270]{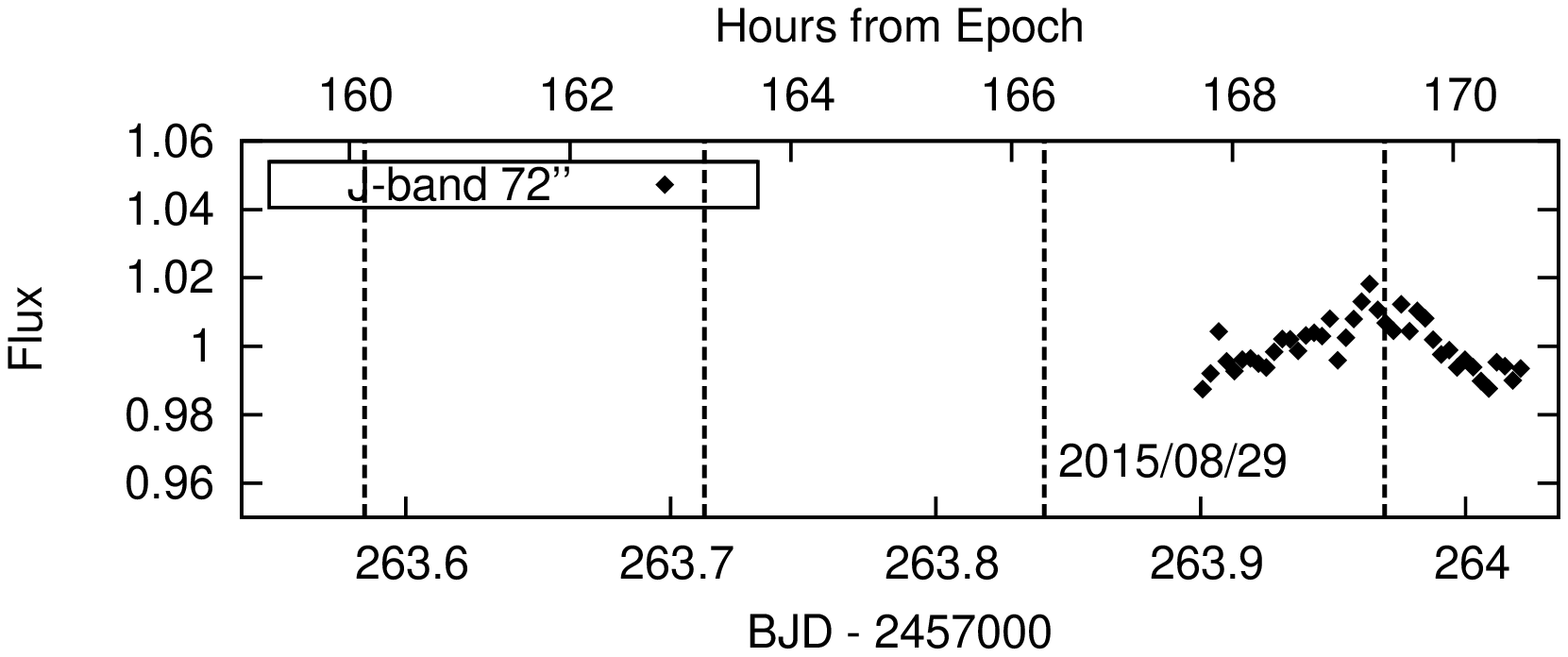}
\includegraphics[scale=0.50, angle = 270]{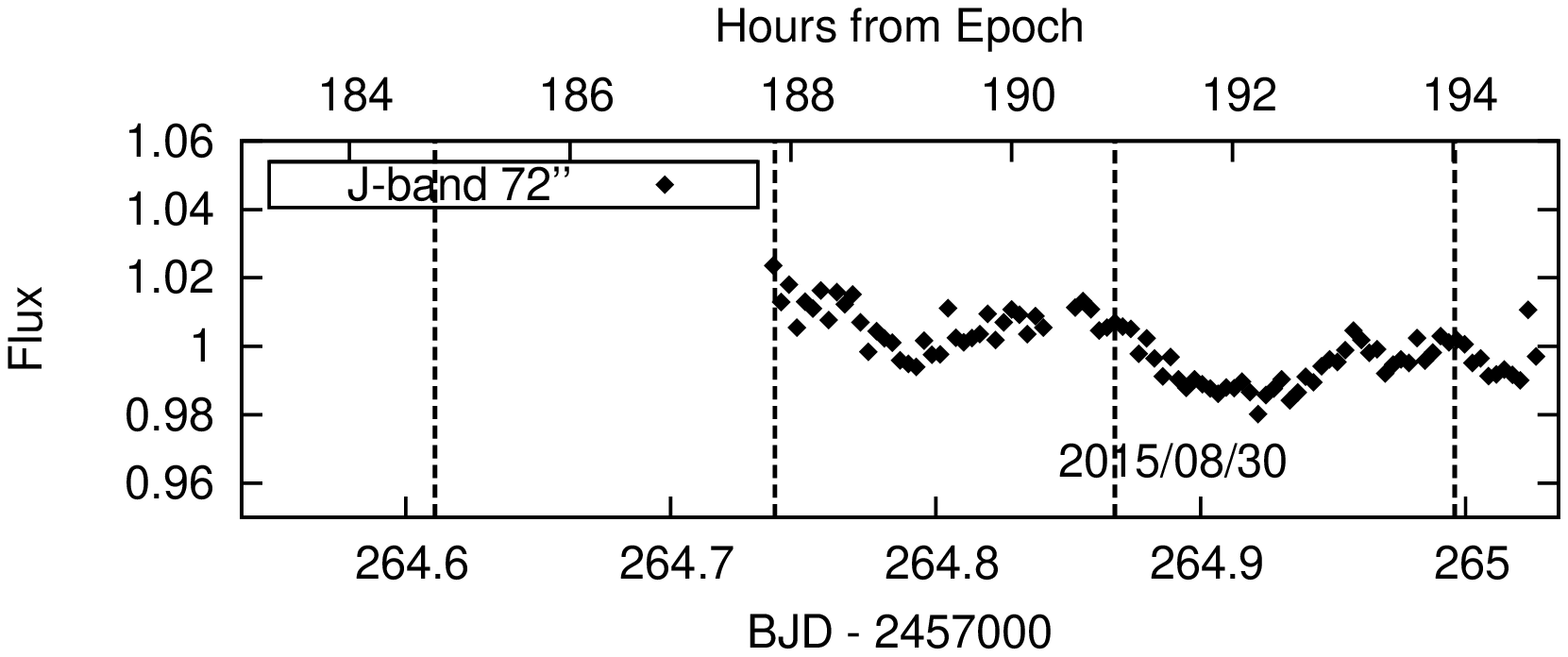}
\includegraphics[scale=0.50, angle = 270]{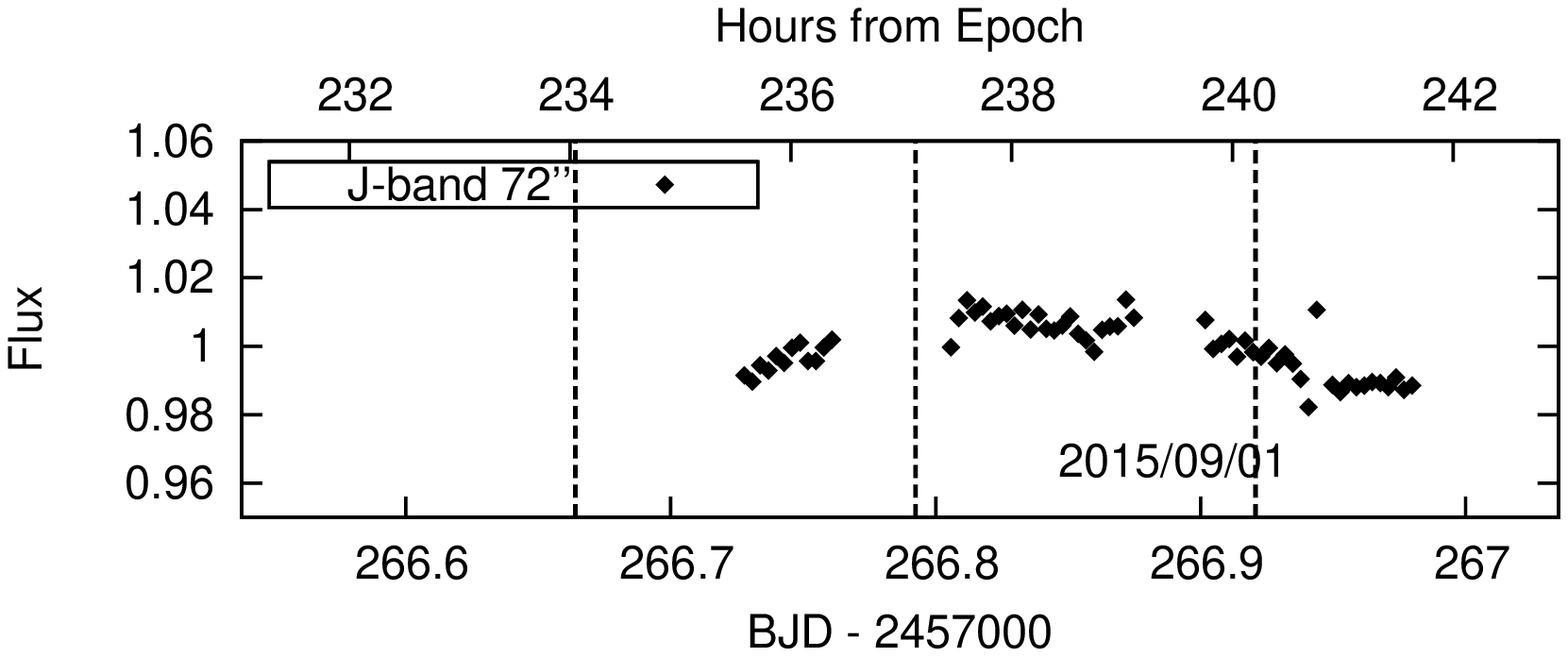}
\includegraphics[scale=0.50, angle = 270]{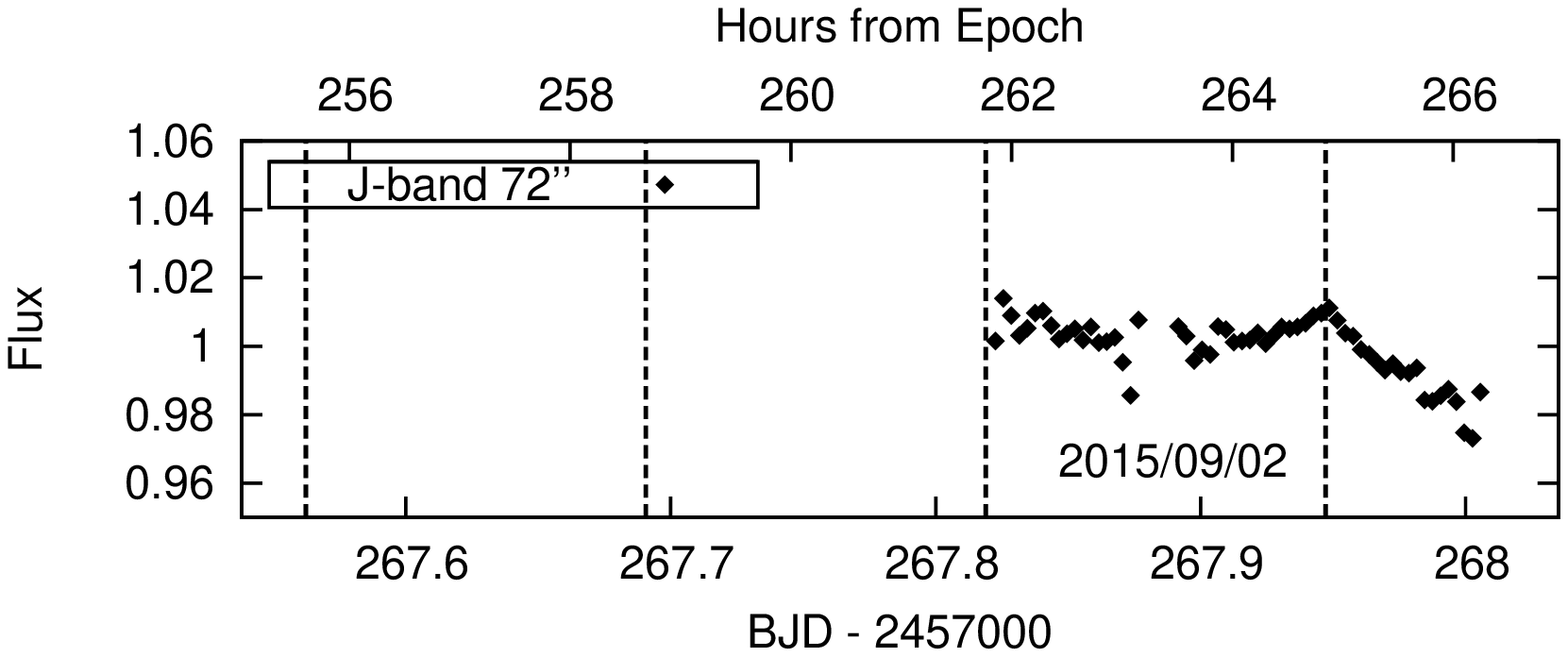}
\includegraphics[scale=0.50, angle = 270]{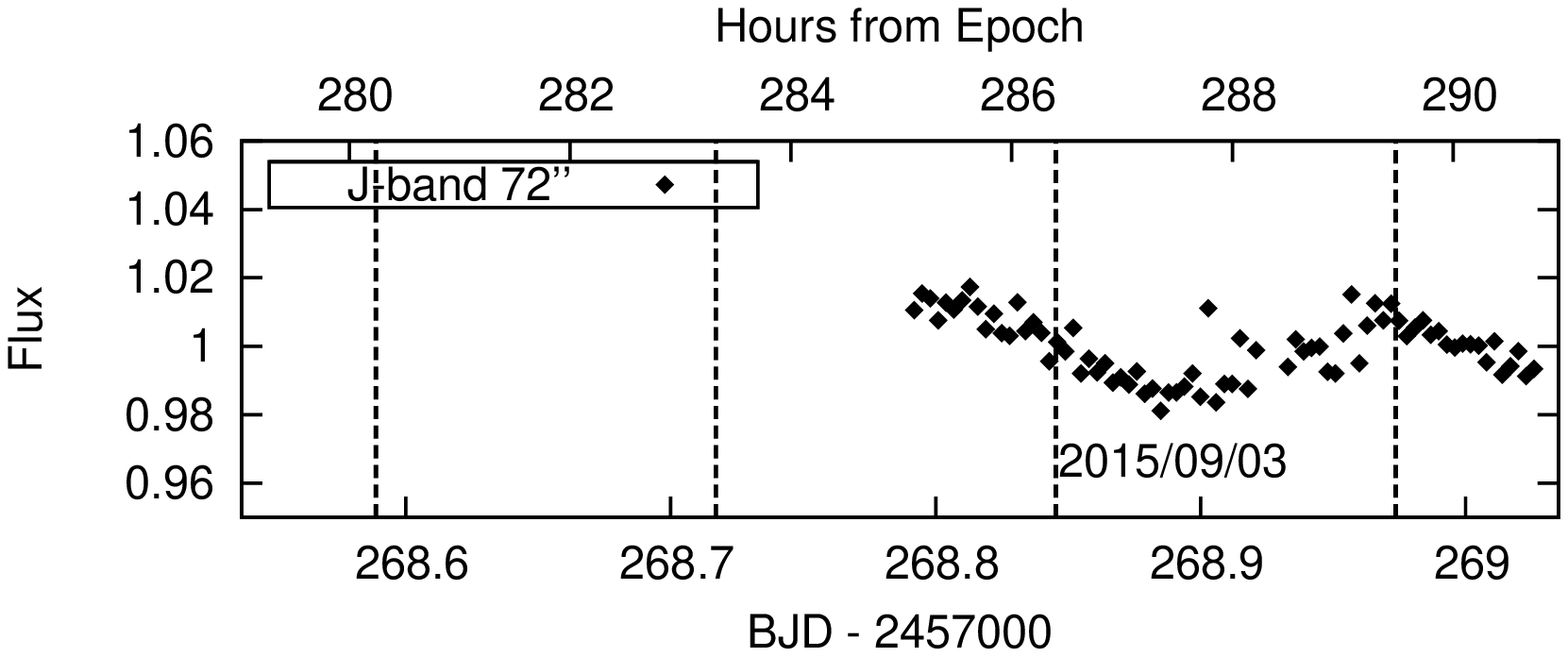}
\includegraphics[scale=0.50, angle = 270]{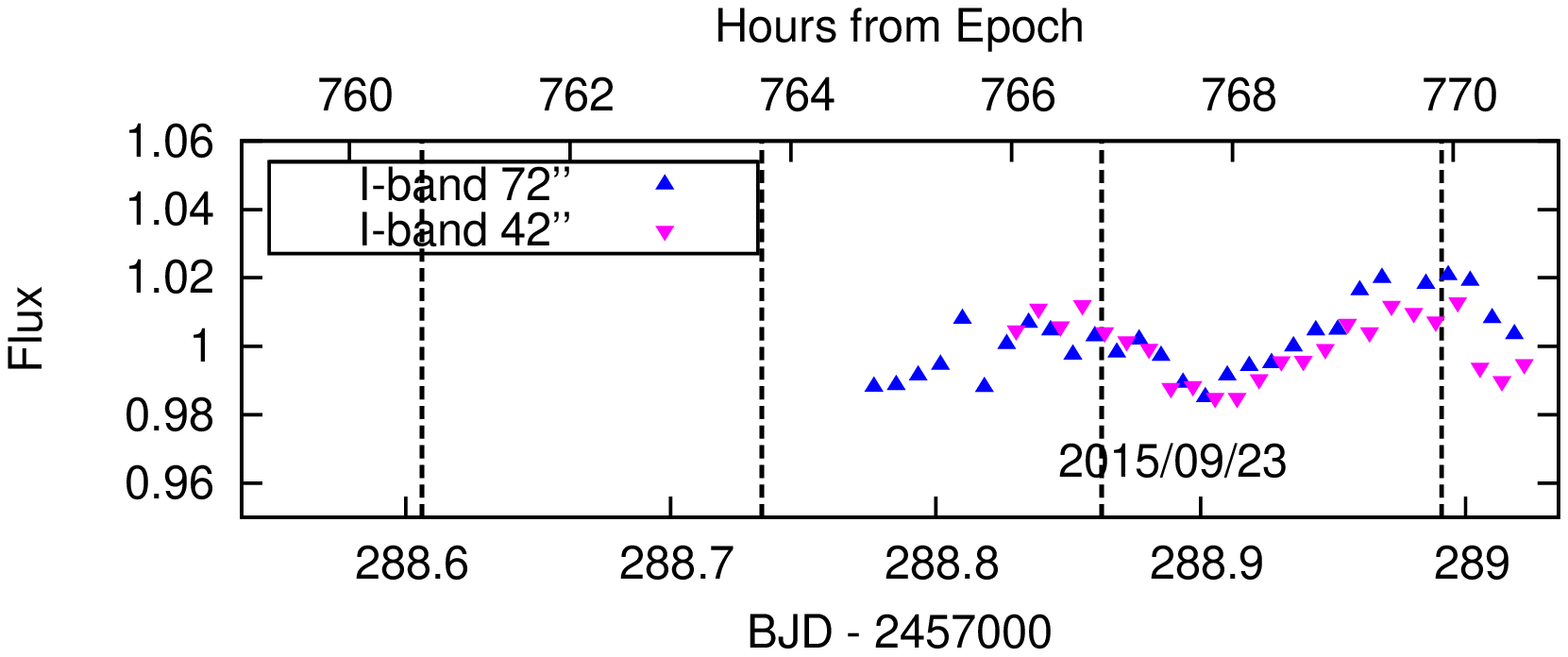}
\includegraphics[scale=0.50, angle = 270]{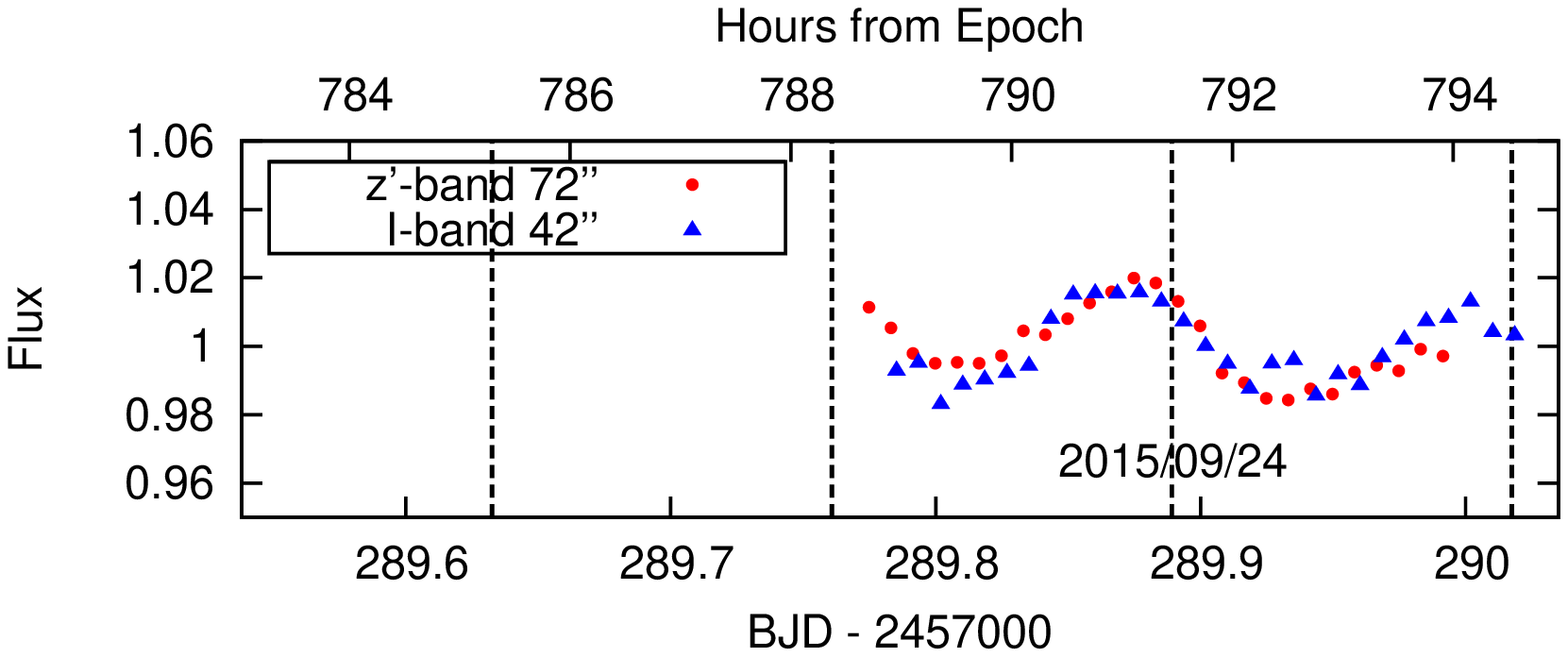}
\caption[]
	{	Photometry of 2MASS 0036+18 on the dates given in the lower-right of each panel (UT)
		with the Perkins 1.8-m (72'') and Hall 1.1-m (42'') telescopes
		at various wavelengths (from the I to the J-band) as indicated in the legend 
		of each panel.
		The vertical dashed lines indicate 
		cycles of the apparent $\sim$\PeriodHoursTwoMassZeroZeroThirtySixAllbandsQuote \ hour rotation period of 2MASS J0036+18,
		compared
		to the apparent flux maximum in our J-band photometry at a
		barycentric julian date (BJD) of: BJD-2457000 $\sim$256.91.
		For clarity we plot our photometry binned every 0.003 $d$ ($\sim$4.3 minutes) for panels featuring 
		single wavelength photometry,
		and binned every 0.008 $d$ ($\sim$12 minutes) for panels featuring multiwavelength photometry.
	}
\label{FigMassJ0036}
\end{figure*}


\begin{figure*}

\includegraphics[scale=0.50, angle = 270]{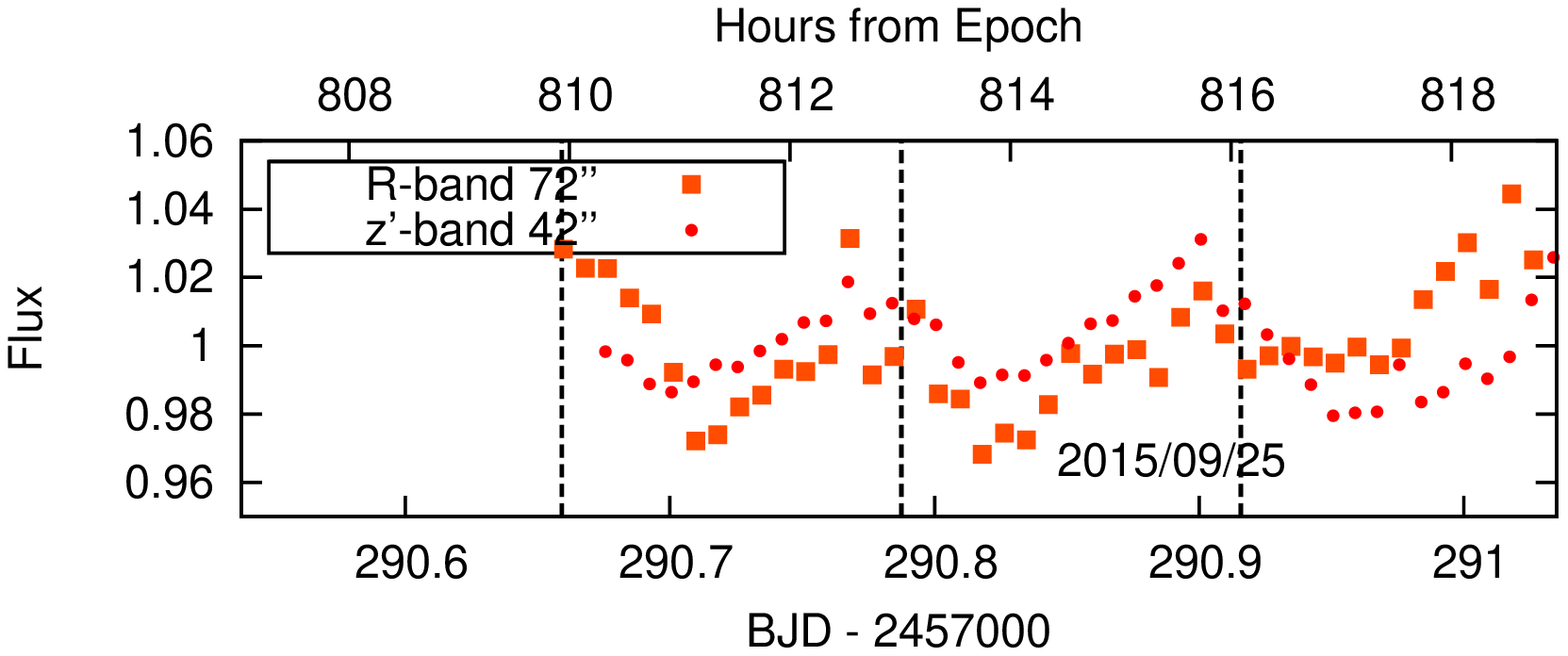}
\includegraphics[scale=0.50, angle = 270]{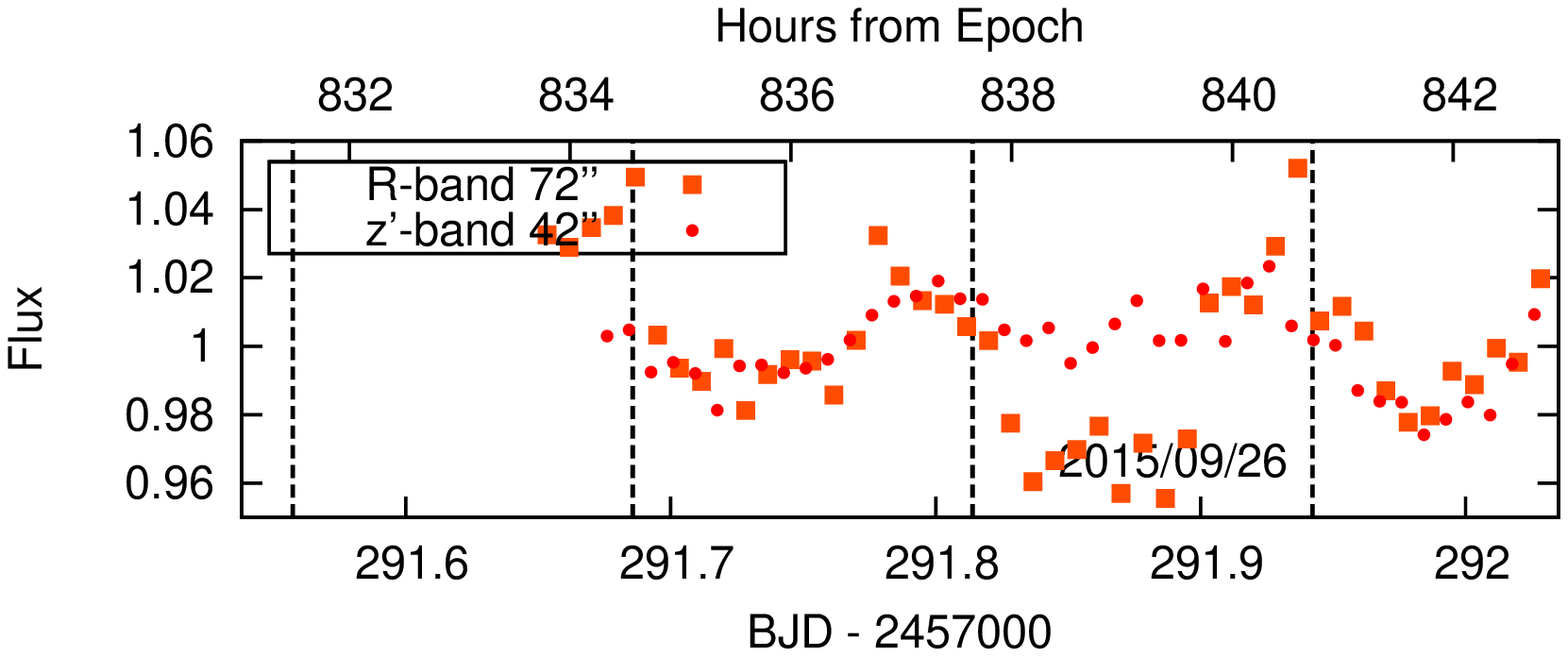}
\includegraphics[scale=0.50, angle = 270]{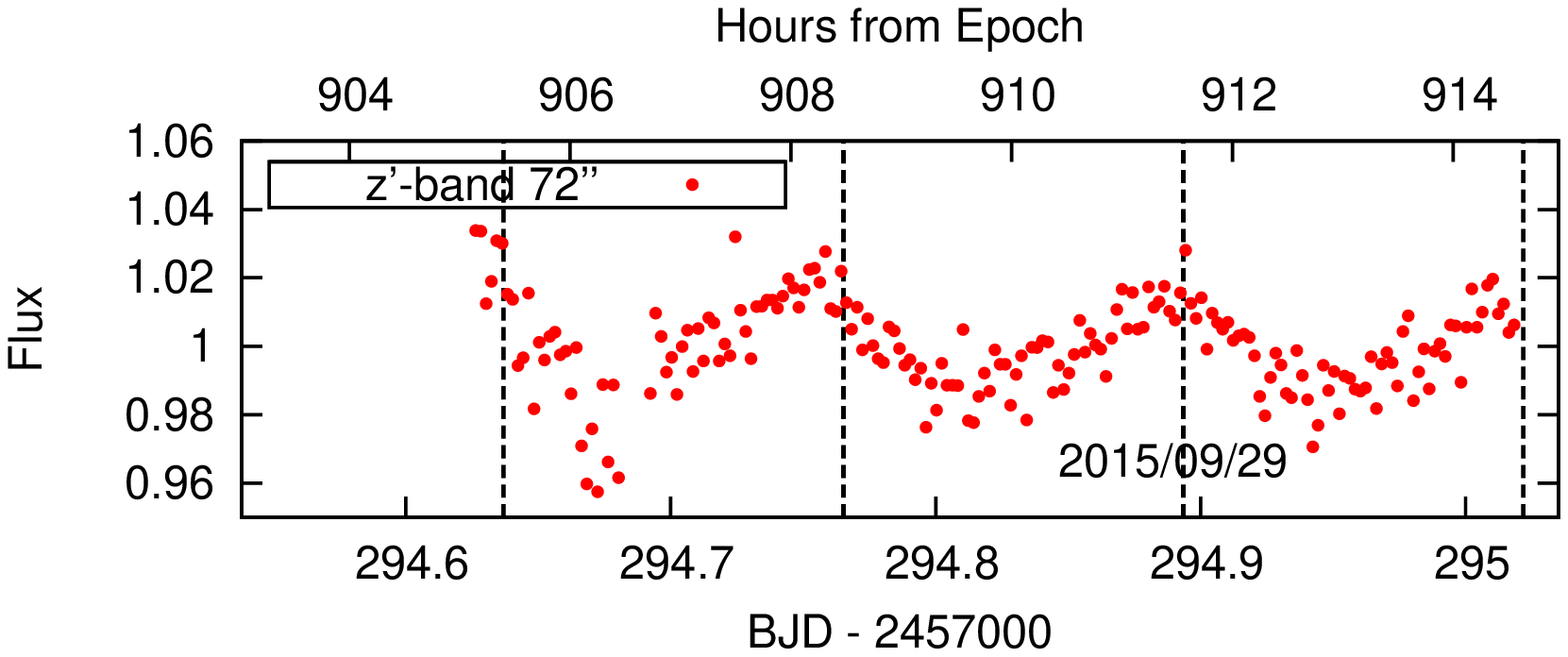}
\includegraphics[scale=0.50, angle = 270]{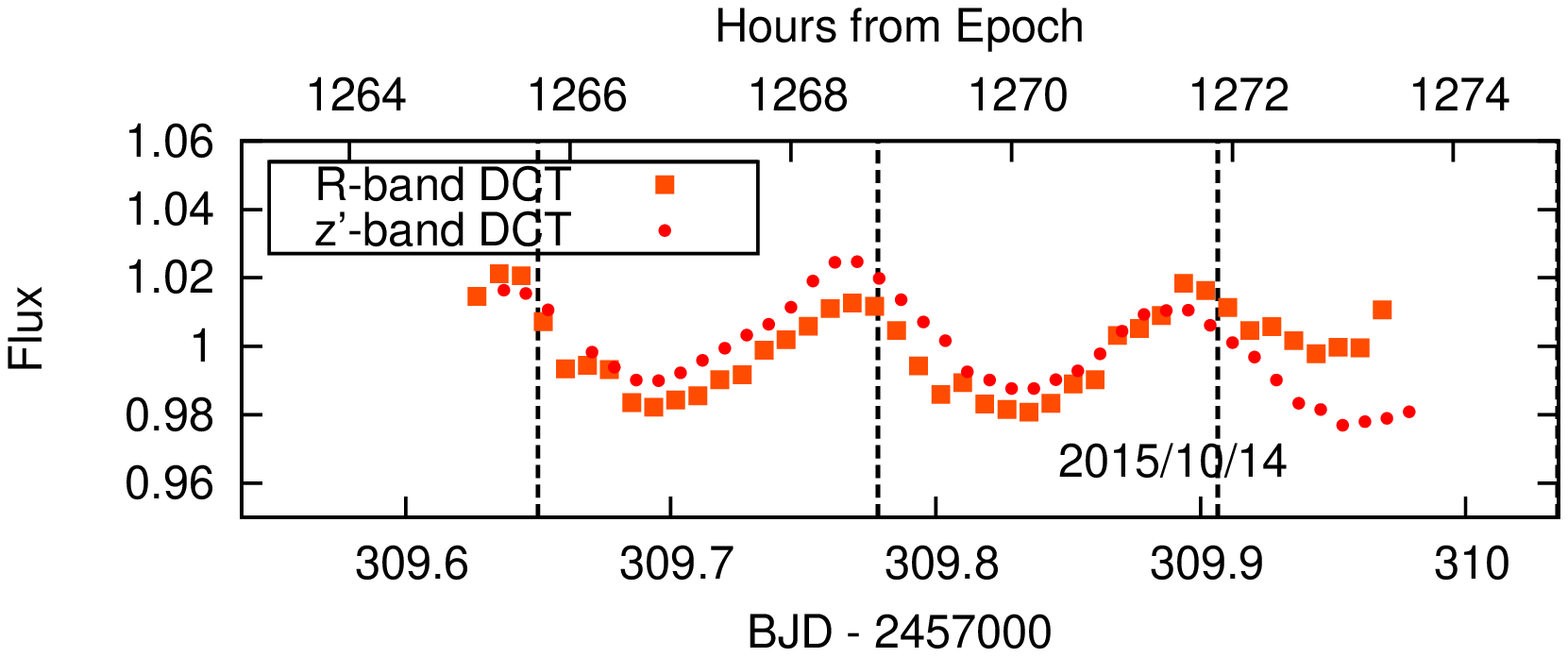}
\includegraphics[scale=0.50, angle = 270]{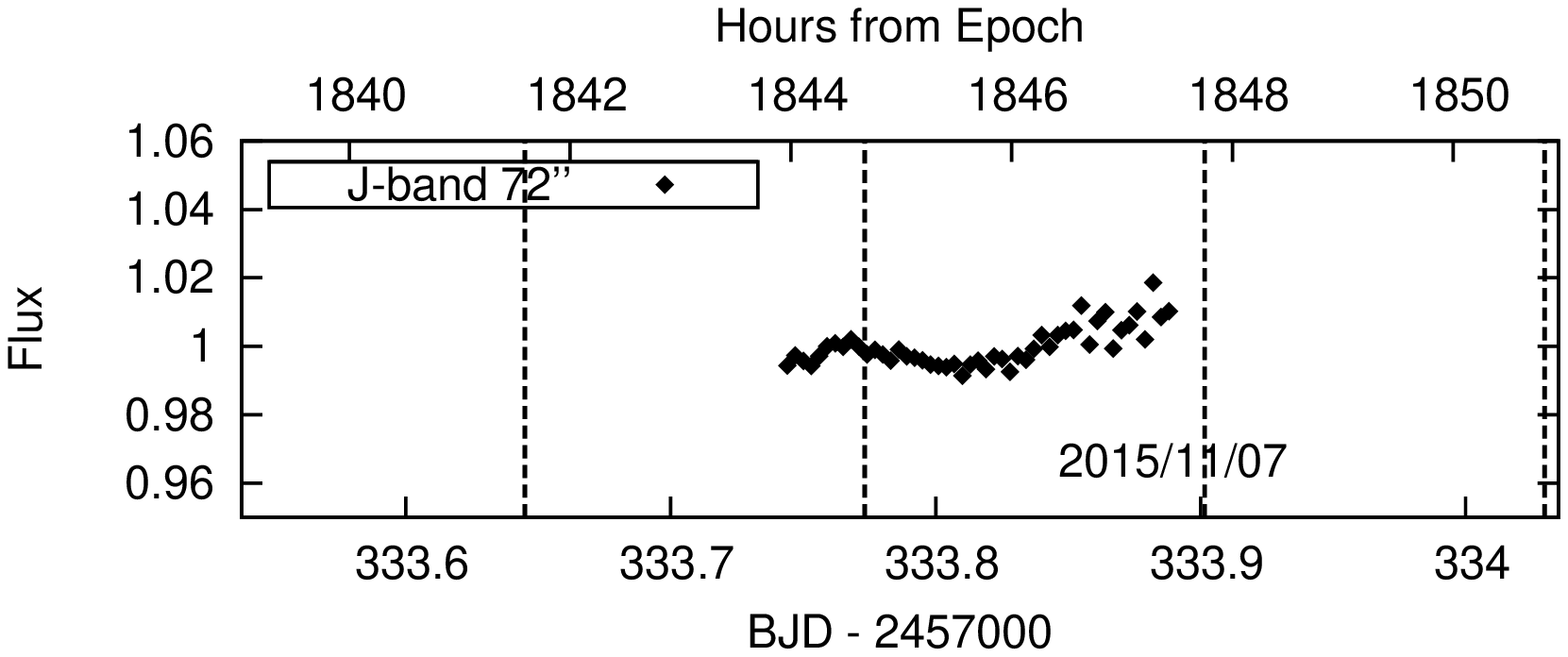}
\includegraphics[scale=0.50, angle = 270]{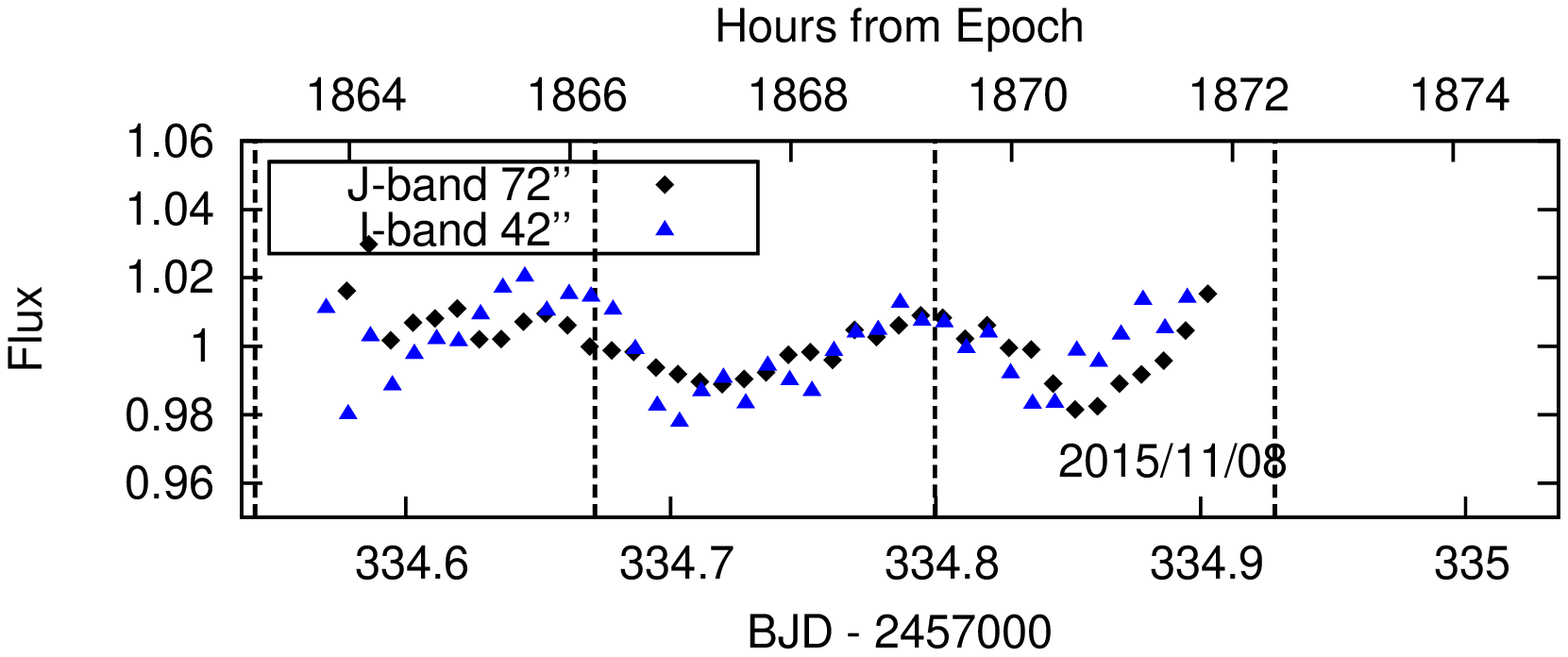}
\includegraphics[scale=0.50, angle = 270]{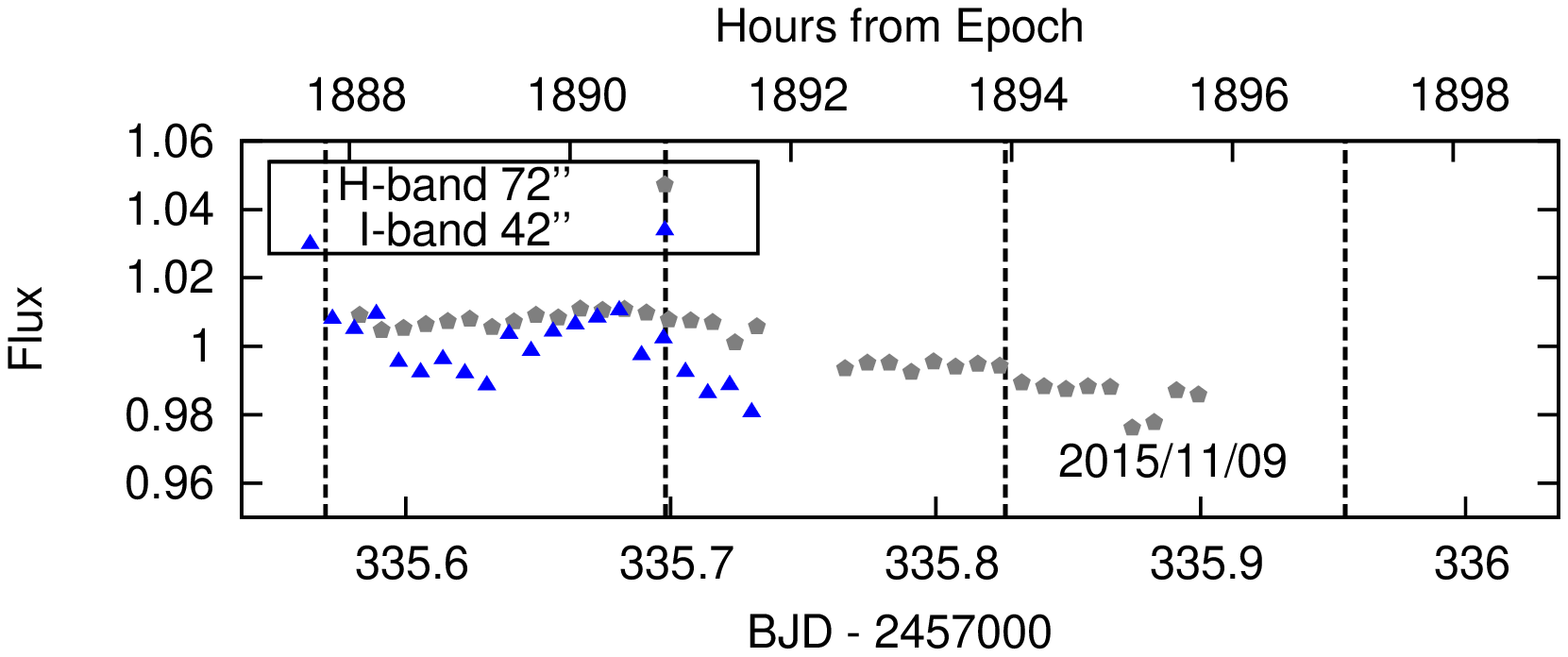}
\includegraphics[scale=0.50, angle = 270]{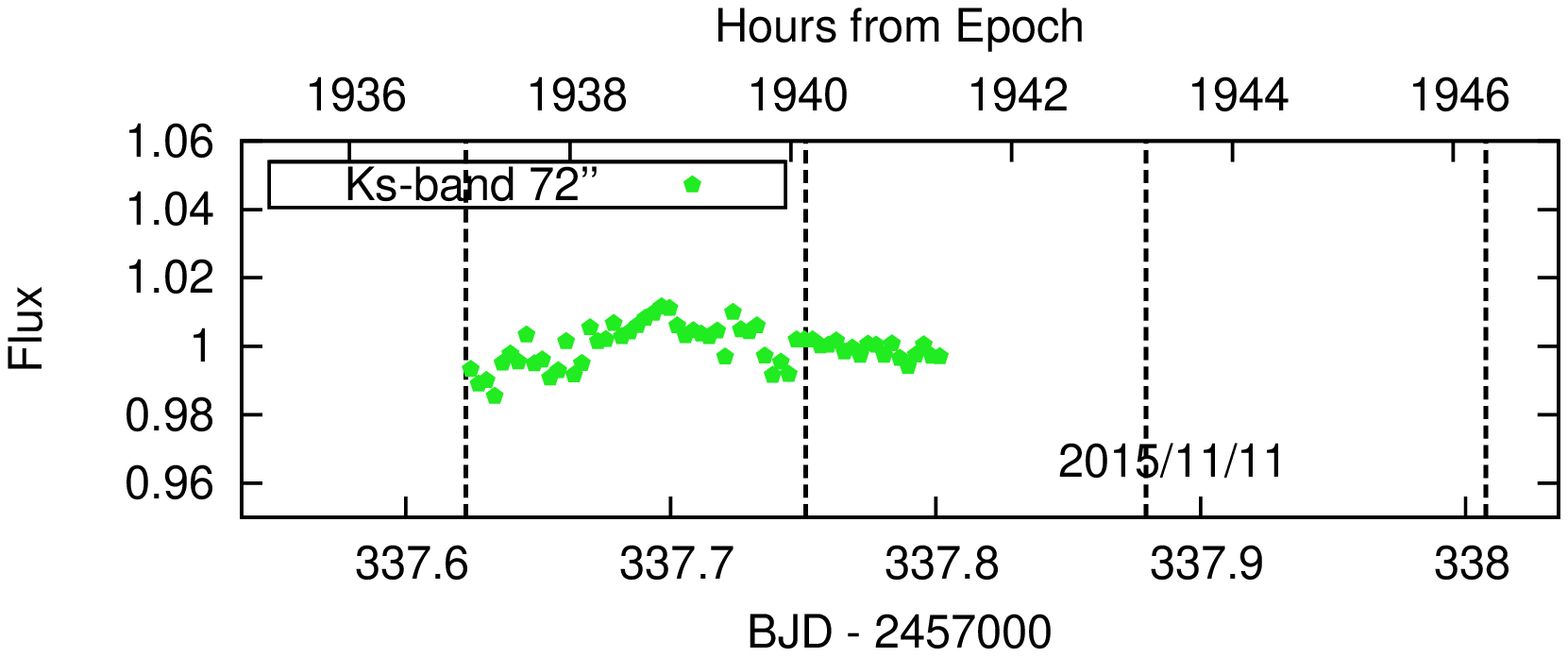}
\includegraphics[scale=0.50, angle = 270]{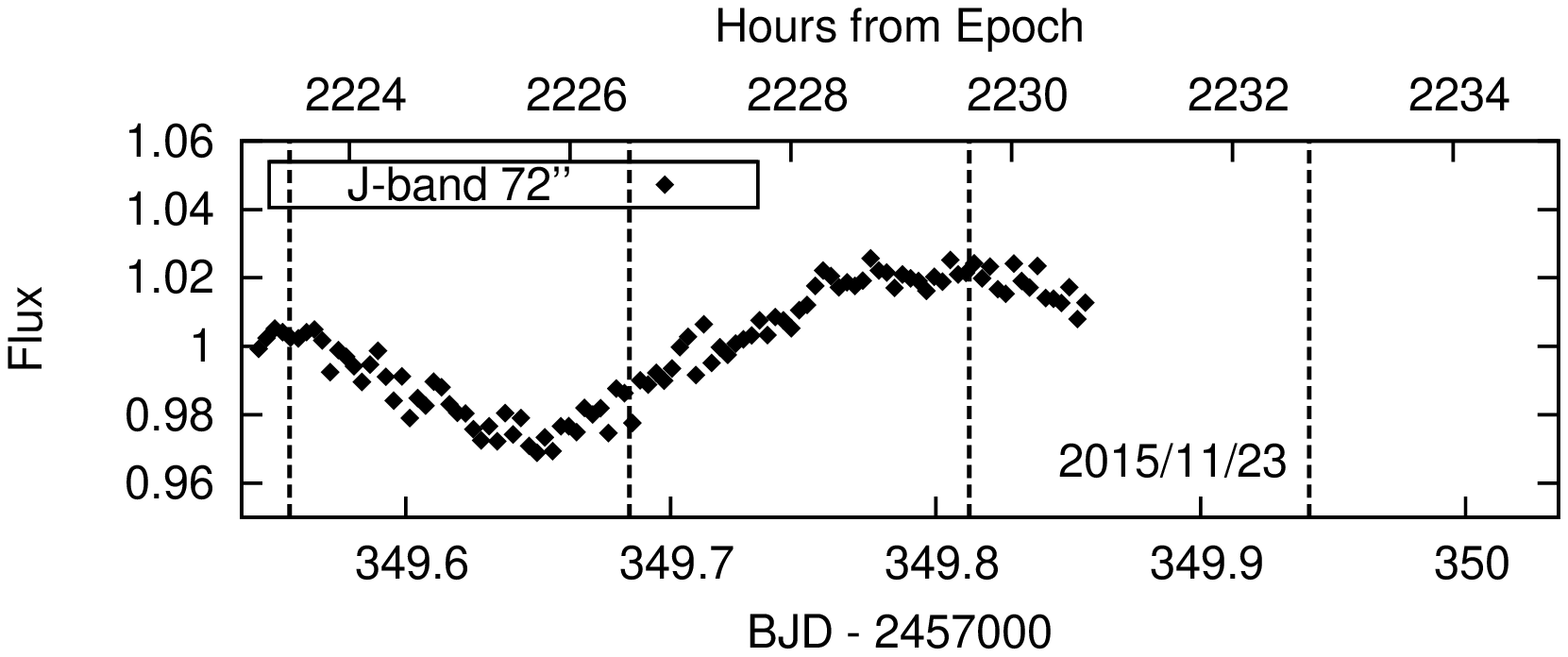}
\includegraphics[scale=0.50, angle = 270]{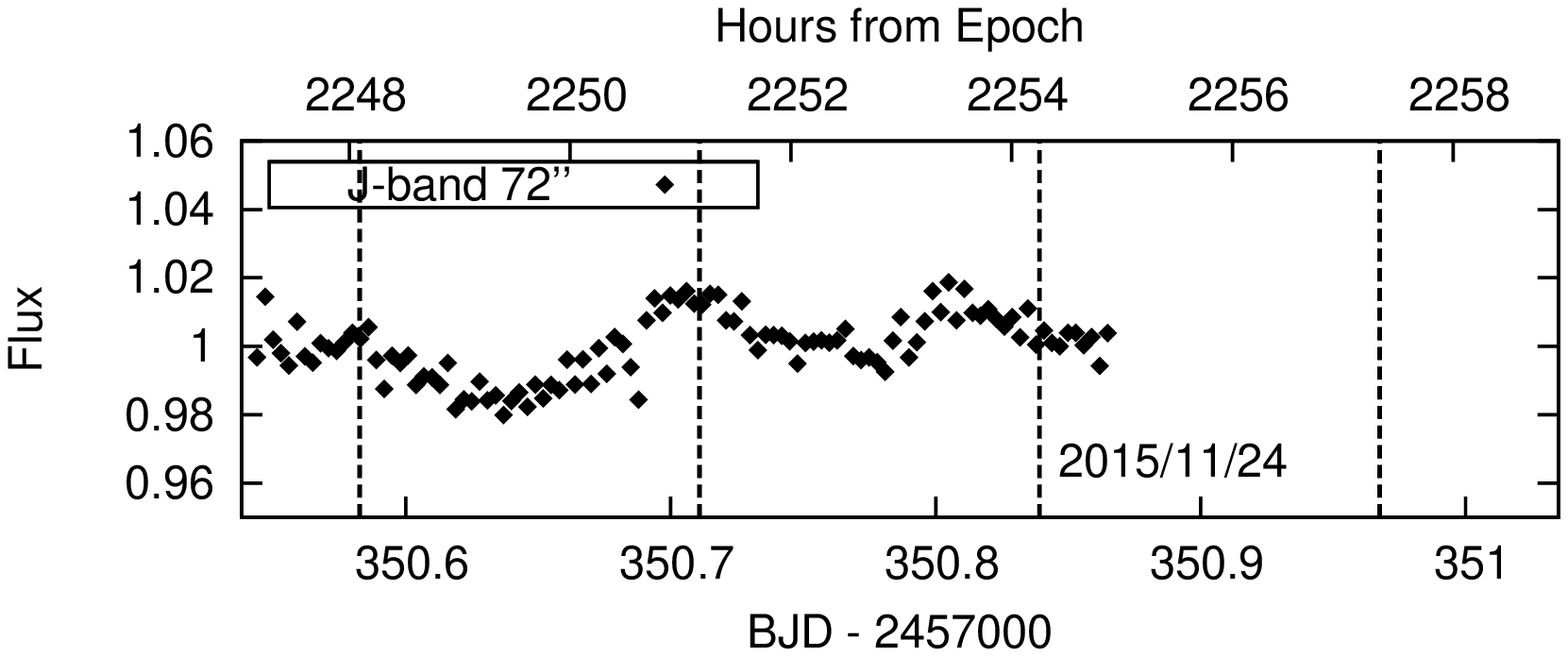}
\includegraphics[scale=0.50, angle = 270]{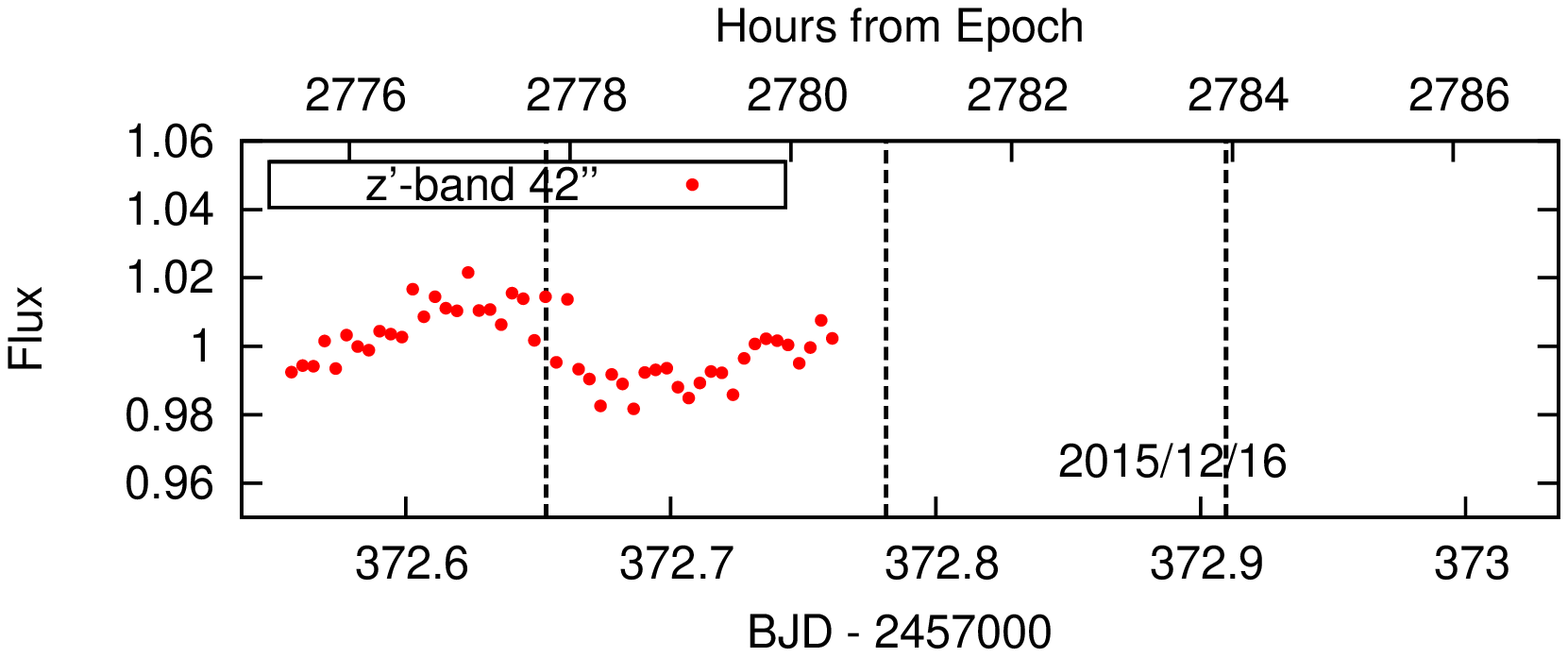}
\includegraphics[scale=0.50, angle = 270]{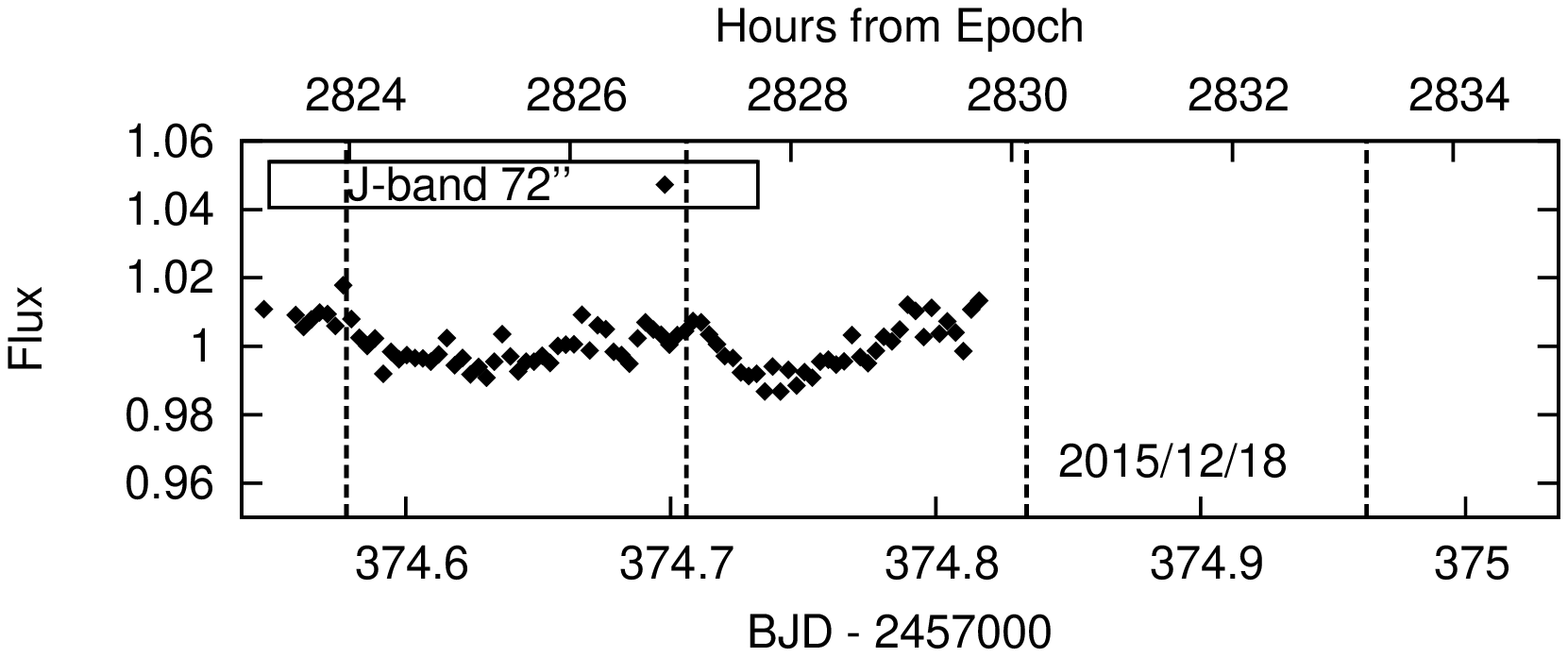}

\caption[]
	{	Photometry of 2MASS J0036+18 on the dates given in the lower-right of each panel (UT)
		with the Discovery Channel 4.3-m, Perkins 1.8-m (72'') and Hall 1.1-m (42'') telescopes 
		at various wavelengths (from the R to the Ks-band) as indicated in the legend 
		of each panel.
		The figure format is otherwise identical to Figure \ref{FigMassJ0036}.
	}
\label{FigMassJ0036PanelTwo}
\end{figure*}

We observed the ultra-cool dwarf 2MASS 0036+18
on  \INSERTNIGHTS \ nights
spread out over $\sim$\INSERTSPREADDAYS \ days.
Observations were conducted using the
Perkins 1.8-m telescope, the Hall 1.1-m telescope,
and the
4.3-m Discovery Channel Telescope (DCT).
On the Perkins telescope our photometry was obtained
using either the PRISM \citep{Janes04} instrument in the optical
and very near-infrared, or
the Mimir instrument \citep{Clemens07} in the near-infrared.
On the Hall telescope we used the NASA42 imager, 
and on the DCT we used
the Large Monolithic Imager (LMI; \citealt{Massey13}).
We summarize our observations in Table \ref{TableObs}.

Whenever possible we attempted to observe the same target with two different telescopes
at the same time at different wavelengths.
For our 2015 October 14 (UTC) observations of 2MASS 0036+18 we achieved multiwavelength observations
of this target by switching the DCT filter wheel back and forth
between the z' and R-band filters. Six 60-second exposures were conducted in R-band, followed by thirteen 20-second in z'-band
and then the process was repeated. In addition to the $\sim$8.5 $s$
overhead for reading out the chip, there was an additional overhead for switching the filter wheel that was usually $\sim$30 $s$.
For our DCT/LMI observations we utilized 4-amplifier read-out and 2x2 pixel binning to improve the duty-cycle.

Our red-optical and very near-infrared data using the Hall/NASA42 CCD
feature significant fringing, due to thin-film interference patterns from the thin CCD and
atmospheric emission lines. To eliminate this obvious fringing we produced night-time fringe frames using
the same band of observation and subtract the fringes from each individual
frame using the technique detailed by \citet{SnodgrassCarry13}.
We noticed that this technique only improved the photometry in a handful of cases when there was significant movement
of the point-spread-function of the target and reference stars due to poor telescope tracking.
We therefore only apply this technique to a few Hall/NASA42 data-sets that we denote in Table \ref{TableObs}.

The DCT/LMI, Hall/NASA42, and Perkins/PRISM data are processed by bias subtracting the data, while
the Perkins/Mimir data are dark-subtracted.
For our
DCT/LMI \& Hall/NASA42 
we use a twilight flat to sky-flat the data,
while for our Perkins/Mimir data we use a dome flat.

We do not perform a non-linearity correction on our Perkins/Mimir data. We attempted
to utilize the non-linearity correction employed by \citet{Clemens07}, but noticed this method simply added noise
to our Perkins/Mimir photometric data-sets.
The peak pixel value in our target and reference stars for our
Perkins/Mimir photometry is often displayed by 2MASS J00360421+1819309,
a star we frequently use as a
reference star.
Although the peak pixel value of 2MASS J00360421+1819309 often reaches ADU values ($\sim$8000)
where non-linearity correction is likely important \citep{Clemens07}, the vast majority of pixels we
utilize in our apertures are at much more modest illumination values. Also as 
we employ differential photometry  the impact of an incorrect non-linearity correction is mitigated; for instance, our 
SIMP J013656.5+093347 \& TVLM 513-46546 Perkins/Mimir photometry \citep{CrollUCDII}
displays very similar light curves utilizing the \citet{Clemens07} non-linearity correction
or no non-linearity correction whatsoever.



Aperture photometry is then performed on all Perkins, Hall and DCT data-sets using
the techniques discussed in \citet{Croll15} and references therein.
The size of the aperture radii, and the inner and outer radii of the annuli we use for sky subtraction
are indicated in Table \ref{TableObs}.
Due to the obvious variability of 2MASS 0036+18, it is difficult to determine the uncertainties on our differential
photometry from the photometry of 2MASS 0036+18 alone; instead the uncertainties
for each observing session are assigned from the RMS of the photometry of reference stars that are similarly bright
as the target.
This technique is employed and described in detail for \citet{CrollWD}.

We plot our multiwavelength DCT, Perkins \& Hall data in Figure \ref{FigMassJ0036} \& 
\ref{FigMassJ0036PanelTwo}.

\section{The Variability of 2MASS 0036+18}
\label{SecAnalysis}

\subsection{The Rotation Period of 2MASS 0036+18}
\begin{figure}
\includegraphics[scale=0.45, angle = 270]{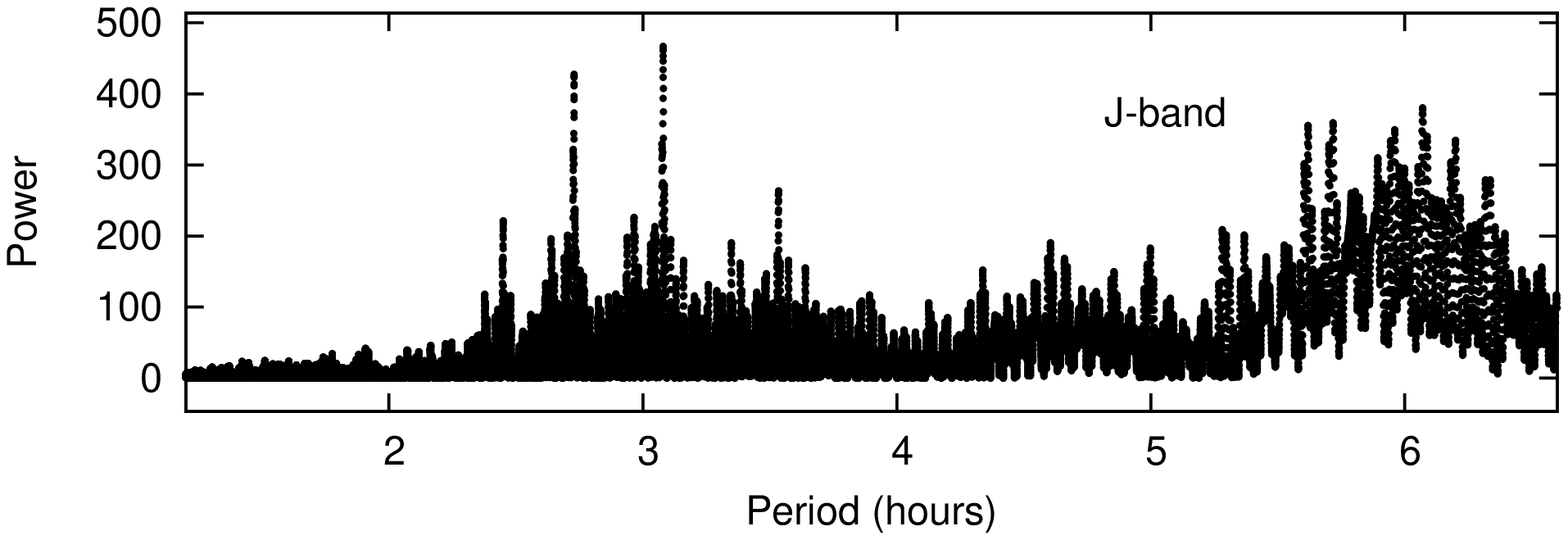}
\includegraphics[scale=0.45, angle = 270]{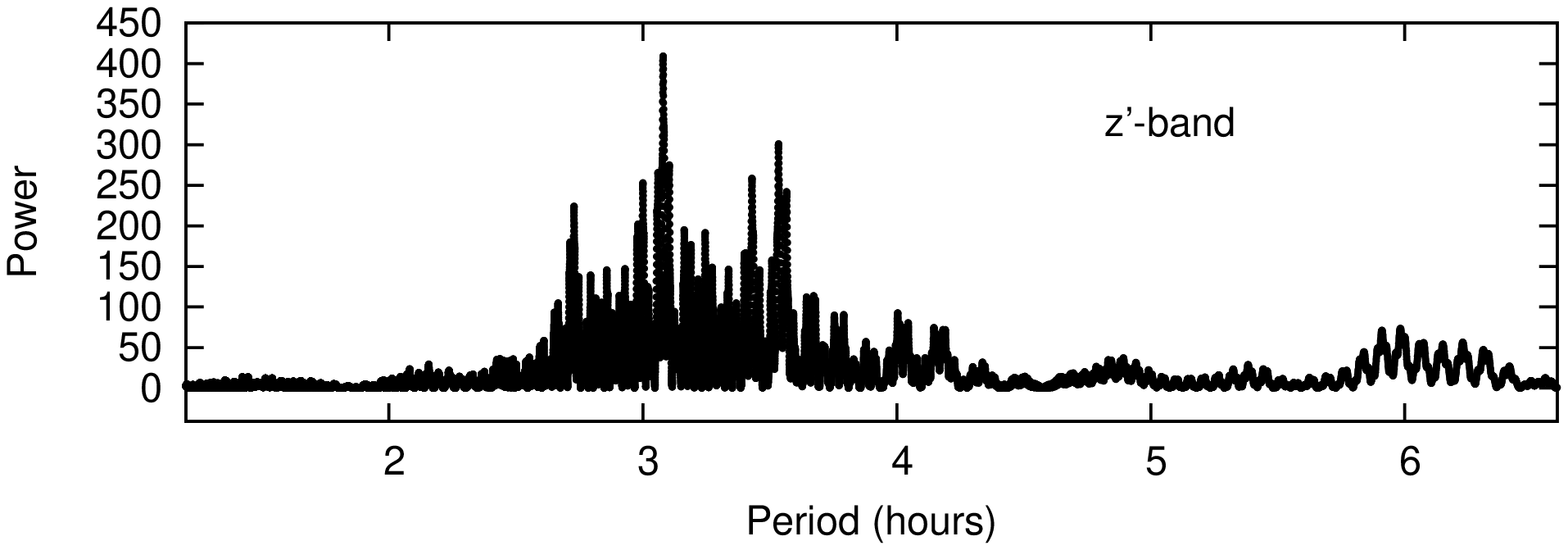}
\includegraphics[scale=0.45, angle = 270]{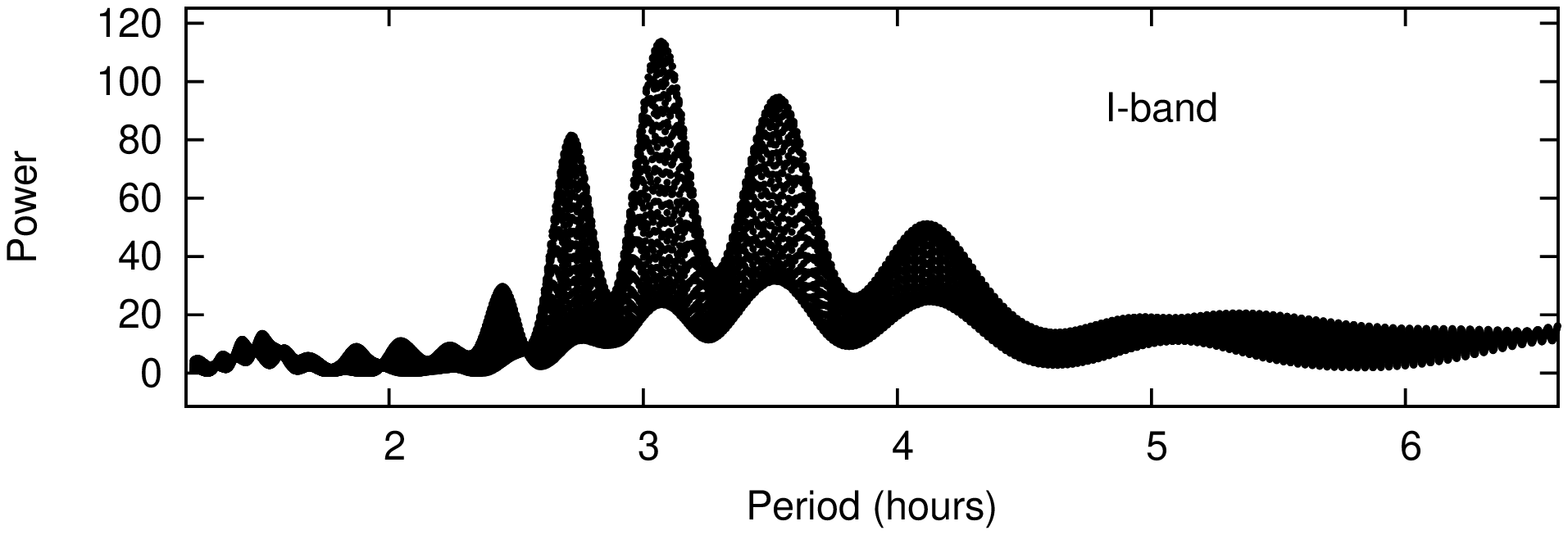}
\includegraphics[scale=0.45, angle = 270]{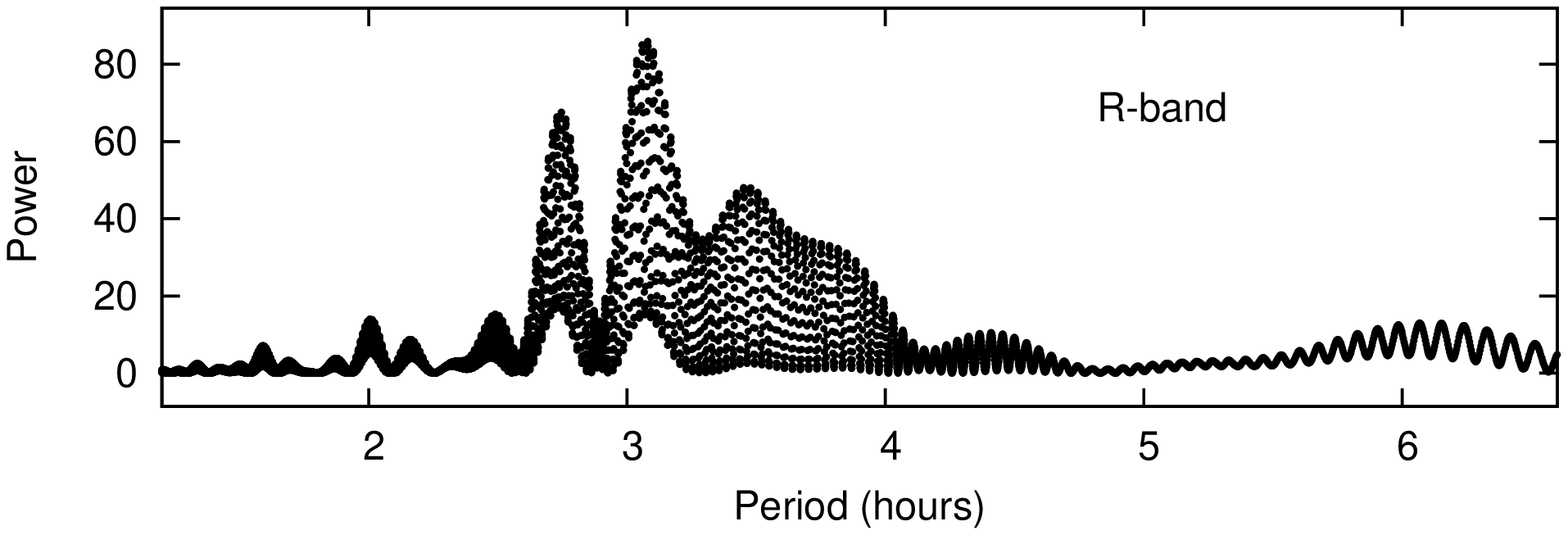}
\includegraphics[scale=0.45, angle = 270]{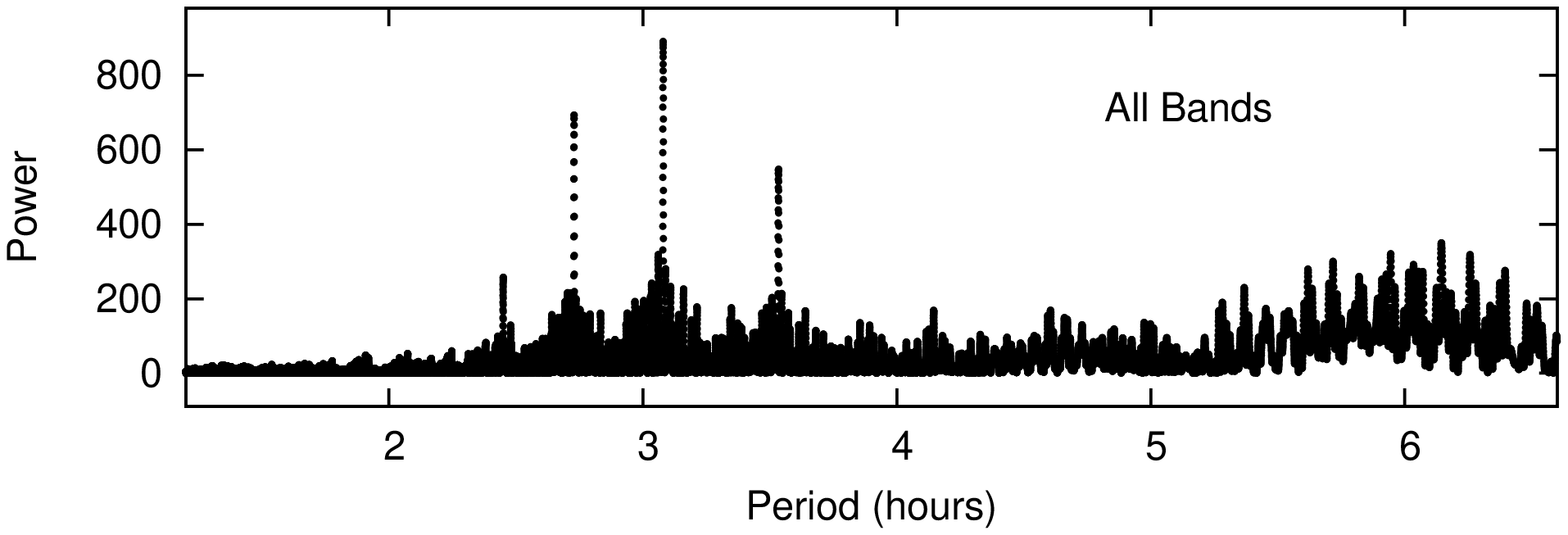}
\caption[]
	{	Lomb-Scargle Periodograms of our photometry of 2MASS 0036+18 in various bands as indicated in the panels, 
		and with all our photometry (bottom). We present the peak periodogram values in Table \ref{TablePeriods},
		which are all approximately $\sim$\PeriodHoursTwoMassZeroZeroThreeSixApprox \ hours.
	}
\label{FigMassJ0036LombPeriodogram}
\end{figure}

\begin{figure*}
\includegraphics[scale=0.90, angle = 270]{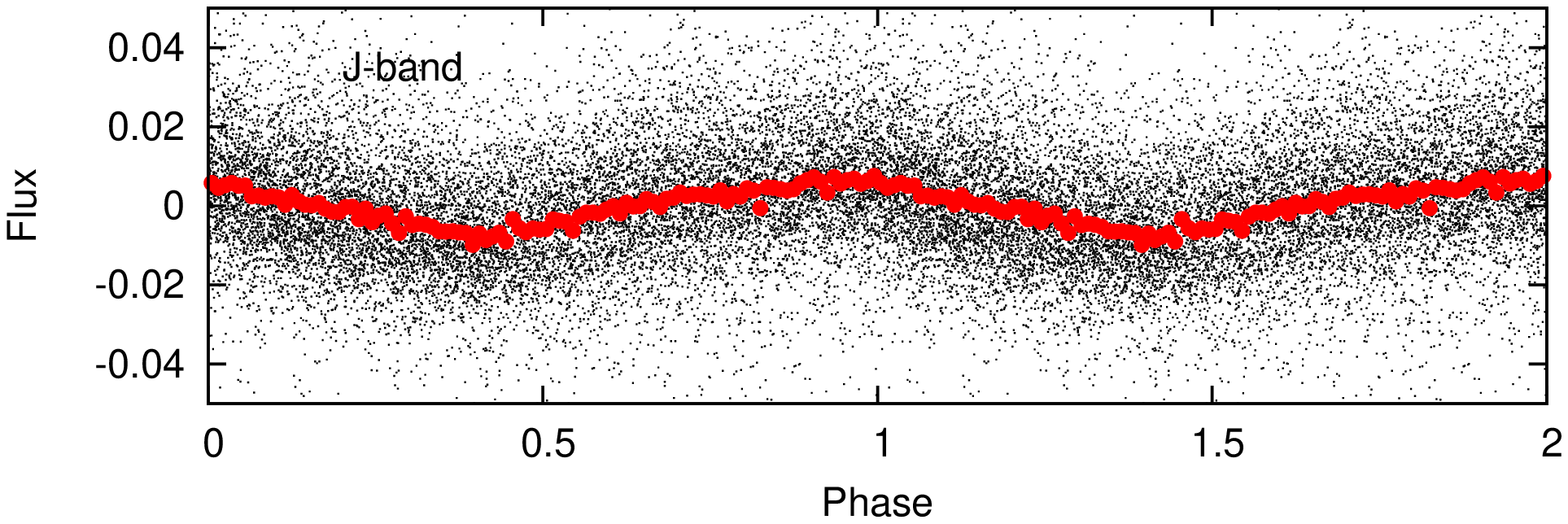}
\includegraphics[scale=0.45, angle = 270]{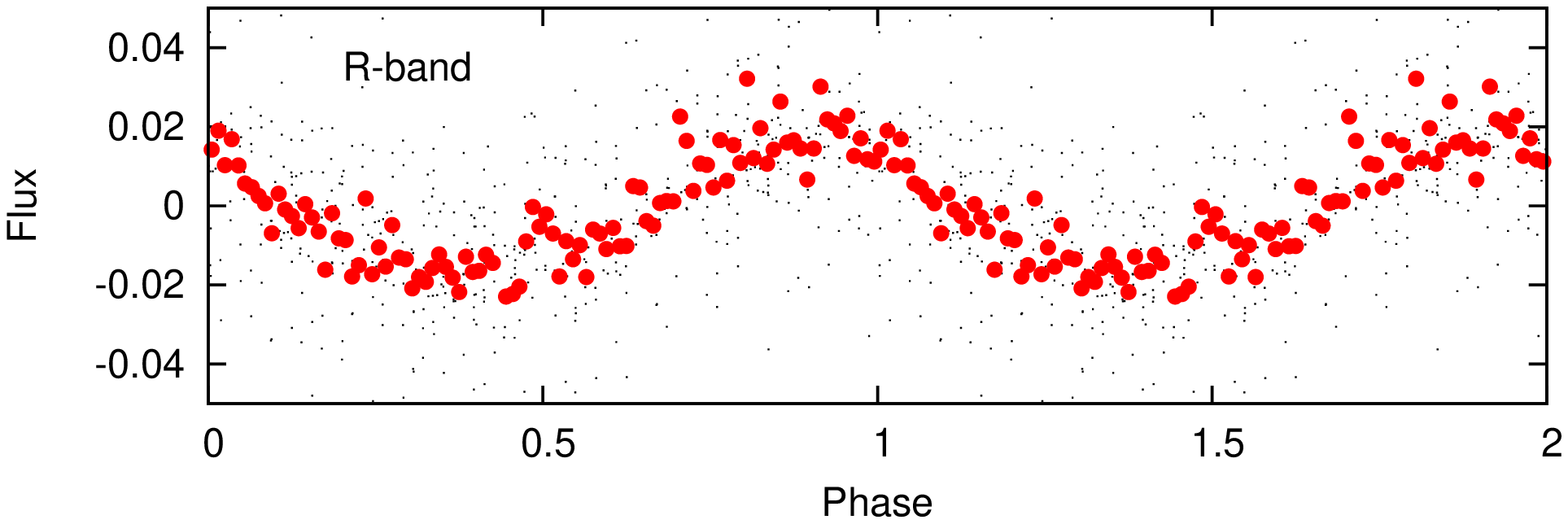}
\includegraphics[scale=0.45, angle = 270]{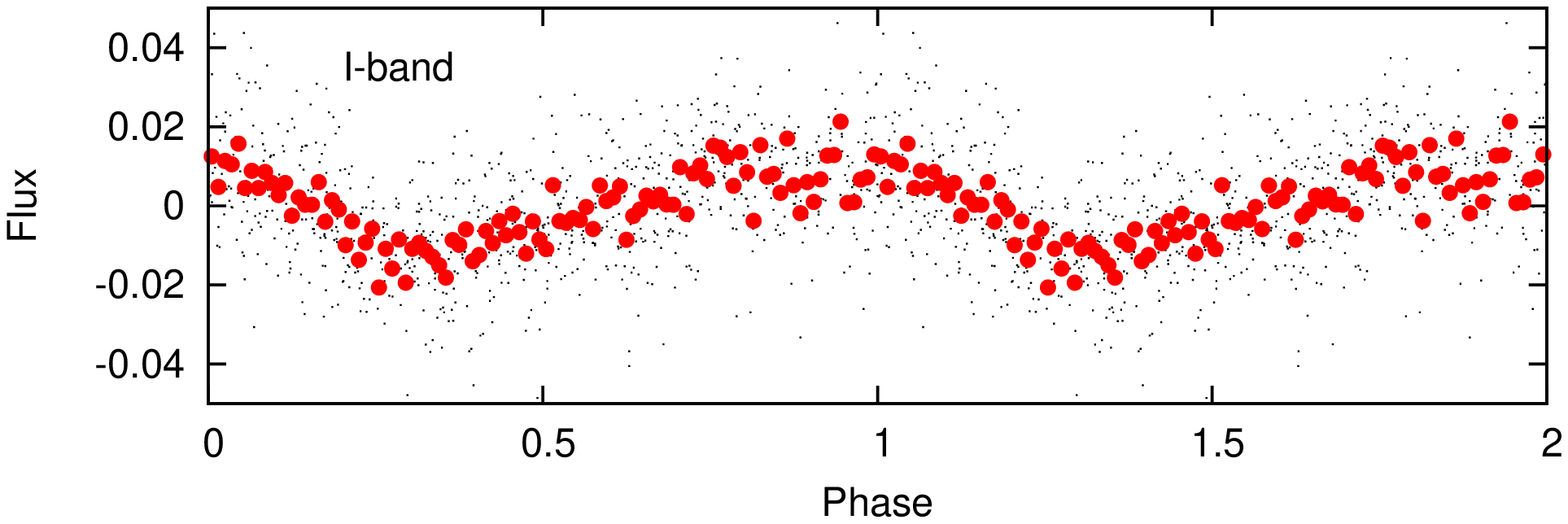}
\includegraphics[scale=0.45, angle = 270]{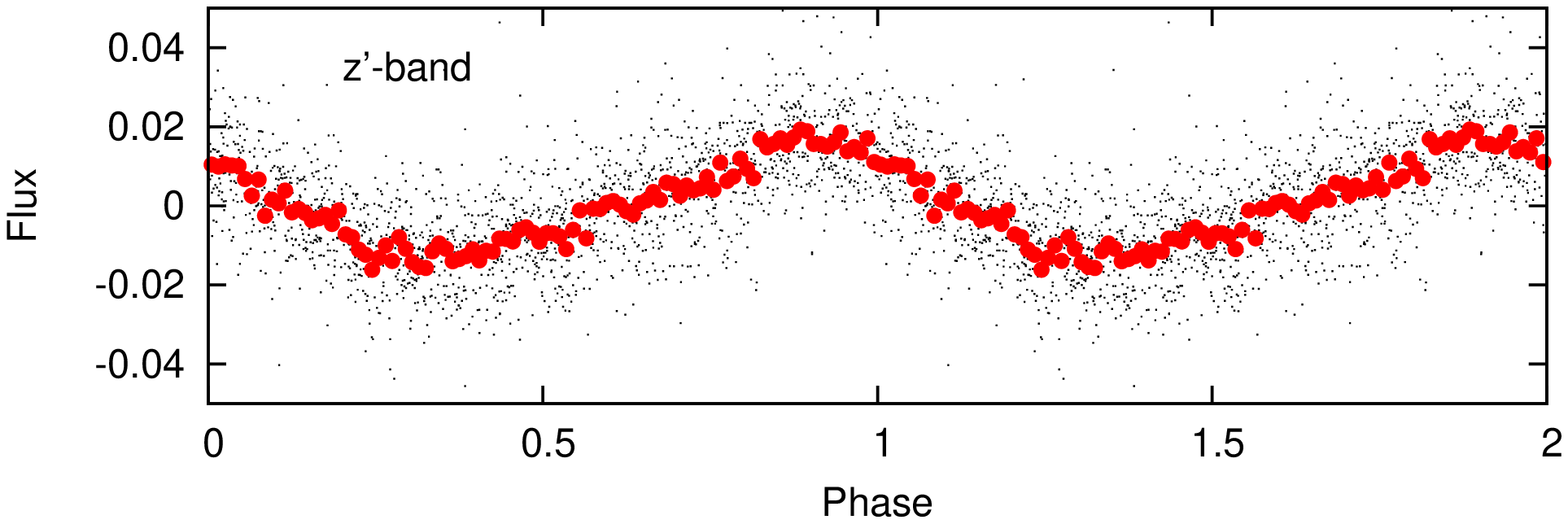}
\includegraphics[scale=0.45, angle = 270]{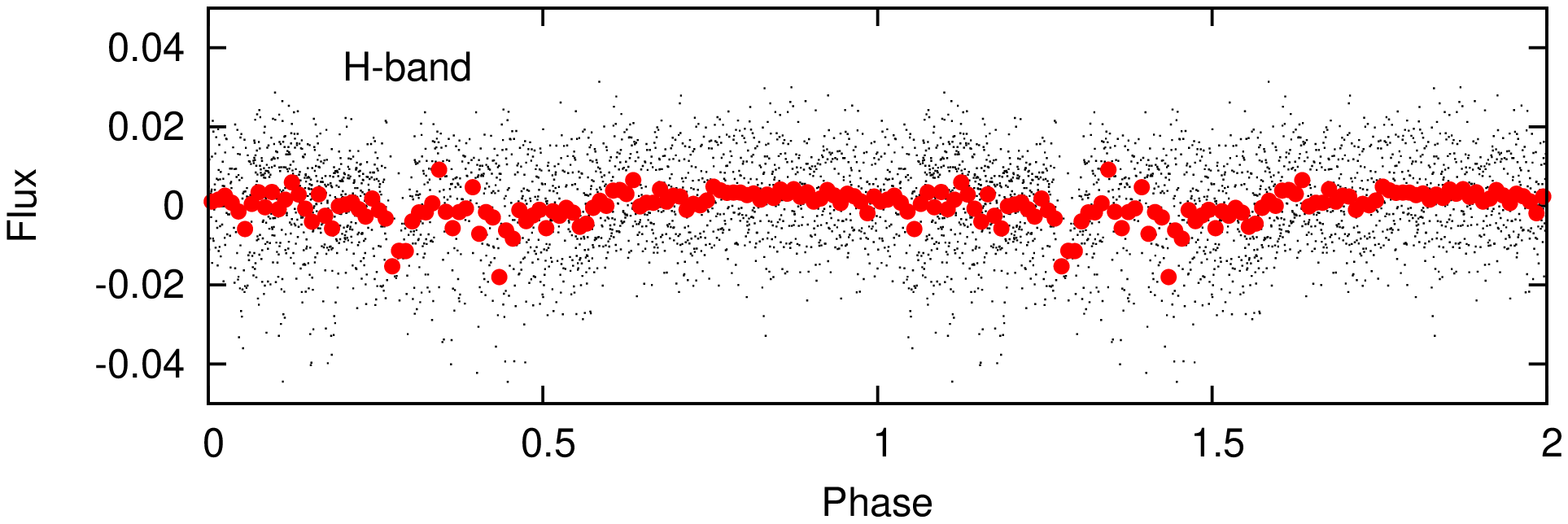}
\includegraphics[scale=0.45, angle = 270]{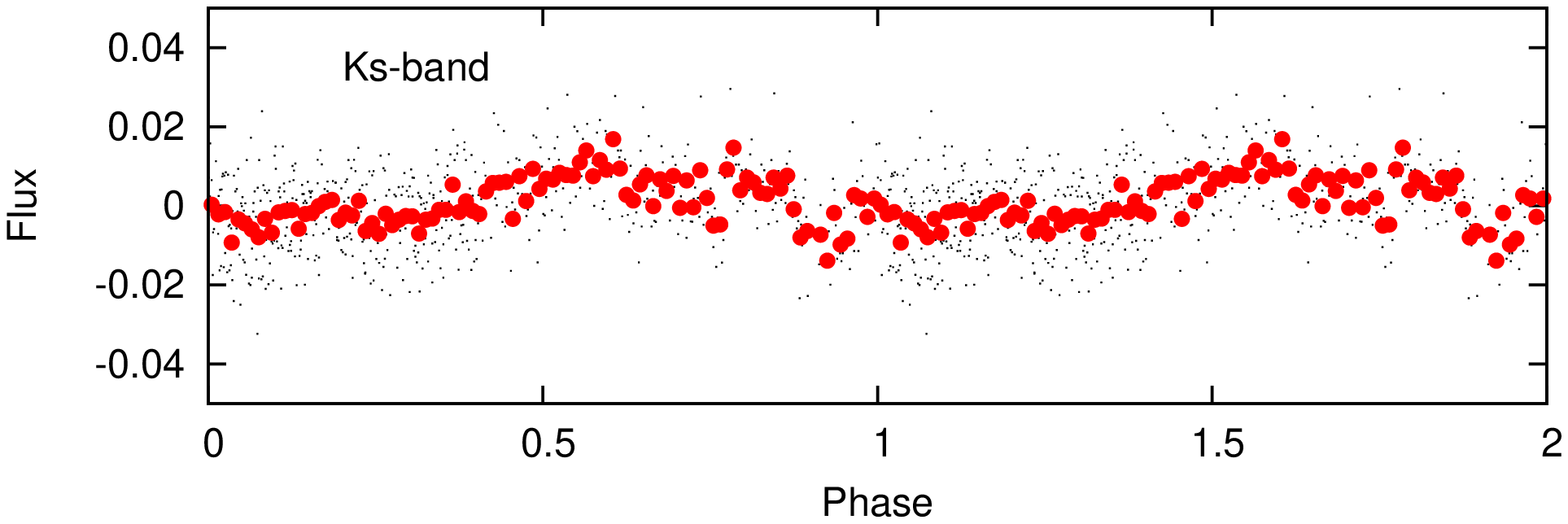}
\includegraphics[scale=0.45, angle = 270]{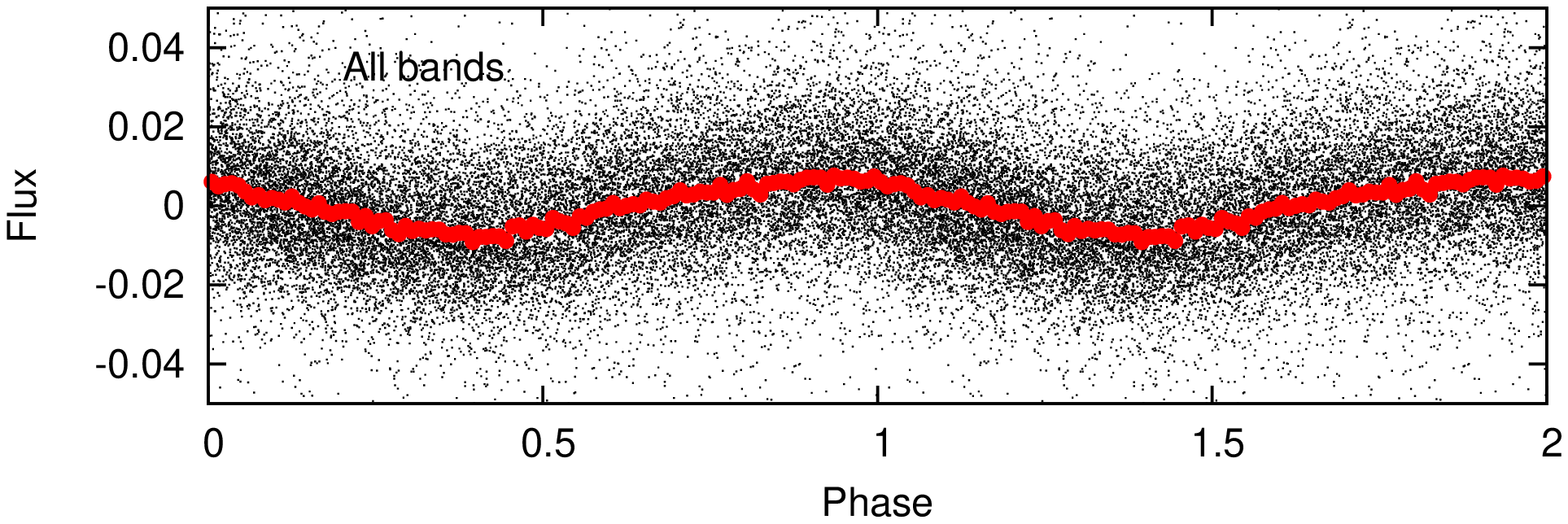}
\caption[]
	{	All our photometry of 2MASS 0036+18 (black points) in various-bands (as indicated in the panels)
		phased to the apparent rotation period of 
		$\sim$\PeriodHoursTwoMassZeroZeroThirtySixAllbandsQuote \ hours.
		For clarity, the phase is plotted up to 2, and therefore two cycles are presented in each panel. 
		Red circles denote the photometry binned every 0.01 in phase.
		Phase 0.0 is defined 
		using the apparent flux maximum in our J-band photometry at BJD-2457000 $\sim$256.91.
		The number of light curves phased together to create each panel are given in Table \ref{TableSinusoidBands}.
	}
\label{FigMassJ0036Phase}
\end{figure*}

\begin{deluxetable}{ccc}
\tablecaption{Lomb-Scargle Periodogram Periods of Various bands of our 2MASS 0036+18 photometry}
\tablehead{
\colhead{Observing} 	& \colhead{Lomb-Scargle}		& \colhead{\# of}		\\
\colhead{Band} 		& \colhead{Peak Period $P$ (hr)}		& \colhead{light curves}	\\
}
\centering
\startdata
R-band		& \PeriodHoursTwoMassZeroZeroThirtySixRband \ $\pm$ \PeriodHoursErrorTwoMassZeroZeroThirtySixRband	 & 3 \\	%
I-band		& \PeriodHoursTwoMassZeroZeroThirtySixIband \ $\pm$ \PeriodHoursErrorTwoMassZeroZeroThirtySixIband	 & 5 \\	%
z-band		& \PeriodHoursTwoMassZeroZeroThirtySixzband \ $\pm$ \PeriodHoursErrorTwoMassZeroZeroThirtySixzband	 & 6 \\	%
J-band		& \PeriodHoursTwoMassZeroZeroThirtySixJband \ $\pm$ \PeriodHoursErrorTwoMassZeroZeroThirtySixJband	 & 13 \\	%
All bands	& \PeriodHoursTwoMassZeroZeroThirtySixAllbands \ $\pm$ \PeriodHoursErrorTwoMassZeroZeroThirtySixAllbands	 & 29\\	%
\enddata
\label{TablePeriods}
\end{deluxetable}

We present Lomb-Scargle periodograms \citep{Lomb76,Scargle82}
of all our data in Figure \ref{FigMassJ0036LombPeriodogram}.
The variable, near-sinusoidal light curves often displayed by 2MASS 0036+18 are well-suited to period finding utilizing a Lomb-Scargle periodogram. 
The apparent rotation period from the Lomb-Scargle periodogram of all our data is 
\PeriodHoursTwoMassZeroZeroThirtySixAllbands \ $\pm$ \PeriodHoursErrorTwoMassZeroZeroThirtySixAllbands \ hours.
We also present Lomb-Scargle periodograms of our individual photometric bands
when there is more than one data-set in Figure \ref{FigMassJ0036LombPeriodogram}.
The periods of the peaks of the Lomb-Scargle periodograms for each individual
wavelength of data are given in Table \ref{TablePeriods}.
The obvious secondary, tertiary, and other peaks in the periodograms of the various wavelengths of observations
are day aliases of the main $\sim$\PeriodHoursTwoMassZeroZeroThirtySixAllbandsQuote \ hour period, due to the nightly observing cadence.
We 
conservatively
estimate the errors on the period using the full width half maximum of the envelope of periodogram values
near the maximum period peak in the periodogram.


The most prominent periods in the Lomb-Scargle periodograms for our various
wavelengths of observations seem to be consistent with one another,
suggesting that if our different wavelengths of observations are penetrating to different depths in the atmosphere
then we are not observing
significant differential rotation at different depths in the atmosphere.
Also, the most prominent period from all our data of
$\sim$\PeriodHoursTwoMassZeroZeroThirtySixAllbands \ $\pm$ \PeriodHoursErrorTwoMassZeroZeroThirtySixAllbands \ hours, agrees well with the 
3.08 $\pm$ 0.05 hour radio period detected for this object by \citet{Hallinan08}.
If the radio period results from the electron cyclotron maser instability from the polar regions of the ultra-cool dwarf \citep{Hallinan08},
then the close similarity between the radio period and our optical/near-infrared period likely place a limit on differential rotation;
if the radio period arises from deep within this ultra-cool dwarf, analogous to where the magnetosphere is believed to be generated
from the core for Jupiter \citep{Higgins96}, then the close correlation with our optical/near-infrared period shows the core magnetosphere
rotates in phase with the photosphere.
Our detected period is also in general agreement 
with the variety of other optical period detections for this object, including:
$\sim$3.0 hours from I-band observations \citep{Lane07}, \&
$\sim$3.0 $\pm$ 0.7 hours from I-band observations \citep{Harding13}.
Lastly our period is consistent with the 2.7 $\pm$ 0.3 hour period estimated from 14 hours of 3.6 \& 4.5 $\mu m$ infrared {\it Spitzer}/IRAC
photometry of 2MASS 0036+18 \citep{Metchev15}; these infrared wavelengths are likely to be probing lower pressure, and therefore higher altitude, regions
of 2MASS 0036+18 than our optical/near-infrared observations, again placing a limit on differential rotation at different depths in the atmosphere.

\subsection{The Phase \& Amplitude of 2MASS 0036+18's Multiwavelength Variability}

\begin{deluxetable}{cccc}
\tablecaption{Sinusoid Fits to Various bands of 2MASS 0036+18 photometry}
\tablehead{
\colhead{Observing} 	& \colhead{Peak-to-Peak}		&  \colhead{Phase}	& \colhead{\# of}		\\
\colhead{Band} 		& \colhead{Amplitude $A$ (\%)}		&  \colhead{($\phi$)}	& \colhead{light curves}	\\
}
\centering
\startdata
R-band		& \SinusoidPeaktoPeakAmplitudePercentageTwoMassZeroZeroThirtySixRband \ $\pm$ \SinusoidPeaktoPeakAmplitudePercentageErrorTwoMassZeroZeroThirtySixRband		& 	\SinusoidPhaseTwoMassZeroZeroThirtySixRband \ $\pm$ \SinusoidPhaseErrorTwoMassZeroZeroThirtySixRband	 & 3 \\	%
I-band		& \SinusoidPeaktoPeakAmplitudePercentageTwoMassZeroZeroThirtySixIband \ $\pm$ \SinusoidPeaktoPeakAmplitudePercentageErrorTwoMassZeroZeroThirtySixIband		& 	\SinusoidPhaseTwoMassZeroZeroThirtySixIband \ $\pm$ \SinusoidPhaseErrorTwoMassZeroZeroThirtySixIband	 & 5 \\	%
z-band		& \SinusoidPeaktoPeakAmplitudePercentageTwoMassZeroZeroThirtySixzband \ $\pm$ \SinusoidPeaktoPeakAmplitudePercentageErrorTwoMassZeroZeroThirtySixzband		& 	\SinusoidPhaseTwoMassZeroZeroThirtySixzband \ $\pm$ \SinusoidPhaseErrorTwoMassZeroZeroThirtySixzband	 & 6 \\	%
J-band		& \SinusoidPeaktoPeakAmplitudePercentageTwoMassZeroZeroThirtySixJband \ $\pm$ \SinusoidPeaktoPeakAmplitudePercentageErrorTwoMassZeroZeroThirtySixJband		& 	\SinusoidPhaseTwoMassZeroZeroThirtySixJband \ $\pm$ \SinusoidPhaseErrorTwoMassZeroZeroThirtySixJband	 & 13 \\	%
H-band		& \SinusoidPeaktoPeakAmplitudePercentageTwoMassZeroZeroThirtySixHband \ $\pm$ \SinusoidPeaktoPeakAmplitudePercentageErrorTwoMassZeroZeroThirtySixHband		& 	\SinusoidPhaseTwoMassZeroZeroThirtySixHband \ $\pm$ \SinusoidPhaseErrorTwoMassZeroZeroThirtySixHband	 & 1 \\	%
Ks-band		& \SinusoidPeaktoPeakAmplitudePercentageTwoMassZeroZeroThirtySixKsband \ $\pm$ \SinusoidPeaktoPeakAmplitudePercentageErrorTwoMassZeroZeroThirtySixKsband	& 	\SinusoidPhaseTwoMassZeroZeroThirtySixKsband \ $\pm$ \SinusoidPhaseErrorTwoMassZeroZeroThirtySixKsband	 & 1 \\	%
All bands	& \SinusoidPeaktoPeakAmplitudePercentageTwoMassZeroZeroThirtySixAllData \ $\pm$ \SinusoidPeaktoPeakAmplitudePercentageErrorTwoMassZeroZeroThirtySixAllData	& 	\SinusoidPhaseTwoMassZeroZeroThirtySixAllData \ $\pm$ \SinusoidPhaseErrorTwoMassZeroZeroThirtySixAllData & 29\\	%
\hline
3.6 $\mu m$ \tablenotemark{a}	& 0.47 $\pm$ 0.05																		& n/a														& 1 \\
4.5 $\mu m$ \tablenotemark{a}	& 0.19 $\pm$ 0.04																		& n/a														& 1 \\
\enddata
\tablenotetext{a}{The amplitudes at these wavelengths are included from 8 hours of 3.6 $\mu m$ \& 6 hours of 4.5 $\mu m$ Spitzer/IRAC data \citep{Metchev15}, respectively.}
\label{TableSinusoidBands}
\end{deluxetable}

\begin{deluxetable}{ccccc}
\tablecaption{Sinusoid Fits to Multiwavelength photometry of 2MASS 0036+18}
\tablehead{
\colhead{Date}	& \colhead{Telescope}	& \colhead{Band} 	& \colhead{Peak-to-Peak}		&  \colhead{Phase}		\\
\colhead{(UTC)}	& \colhead{}		& \colhead{} 		& \colhead{Amplitude $A$ (\%)}		&  \colhead{($\phi$)}	\\
}
\centering
\startdata
2015/09/23	& Perkins		& I		& \SinusoidPeaktoPeakAmplitudePercentageTwoMassZeroZeroThirtySixFifteenSeptemberTwentyThreePerkinsIband \ $\pm$ \SinusoidPeaktoPeakAmplitudePercentageErrorTwoMassZeroZeroThirtySixFifteenSeptemberTwentyThreePerkinsIband		& 	\SinusoidPhaseTwoMassZeroZeroThirtySixFifteenSeptemberTwentyThreePerkinsIband \ $\pm$ \SinusoidPhaseErrorTwoMassZeroZeroThirtySixFifteenSeptemberTwentyThreePerkinsIband	 \\	%
2015/09/23	& Hall			& I		& \SinusoidPeaktoPeakAmplitudePercentageTwoMassZeroZeroThirtySixFifteenSeptemberTwentyThreeHallIband \ $\pm$ \SinusoidPeaktoPeakAmplitudePercentageErrorTwoMassZeroZeroThirtySixFifteenSeptemberTwentyThreeHallIband		& 	\SinusoidPhaseTwoMassZeroZeroThirtySixFifteenSeptemberTwentyThreeHallIband \ $\pm$ \SinusoidPhaseErrorTwoMassZeroZeroThirtySixFifteenSeptemberTwentyThreeHallIband	 \\	%
\hline
2015/09/24	& Perkins		& z'		& \SinusoidPeaktoPeakAmplitudePercentageTwoMassZeroZeroThirtySixFifteenSeptemberTwentyFourzband \ $\pm$ \SinusoidPeaktoPeakAmplitudePercentageErrorTwoMassZeroZeroThirtySixFifteenSeptemberTwentyFourzband		& 	\SinusoidPhaseTwoMassZeroZeroThirtySixFifteenSeptemberTwentyFourzband \ $\pm$ \SinusoidPhaseErrorTwoMassZeroZeroThirtySixFifteenSeptemberTwentyFourzband	 \\	%
2015/09/24	& Hall			& I		& \SinusoidPeaktoPeakAmplitudePercentageTwoMassZeroZeroThirtySixFifteenSeptemberTwentyFourIband \ $\pm$ \SinusoidPeaktoPeakAmplitudePercentageErrorTwoMassZeroZeroThirtySixFifteenSeptemberTwentyFourIband		& 	\SinusoidPhaseTwoMassZeroZeroThirtySixFifteenSeptemberTwentyFourIband \ $\pm$ \SinusoidPhaseErrorTwoMassZeroZeroThirtySixFifteenSeptemberTwentyFourIband	 \\	%

\hline
2015/09/25	& Perkins		& R		& \SinusoidPeaktoPeakAmplitudePercentageTwoMassZeroZeroThirtySixFifteenSeptemberTwentyFiveRband \ $\pm$ \SinusoidPeaktoPeakAmplitudePercentageErrorTwoMassZeroZeroThirtySixFifteenSeptemberTwentyFiveRband		& 	\SinusoidPhaseTwoMassZeroZeroThirtySixFifteenSeptemberTwentyFiveRband \ $\pm$ \SinusoidPhaseErrorTwoMassZeroZeroThirtySixFifteenSeptemberTwentyFiveRband	 \\	%
2015/09/25	& Hall			& z'		& \SinusoidPeaktoPeakAmplitudePercentageTwoMassZeroZeroThirtySixFifteenSeptemberTwentyFivezband \ $\pm$ \SinusoidPeaktoPeakAmplitudePercentageErrorTwoMassZeroZeroThirtySixFifteenSeptemberTwentyFivezband		& 	\SinusoidPhaseTwoMassZeroZeroThirtySixFifteenSeptemberTwentyFivezband \ $\pm$ \SinusoidPhaseErrorTwoMassZeroZeroThirtySixFifteenSeptemberTwentyFivezband	 \\	%
\hline
2015/09/26	& Perkins		& R		& \SinusoidPeaktoPeakAmplitudePercentageTwoMassZeroZeroThirtySixFifteenSeptemberTwentySixRband \ $\pm$ \SinusoidPeaktoPeakAmplitudePercentageErrorTwoMassZeroZeroThirtySixFifteenSeptemberTwentySixRband		& 	\SinusoidPhaseTwoMassZeroZeroThirtySixFifteenSeptemberTwentySixRband \ $\pm$ \SinusoidPhaseErrorTwoMassZeroZeroThirtySixFifteenSeptemberTwentySixRband	 \\	%
2015/09/26	& Hall			& z'		& \SinusoidPeaktoPeakAmplitudePercentageTwoMassZeroZeroThirtySixFifteenSeptemberTwentySixzband \ $\pm$ \SinusoidPeaktoPeakAmplitudePercentageErrorTwoMassZeroZeroThirtySixFifteenSeptemberTwentySixzband		& 	\SinusoidPhaseTwoMassZeroZeroThirtySixFifteenSeptemberTwentySixzband \ $\pm$ \SinusoidPhaseErrorTwoMassZeroZeroThirtySixFifteenSeptemberTwentySixzband	 \\	%

\hline
2015/10/14	& DCT			& R		& \SinusoidPeaktoPeakAmplitudePercentageTwoMassZeroZeroThirtySixFifteenOctoberFourteenRband \ $\pm$ \SinusoidPeaktoPeakAmplitudePercentageErrorTwoMassZeroZeroThirtySixFifteenOctoberFourteenRband		& 	\SinusoidPhaseTwoMassZeroZeroThirtySixFifteenOctoberFourteenRband \ $\pm$ \SinusoidPhaseErrorTwoMassZeroZeroThirtySixFifteenOctoberFourteenRband	 \\	%
2015/10/14	& DCT			& z'		& \SinusoidPeaktoPeakAmplitudePercentageTwoMassZeroZeroThirtySixFifteenOctoberFourteenzband \ $\pm$ \SinusoidPeaktoPeakAmplitudePercentageErrorTwoMassZeroZeroThirtySixFifteenOctoberFourteenzband		& 	\SinusoidPhaseTwoMassZeroZeroThirtySixFifteenOctoberFourteenzband \ $\pm$ \SinusoidPhaseErrorTwoMassZeroZeroThirtySixFifteenOctoberFourteenzband	 \\	%
\hline
2015/11/08	& Perkins		& J		& \SinusoidPeaktoPeakAmplitudePercentageTwoMassZeroZeroThirtySixFifteenNovemberEightJband \ $\pm$ \SinusoidPeaktoPeakAmplitudePercentageErrorTwoMassZeroZeroThirtySixFifteenNovemberEightJband		& 	\SinusoidPhaseTwoMassZeroZeroThirtySixFifteenNovemberEightJband \ $\pm$ \SinusoidPhaseErrorTwoMassZeroZeroThirtySixFifteenNovemberEightJband	 \\	%
2015/11/08	& Hall			& I		& \SinusoidPeaktoPeakAmplitudePercentageTwoMassZeroZeroThirtySixFifteenNovemberEightIband \ $\pm$ \SinusoidPeaktoPeakAmplitudePercentageErrorTwoMassZeroZeroThirtySixFifteenNovemberEightIband		& 	\SinusoidPhaseTwoMassZeroZeroThirtySixFifteenNovemberEightIband \ $\pm$ \SinusoidPhaseErrorTwoMassZeroZeroThirtySixFifteenNovemberEightIband	 \\	%
\hline
2015/11/09	& Perkins		& H		& \SinusoidPeaktoPeakAmplitudePercentageTwoMassZeroZeroThirtySixFifteenNovemberNineHband \ $\pm$ \SinusoidPeaktoPeakAmplitudePercentageErrorTwoMassZeroZeroThirtySixFifteenNovemberNineHband		& 	\SinusoidPhaseTwoMassZeroZeroThirtySixFifteenNovemberNineHband \ $\pm$ \SinusoidPhaseErrorTwoMassZeroZeroThirtySixFifteenNovemberNineHband	 \\	%
2015/11/09	& Hall			& I		& \SinusoidPeaktoPeakAmplitudePercentageTwoMassZeroZeroThirtySixFifteenNovemberNineIband \ $\pm$ \SinusoidPeaktoPeakAmplitudePercentageErrorTwoMassZeroZeroThirtySixFifteenNovemberNineIband		& 	\SinusoidPhaseTwoMassZeroZeroThirtySixFifteenNovemberNineIband \ $\pm$ \SinusoidPhaseErrorTwoMassZeroZeroThirtySixFifteenNovemberNineIband	 \\	%

\enddata
\label{TableSinusoidIndividualLightCurves}
\end{deluxetable}

We phase  
our data in various bands to the apparent rotation period from all our data
of $\sim$\PeriodHoursTwoMassZeroZeroThirtySixAllbandsQuote \ hours in Figure \ref{FigMassJ0036Phase}.
Zero phase is defined at BJD-2457000 = 256.91 (corresponding to an apparent flux maximum in our J-band photometry).
We fit sinusoids with this period to the photometry
in various bands, allowing the peak-to-peak amplitude, $A$, and phase, $\phi$, 
to vary for each individual band; 
these results are given in Table \ref{TableSinusoidBands}.
$\phi$ is defined from 0 - 1 and
$\phi$ = 0 denotes the flux maximum of the sinusoid.
In Table \ref{TableSinusoidBands} we also include 
the peak-to-peak amplitudes returned recently for 2MASS 0036+18 at the 3.6 $\mu m$ and 4.5 $\mu m$ Spitzer/IRAC wavelengths \citep{Metchev15}.
With the caveat that our data and the \citet{Metchev15} data were obtained many months apart, in general it is apparent that
the amplitude of variability decreases as one moves to longer wavelengths.

There is also a statistically significant difference in $\phi$ values between the J-band photometry and some of
the other photometric bands (z'-band for instance). As these light curves represent averages over up to 
$\sim$\INSERTSPREADDAYS \ days of data, it is not clear whether these phase differences represent statistically significant
phase offsets between various wavelengths of observations. Such apparent
phase offsets 
may be caused by evolution in the longitude of starspot features from night-to-night and week to week,
or a small mis-estimate of the true long-term period
propagated over a great many rotation periods.
An arguably superior way to determine whether there are statistically significant phase offsets
between the near-sinusoidal variability at various wavelength of observations is to 
search for such phase offsets in multiwavelength data taken simultaneously.


\begin{figure}
\includegraphics[scale=0.45, angle = 270]{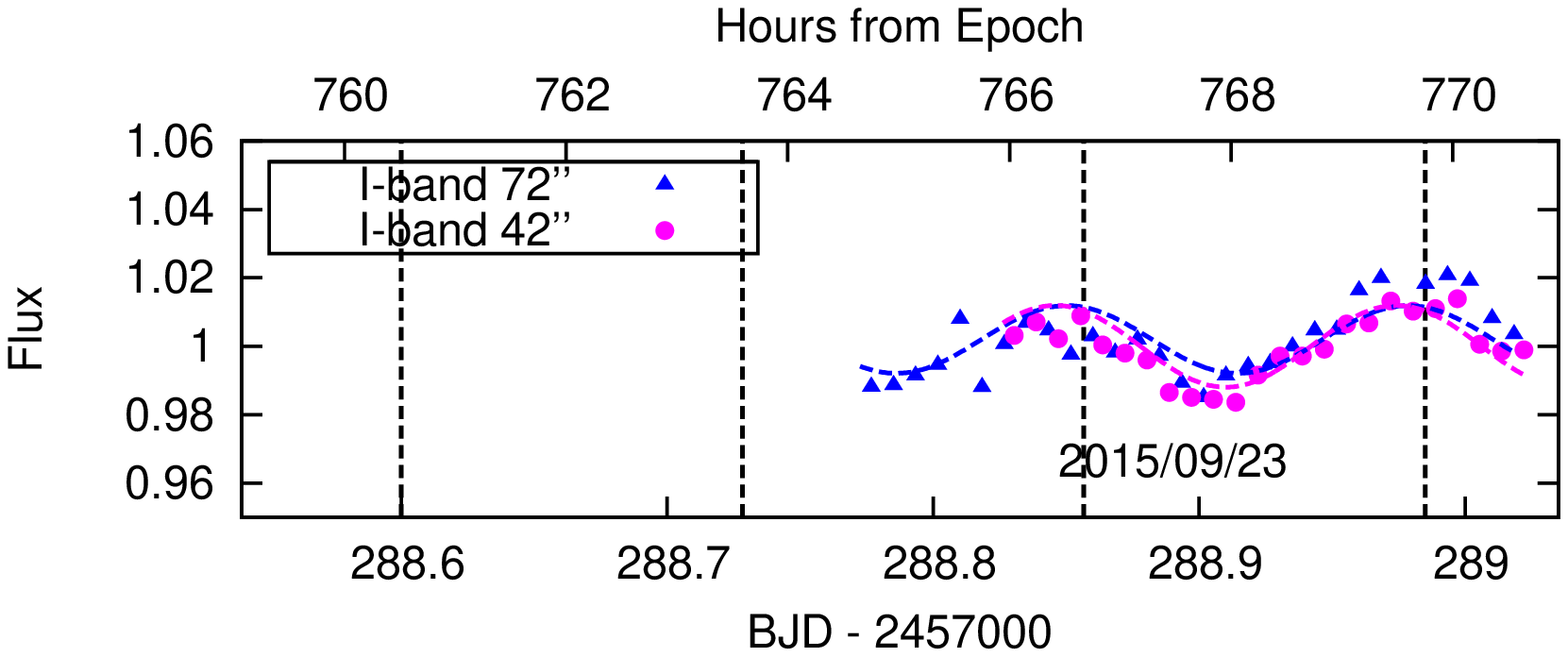}
\includegraphics[scale=0.45, angle = 270]{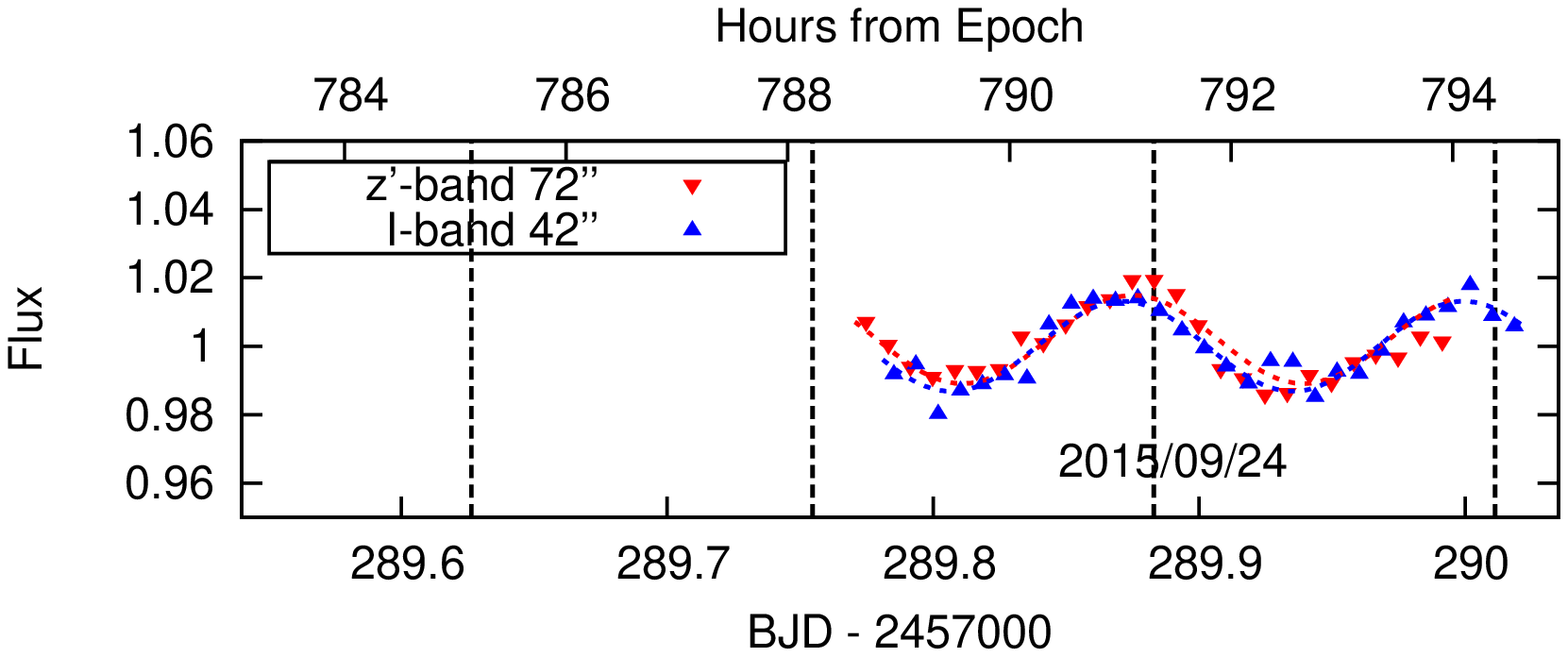}
\includegraphics[scale=0.45, angle = 270]{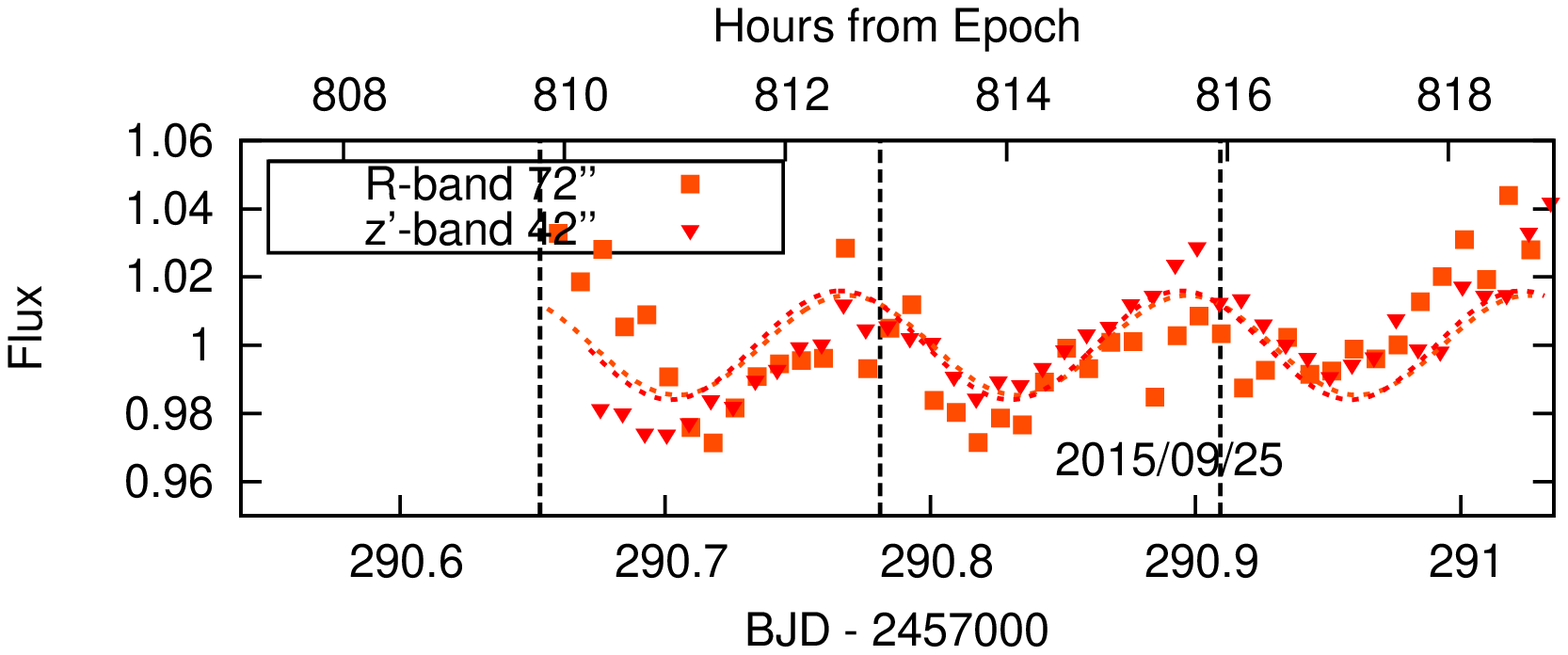}
\includegraphics[scale=0.45, angle = 270]{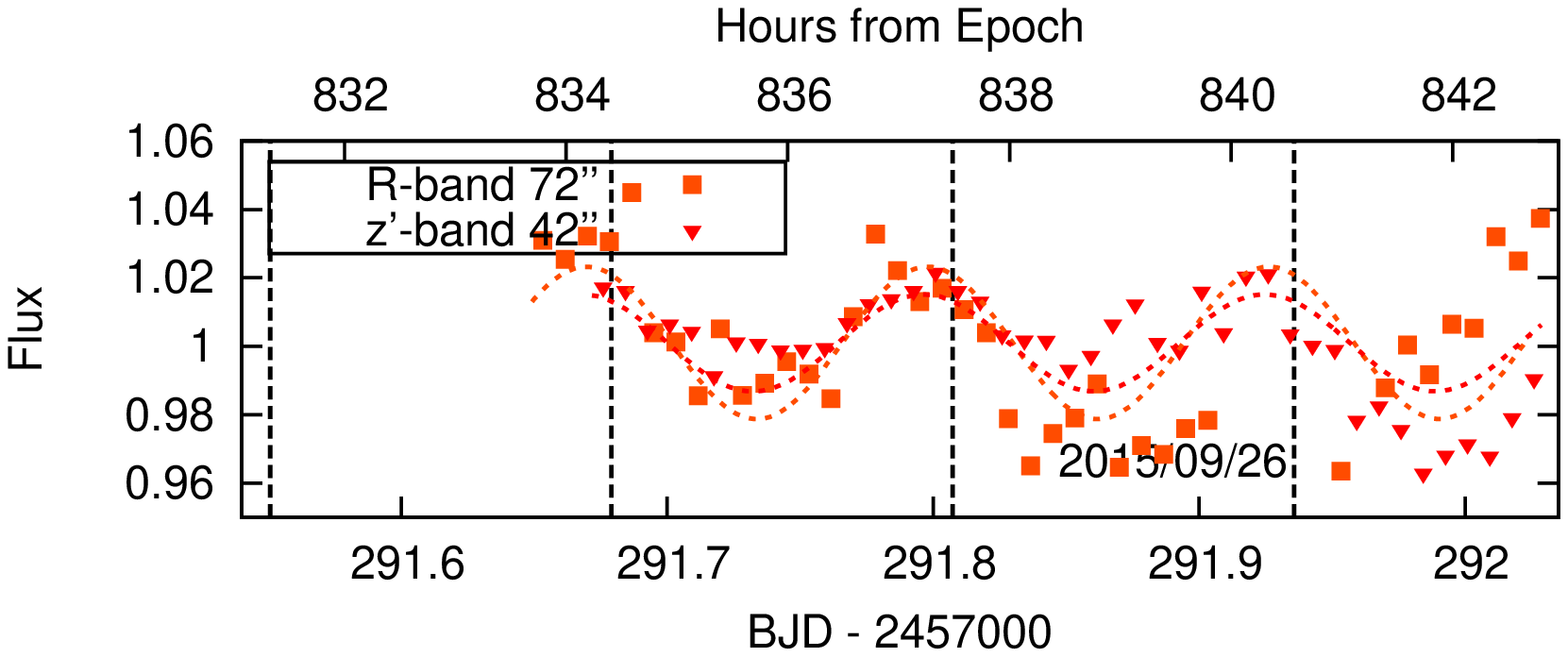}
\includegraphics[scale=0.45, angle = 270]{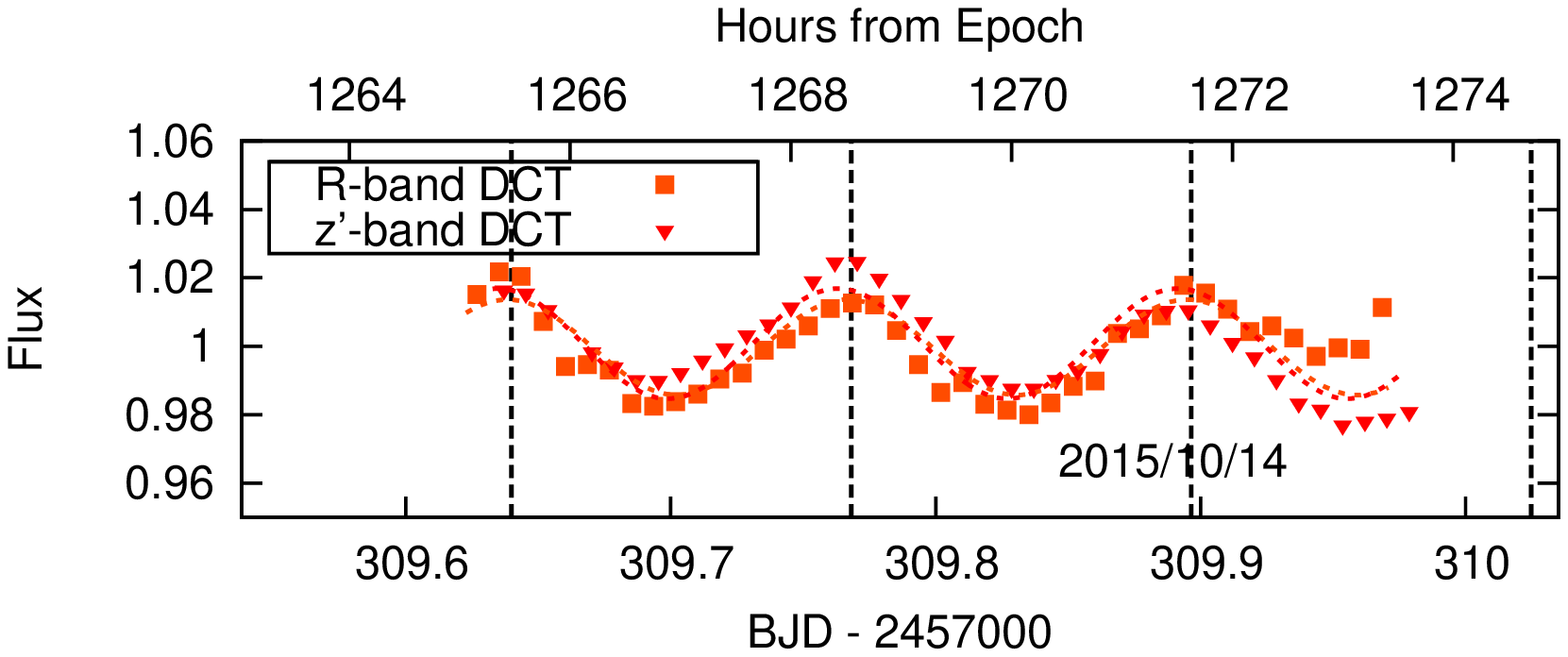}
\includegraphics[scale=0.45, angle = 270]{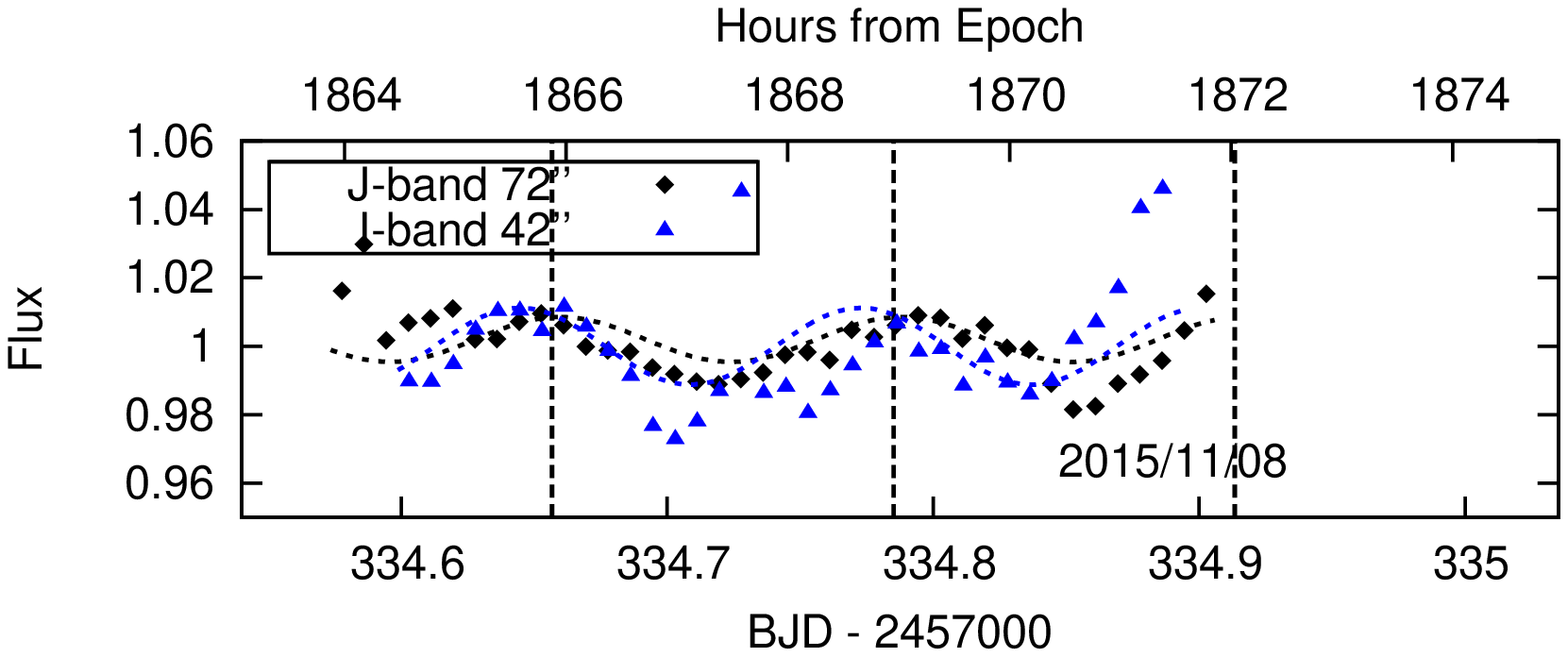}
\includegraphics[scale=0.45, angle = 270]{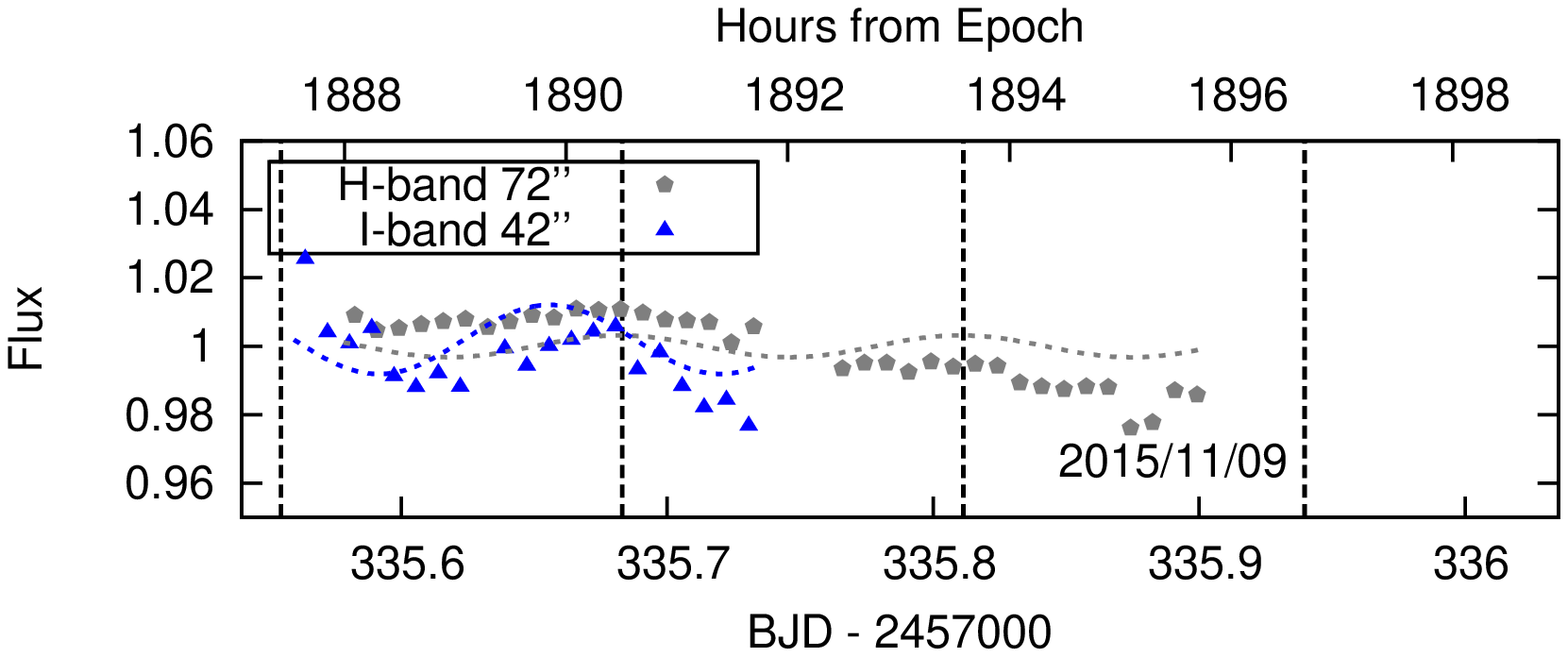}
\caption[]
	{	Sinusoid fits (dotted lines) to the simultaneous, multiwavelength
		Perkins, Hall \& DCT photometry of 2MASS 0036+18 in various bands.
		The colour of the dashed lines corresponds to the wavelength of observations
		as denoted by the colours of the data-points 
		as indicated in the legend.
		The figure format is otherwise identical to Figures \ref{FigMassJ0036} \& \ref{FigMassJ0036PanelTwo}.
	}
\label{FigMassJ0036JointSinusoids}
\end{figure}

 Thus, we also fit our simultaneous multiwavelength light curves from a single night with sinusoidal fits.
The associated amplitudes, $A$, and
phases, $\phi$, of the sinusoidal fits to each wavelength of photometry are shown in
Figure \ref{FigMassJ0036JointSinusoids}, and Table \ref{TableSinusoidIndividualLightCurves}. 
Most of the phase offsets between the various simultaneous wavelengths of photometry are not significant at the 3$\sigma$ level.
Our 2015 October 14th DCT light curve has an apparently significant phase offset; however, 
if we scale the reduced chi-squared of our sinusoidal fit to this light curve to 1 -- to take into account the fact that the 
variability of 2MASS 0036+18 on this night may not be perfectly sinusoidal -- then 
the phase offset is not significant at the 3$\sigma$ level between the R-band and z'-band photometry.


\begin{figure}
\includegraphics[scale=0.65, angle = 270]{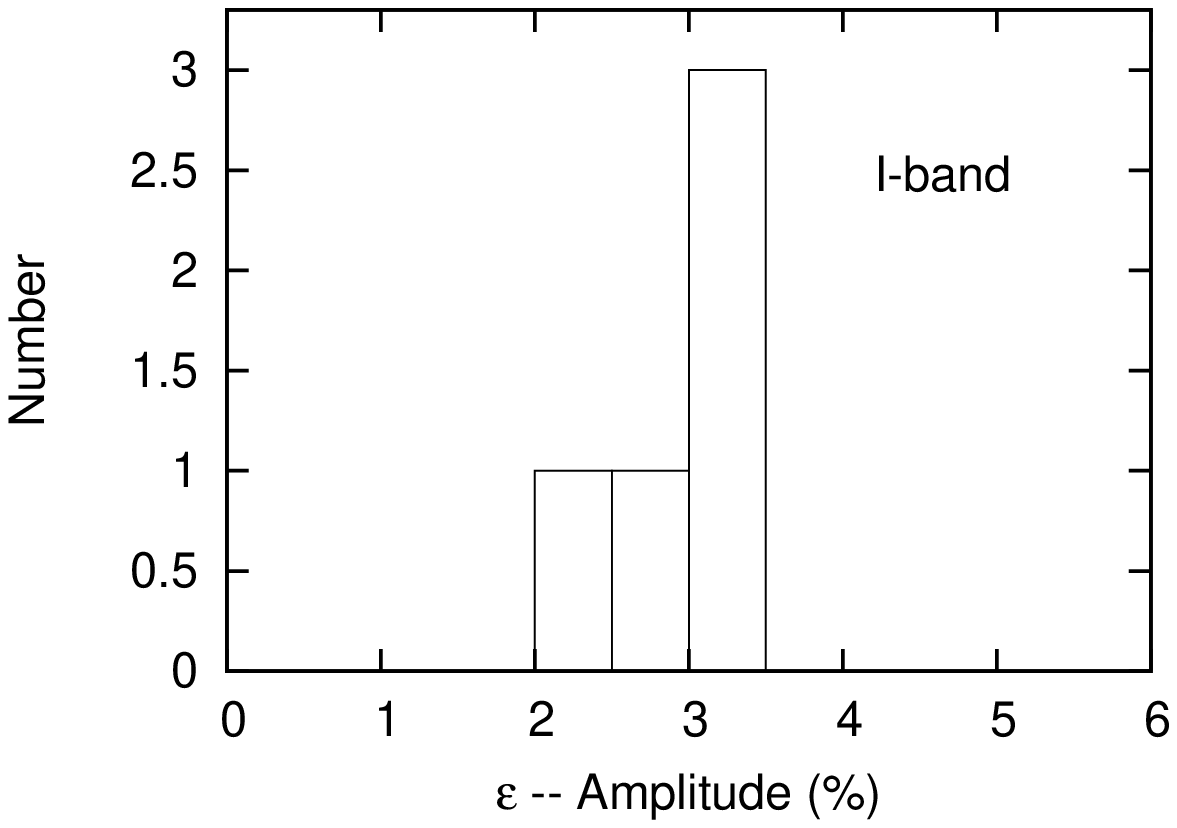}
\includegraphics[scale=0.65, angle = 270]{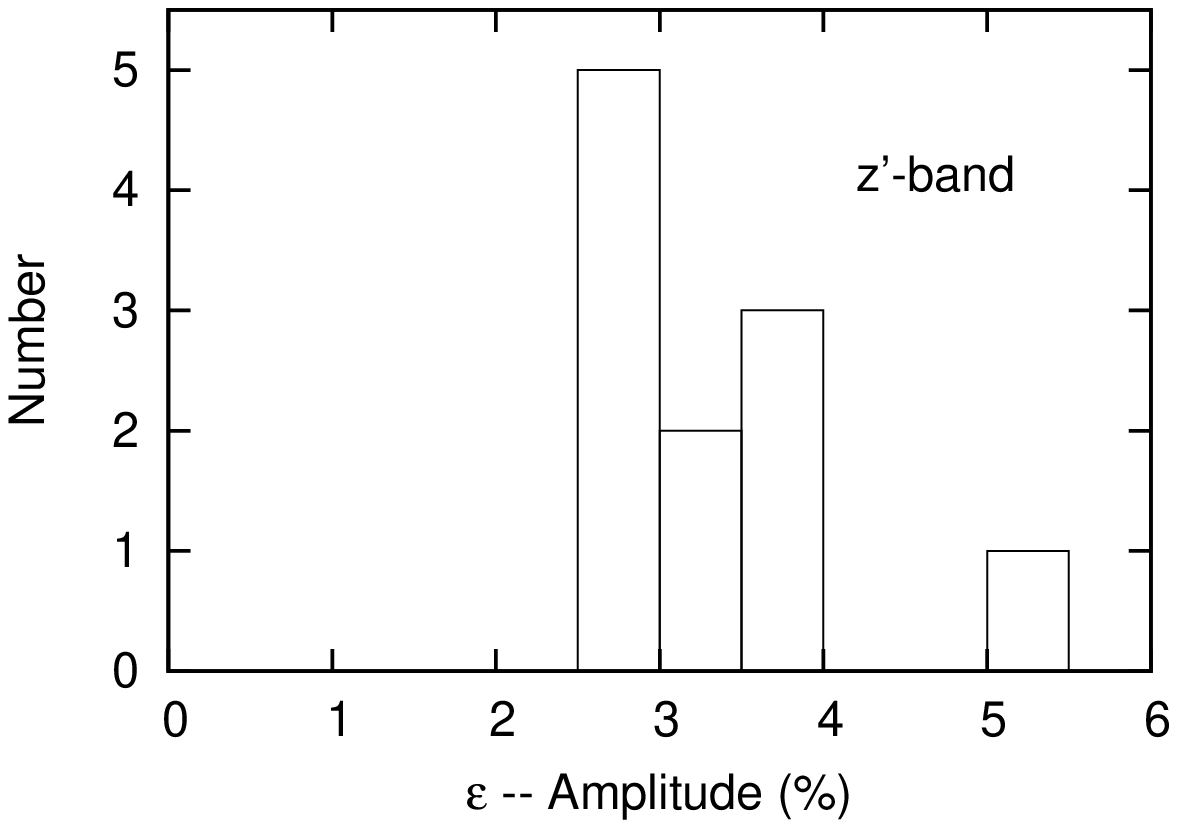}
\includegraphics[scale=0.65, angle = 270]{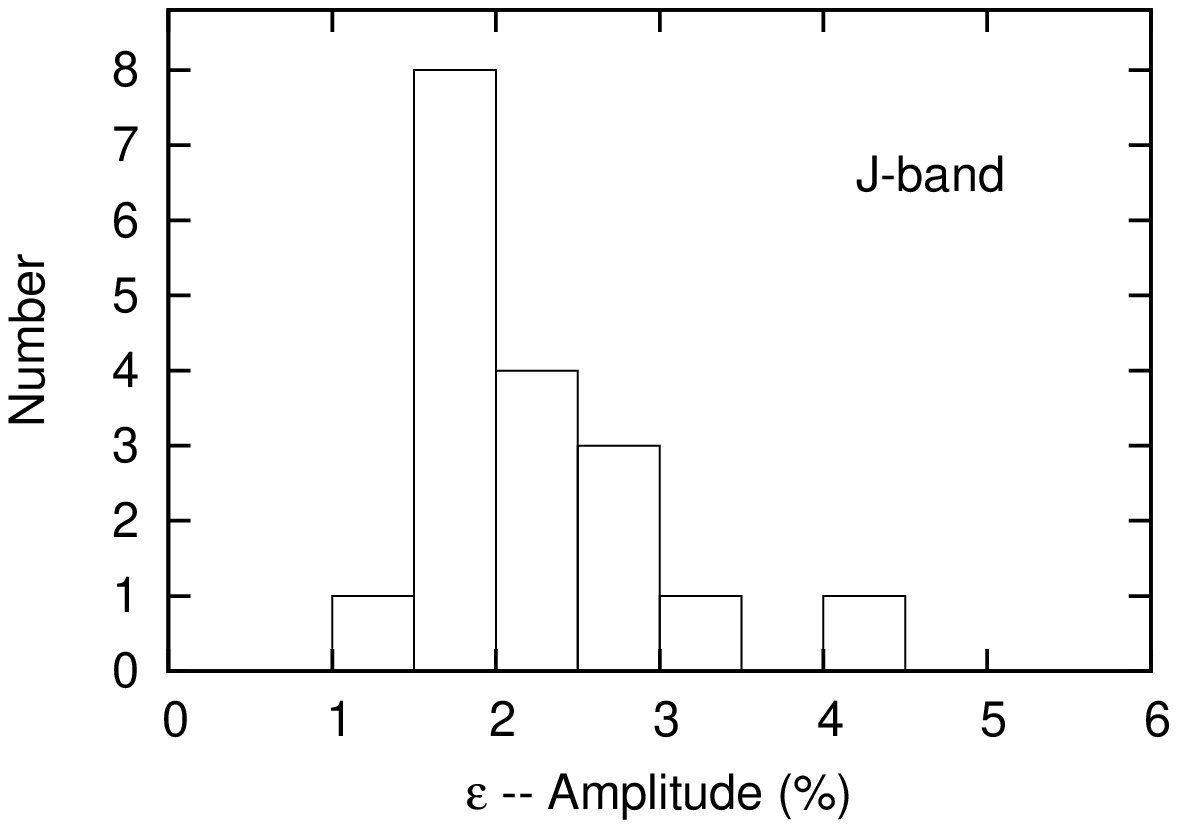}
\caption[]
	{	Histogram of the peak-to-peak amplitudes per complete rotation period, $\epsilon$,
		of our photometry of 2MASS 0036+18 at various wavelengths, as indicated in the panels.
	}
\label{FigMassJ0036HistogramAmplitude}
\end{figure}

We also investigate the evolution, or lack thereof, of the light curve of 
2MASS 0036+18 with time. 
Specifically we investigate the 
change in amplitude of the variability 
of 2MASS 0036+18
from rotation period to rotation period, and night-to-night.
For each complete rotation period of 2MASS 0036+18 we determine the peak-to-peak amplitude, $\epsilon$,
of the observed variability of that rotation period. To approximate $\epsilon$ we first 
exclude significant outliers (we cut out all data greater than 4 standard deviations from the mean);
then we time-bin the data,
by taking a running average of the photometry every \DataRunningDays \ days (or $\sim$\DataRunningMinutes \ minutes).
We then define $\epsilon$ as the difference between the maximum and minimum points of the time-binned
data observed over each rotation period.
We calculate $\epsilon$ for each complete rotation period
that we have observed; 
a complete rotation period
is defined when we have observed a full rotation period
after the start of each night of observations (we also exclude
$\sim$4 minutes at the start and end of each night of
observations to avoid possibly, systematically biased photometry).
We include rotation periods with small gaps of data missing due to clouds, or other telescope/instrument failures.
We display a histogram of the peak-to-peak amplitude values, $\epsilon$, for each complete rotation period
for our J-band, z'-band \& I-band photometry in Figure \ref{FigMassJ0036HistogramAmplitude}.
Our J-band, z'-band \& I-band histograms of our 2MASS 0036+18 photometry indicate that the 
variability we observe from 2MASS 0036+18 displays relatively constant amplitudes at these 
wavelengths; from rotation period to rotation period we appear to be observing modest amplitude evolution,
but not nearly as significant 
as for our similar photometry of the L/T transition brown dwarf 
SIMP J013656.5+093347 \citep{CrollUCDII}.
This lack of significant amplitude evolution appears to be consistent
for the duration of our Fall 2015 observations of 2MASS 0036+18
(with the exception of our 2015 November 23 light curve; Section \ref{SecNovemberTwentyThirdJBand}).
The much more modest evolution in amplitude displayed by 2MASS 0036+18 compared to L/T transition brown dwarfs,
suggests different variability properties on this L3.5 dwarf than brown dwarfs at the L/T transition.

We encourage continuing optical and near-infrared monitoring of this ultra-cool dwarf to determine how the variability amplitudes
evolve in future observing sessions.


\subsection{Comparison to $v$$\sin$$i$}
\label{SecVSinI}

There have been a range of $v$$\sin$$i$ measurements for 2MASS 0036+18. We will henceforth utilize the
$v$$\sin$$i$ measurement 
quoted in \citet{Crossfield14},
$v$$\sin$$i$ = 40.0 $\pm$ 2.0 $km$/$s$, a weighted mean of a number of 
different $v$$\sin$$i$ measurements \citep{Jones05,Zapatero06,ReinersBasri08}
for this ultra-cool dwarf.
Using the inferred radius 
of 2MASS 0036+18 of 0.09 $\pm$ 0.01 $R_{\odot}$ \citep{Dahn02}, and our 
inferred rotation period of
$P_{rot}$ = \PeriodHoursTwoMassZeroZeroThirtySixAllbands \ $\pm$ \PeriodHoursErrorTwoMassZeroZeroThirtySixAllbands \ $hr$, this
suggests a near edge-on inclination angle:
$\sin$$i$ = 1.13 $\pm$ 0.14, or $i$ $>$ 80$^o$ (1$\sigma$ error).

\subsection{Our 2015 November 23rd J-band light curve}
\label{SecNovemberTwentyThirdJBand}

 Our 2015 November 23rd J-band light curve of 2MASS 0036+18 displays a significantly longer period
than our other light curves of this object.
The Lomb-Scargle periodogram of this light curve
indicates a period of 
\PeriodHoursTwoMassZeroZeroThirtySixJbandFifteenNovemberTwentyThird $\pm$ \PeriodHoursErrorTwoMassZeroZeroThirtySixJbandFifteenNovemberTwentyThird 
\ hours, with a peak-to-peak amplitude of $\sim$4.6 \%. 
A period
approximately twice that of the rotation
period we reach with the rest of our data
(\PeriodHoursTwoMassZeroZeroThirtySixAllbands \ $\pm$ \PeriodHoursErrorTwoMassZeroZeroThirtySixAllbands \ hr)
indicates the possibility that the actual rotation period is twice our inferred value, and that
the variability we have been observing is due to two hot/cool spots on different sides of the ultra-cool
dwarf. However, a $\sim$\PeriodHoursTwoMassZeroZeroThirtySixJbandFifteenNovemberTwentyThird \ hour period would
result in a 
$v$$\sin$$i$ estimate of $\sim$14.6 $km$/$s$ 
(assuming an edge-on inclination angle), compared
to the much larger measured value (40.0 $\pm$ 2.0 $km$/$s$; \citealt{Crossfield14}). We therefore do not find this possibility
of a longer $\sim$\PeriodHoursTwoMassZeroZeroThirtySixJbandFifteenNovemberTwentyThird \ hour
rotation period for 2MASS 0036+18 compelling.

Another possibility is that the variability on that night is driven by another mechanism, unrelated
to the rotation period of that ultra-cool dwarf. 
The significant growth of a spot or cloud clearing that is visible for a complete rotation period
(e.g. a polar spot for edge-on inclination angles), or a multi-hour flare, could possibly explain our
2015 November 23rd J-band light curve. Significant growth (or decline) in the clouds that envelope 2MASS 0036+18, could also explain the observed
variability, and mask the true
rotation period of 2MASS 0036+18 (as suggested by \citealt{Gelino02}) on that evening.

The last possibility is that systematic errors have affected the light curve
on that evening. However, the weather was clear that evening, and we have ensured
that the variability we observe on that evening is not due to a single reference star.
Reference stars of similar brightness to 2MASS 0036+18, if analyzed utilizing the same techniques we apply to
2MASS 0036+18, do not display prominent variability on that evening,
and therefore we have no strong evidence in favour that systematics affect our light curve.
For this reason, we encourage further optical and near-infrared monitoring of this star to indicate whether
variable light curves displaying periods
longer than the 
rotation period we infer from our photometry
(\PeriodHoursTwoMassZeroZeroThirtySixAllbands \ $\pm$ \PeriodHoursErrorTwoMassZeroZeroThirtySixAllbands \ hr)
are observed for this ultra-cool dwarf.

\section{A limit on Optical/Infrared Flares}
\label{SecFlares}


Our optical and near-infrared photometry of 2MASS 0036+18 also allows us to place a limit on the frequency of flares 
exhibited by this ultra-cool dwarf. 2MASS 0036+18 was observed to display a 
flaring event for $\sim$20 minutes from 3 hours of radio observations \citep{Berger02}; $\sim$18 hours of follow-up radio
observations of 2MASS 0036+18 by \citet{Berger05} did not detect flares, and placed a limit on radio flaring
occurrences of $<$0.04 per hour.
We do not see obvious signs of significant flaring activity in our 
\INSERTNIGHTS \ nights of optical and near-infrared photometry that consists of $\sim$\TOTALHOURS \ hours of photometry.
Inspection of the unbinned light curves suggests that we can conservatively 
rule-out\footnote{Although please see the discussion in Section \ref{SecNovemberTwentyThirdJBand} about our 2015 November 23rd J-band light curve.}
flares lasting twenty minutes or more that increase
the optical or infrared flux by more than 10\% in all our 
$\sim$\TOTALHOURS \ hours of optical and near-infrared photometry, consisting of 
$\sim$\TOTALNONOVERLAPPINGHOURS \ hours of total, non-overlapping hours
(excluding time when two telescopes are observing 2MASS 0036+18
at the same time).
More stringent limits could likely be set using our photometric data-sets by applying a more detailed inspection
of the light curves, including injection
and recovery of putative flare events.
We note that our limit on flares assumes that the persistent 
$\sim$\PeriodHoursTwoMassZeroZeroThirtySixAllbandsQuote \ hour
variability we observe is not due to periodic flaring --
a possibility suggested by \citet{Berger05}.

Our limit on optical and near-infrared flares is not stringent enough to suggest that
the flare-rate of 
2MASS 0036+18 is dissimilar to that of the L1 dwarf WISEP J190648.47+401106.8 (hereafter W1906+40); 
\citet{Gizis13} presented 15 months of {\it Kepler} white light photometry of this ultra-cool dwarf, and 
suggested that the rate of energetic flares ($>$10$^{31}$ erg) was approximately 1-2 per month (or $\sim$730 hours).
The observed flares of W1906+40 (with energies $>$ 10$^{31}$ erg) 
increased the {\it Kepler} flux by up to four times the mean flux for durations of up to 1 hour.
Simultaneous Gemini spectra suggested flare temperatures of 8000 $\pm$ 2000 $K$ of W1906+40, meaning
that the impact of flares of such temperatures would be diluted in our red optical and near-infrared photometry of 2MASS 0036+18,
compared to the {\it Kepler} bandpass for \citet{Gizis13} photometry of W1906+40.
Nonetheless our much more stringent limit on flares, compared to the observed more frequent radio flares \citep{Berger02,Berger05},
suggest that either the \citet{Berger02} observed radio flare occurs rarely,
or that there is little correlation between radio flares and those observed in the optical \& near-infrared
(a conclusion similar to that reached from simultaneous optical \& radio monitoring of
two M8.5 dwarfs; \citealt{Berger08a,Berger08b}).

\section{Searching for Earth-sized planets in the Habitable Zone}
\label{SecPlanets}

\begin{figure}
\includegraphics[scale=0.52, angle = 270]{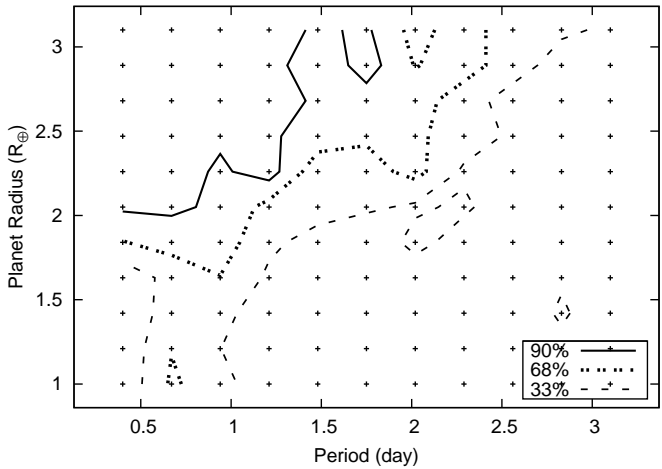}
\caption[]
	{	Percentage of inserted edge-on ($i_p$=90$^o$) transits (as indicated in the legend)
		that are recovered as a function of planet radius and orbital period using our 
		multiwavelength ground-based photometry of 2MASS 0036+18.
		The small tick marks indicate the periods and radii at which
		signals from transiting planets were injected in the data. 
		
	}
\label{FigInsertedTransits}
\end{figure}

Our multi-night, multiwavelength light curves also enable a search for transiting Earth-sized planets in the habitable
zone of this brown dwarf.
The inferred radius 
of 2MASS 0036+18 (0.09 $\pm$ 0.01 $R_{\odot}$; \citealt{Dahn02}),
means that even Earth-sized planets will display
$\sim$1\% transit depths. Using the inferred effective temperature
of $\sim$1700 K \citep{Cushing08}, planets with periods of a day or less to a few days
will be sufficiently warm such that liquid water might be able to exist on these planet's surface
(depending on the effects of tidal heating and the influence of various atmospheric
compositions and albedos; \citealt{Bolmont11}; \citealt{BarnesHeller13}; \citealt{Zsom13}).


Further motivation for searching for nearby companions in this system is provided by \citet{Berger05} and \citet{Pineda16}; these authors suggest
the possibility that the observed radio emission from 2MASS 0036+18 
could result from tidal interactions with a close companion, analogous to Jupiter's interactions with its moon
Io, or Saturn's interaction with its moon Enceladus.
\citet{Berger05} originally suggested the putative companion would have a $\sim$3 hour orbital period and tidal interactions would
explain the periodicity in the
radio observations of this system; \citet{Hallinan08}, however, explained the radio emissions as being due to the electron cyclotron maser instability.
Nonetheless, \citet{Pineda16} recently suggested that the observed 
H$\alpha$ modulation of 2MASS 0036+18 could be induced by tidal interactions with a longer period planetary-mass companion.

 Also, 
one might expect that if there is a planet in the system
that it may very well lie along, or close to, the line of sight.
This is because: (i) first of all, the comparison between the $v$$\sin$$i$ measurement 
and the rotation period 
for this ultra-cool dwarf
suggests a near edge-on rotation angle, (ii) second, the conclusion of \citet{Albrecht12} 
suggests that lower temperature stars seem to host spin-orbit aligned planets much more frequently than hotter stars, and
this trend may continue into the sub-stellar regime.

To determine our ability to detect transiting, rocky exoplanets with our ground-based
multiwavelength light curves of 
2MASS 0036+18
we inject transits of simulated
exoplanets into these light curves and test our ability to recover these planets.
We use similar methodology to that presented in \citet{Croll07,Croll07b}.
We judge an inserted transit to be detected if the phase, $\phi_T$,
and period, $P_T$, returned by the transit search algorithm
were sufficient close to the input parameters, $P_{inp}$ \& $\phi_{inp}$.
The detected period had to satisfy the following criterion: $|P_T/P_{inp} - 1|$ $<$ 1\%.
If the transit routine returned 
an obvious harmonic of the orbital period (and the associated phase was incorrect), these periods
were also judged to be detected; values up to four times the inserted period ($P_T$ $\sim$ 4 $\times$$P_{inp}$),
or down to one-quarter the inserted period ($P_T$ $\sim$ $\frac{1}{4}$ $\times$ $P_{inp}$) were therefore accepted.

The results of these injection and recovery of transit simulations are presented in Figure \ref{FigInsertedTransits}.
Our Monte Carlo simulations feature injected transits at \InsertRadius \ radii points and \InsertPeriod \ period
values, as indicated by the tick marks
in Figure \ref{FigInsertedTransits}.
For each radius, $R_{inp}$, and period, $P_{inp}$, value that transits were injected into the light curve
for our Monte Carlo simulations, we repeat the injection and recovery for \InsertPhase \ random phase values at that radius and
period value. 

We note that much more stringent limits can likely be set from this photometry for our Monte Carlo tests by utilizing techniques to
speficially detect single transit events. We have ruled out such single transit events by visual inspection for our photometry of 
2MASS 0036+18, but have not performed such inspection for our Monte Carlo data-sets.

We lastly note that we detect no compelling transiting planet signals in our ground-based photometry of this system.
We therefore believe we can conservatively rule out edge-on transiting planets in this system 
with radii equal to or greater than those given in Figure \ref{FigInsertedTransits} at the associated periods.
As we have inspected our light curves for single transit events, we can likely set more stringent limits;
we therefore believe that our photometry is sensitive to transiting planets as large or larger than super-Earths
for most of the habitable
zone around this ultra-cool dwarf.


\section{The Physical Mechanism Driving Variability on 2MASS 0036+18}
\label{SecDiscussion}

Our 
\INSERTNIGHTS \ nights of photometry, including 
\INSERTNIGHTSSIMULTANEOUS \ nights of multiwavelength photometry, of
2MASS 0036+18 display a lack of obvious phase shifts between the variability observed simultaneously
at different wavelengths, and variability amplitudes that generally decrease with increasing wavelength.
Such behaviour is expected for starspot modulation, and is less consistent with
variability driven by gaps in clouds, temperature variations, or aurorae, unless
these other mechanisms only cause multiwavelength phase variations as small or smaller than what 
we have observed (Table \ref{TableSinusoidIndividualLightCurves}).
Starspot modulation causing the variability of 2MASS 0036+18 
is also buttressed by the recent detection of 
H$\alpha$ from 2MASS 0036+18 \citep{Pineda16}.
In Section \ref{SecStarSpotz} we present photometric starspot models that demonstrate
that cool or hot spots can reproduce
the profile of variability that we have generally observed for 2MASS 0036+18.
In Section \ref{SecSpotsTemperature} we constrain the temperature of starspots
on this ultra-cool dwarf, and demonstrate that
spots $\sim$100 $K$ cooler or hotter than the photosphere of 2MASS 0036+18 can explain
the observed variability. In Section \ref{SecClouds} we constrain the characteristics of clouds that are consistent
with the variability that we observe on 2MASS 0036+18; gaps in clouds that reveal hotter layers underneath
and span multiple pressure layers can explain the observed variability.

\subsection{Starspot model for 2MASS 0036+18}
\label{SecStarSpotz}

\begin{figure}
\centering
\includegraphics[scale = 0.045, angle = 0]{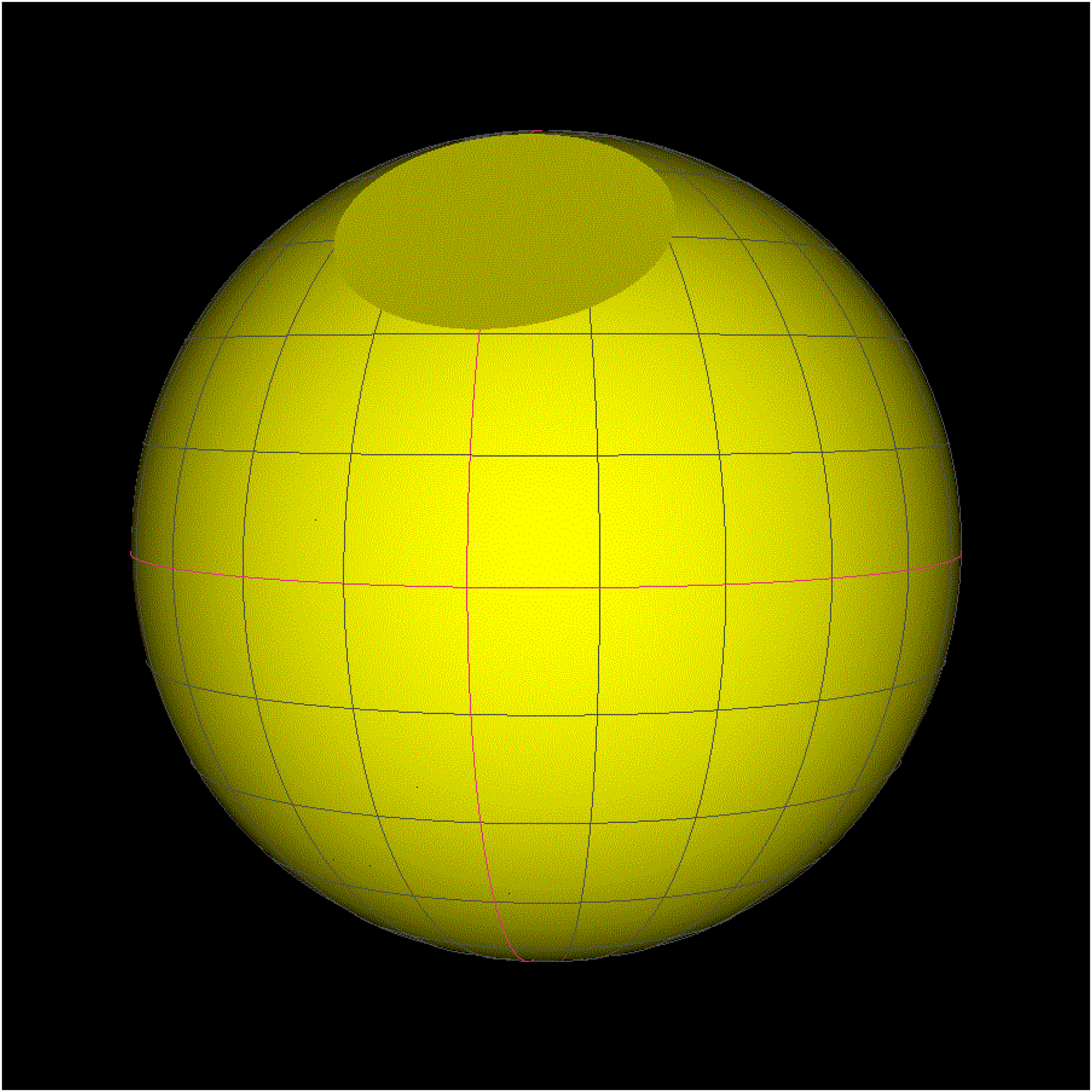}
\includegraphics[scale = 0.045, angle = 0]{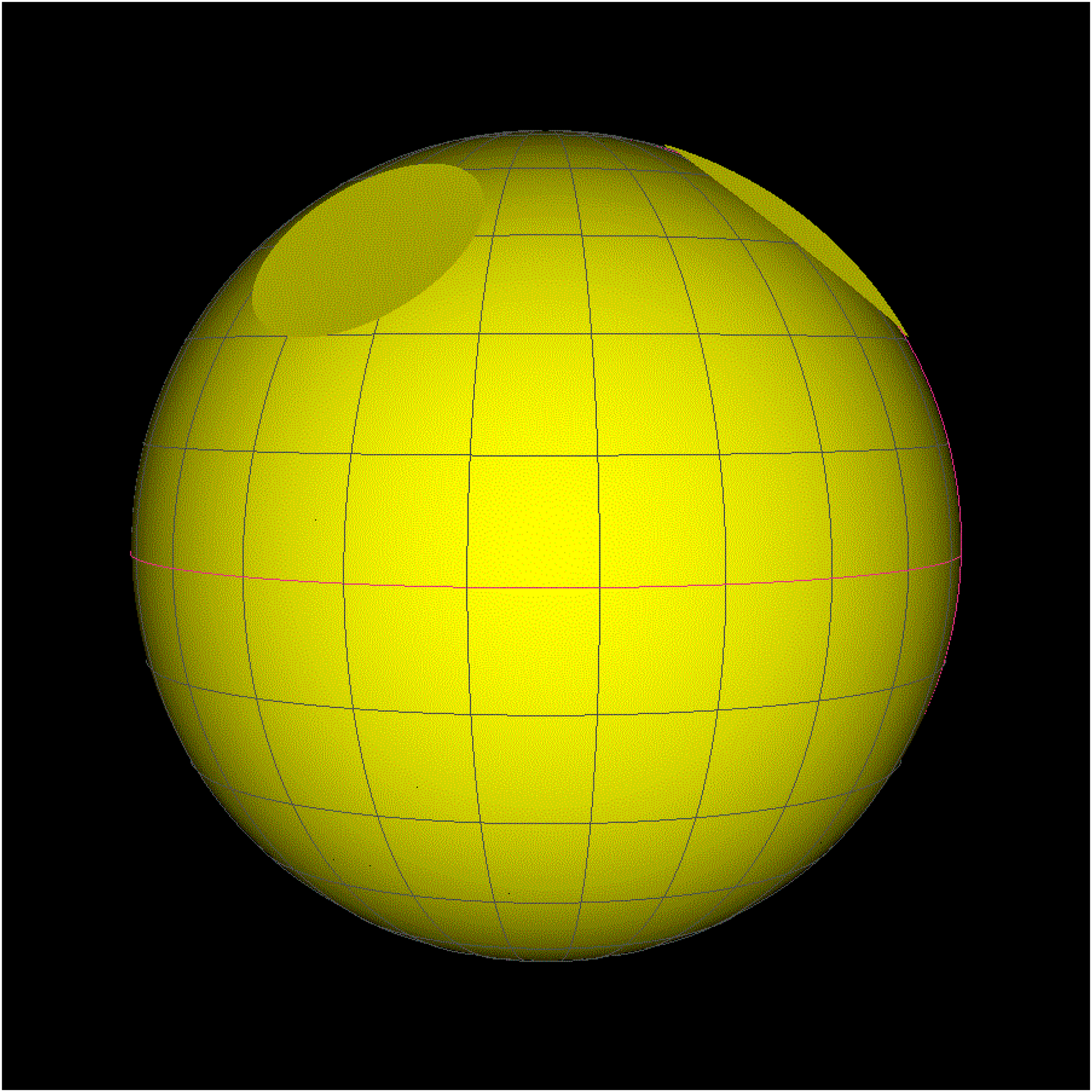}
\includegraphics[scale = 0.045, angle = 0]{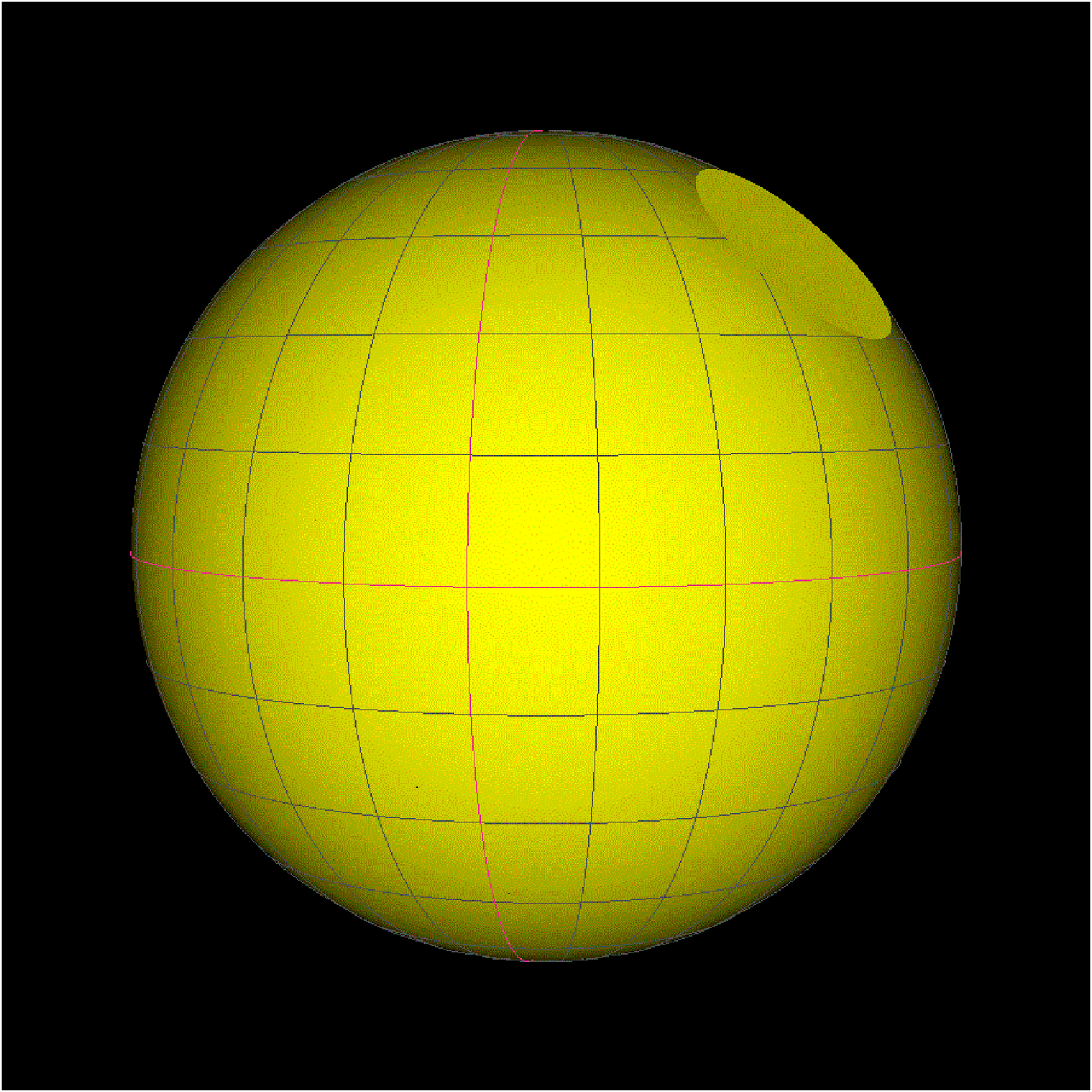}
\includegraphics[scale = 0.045, angle = 0]{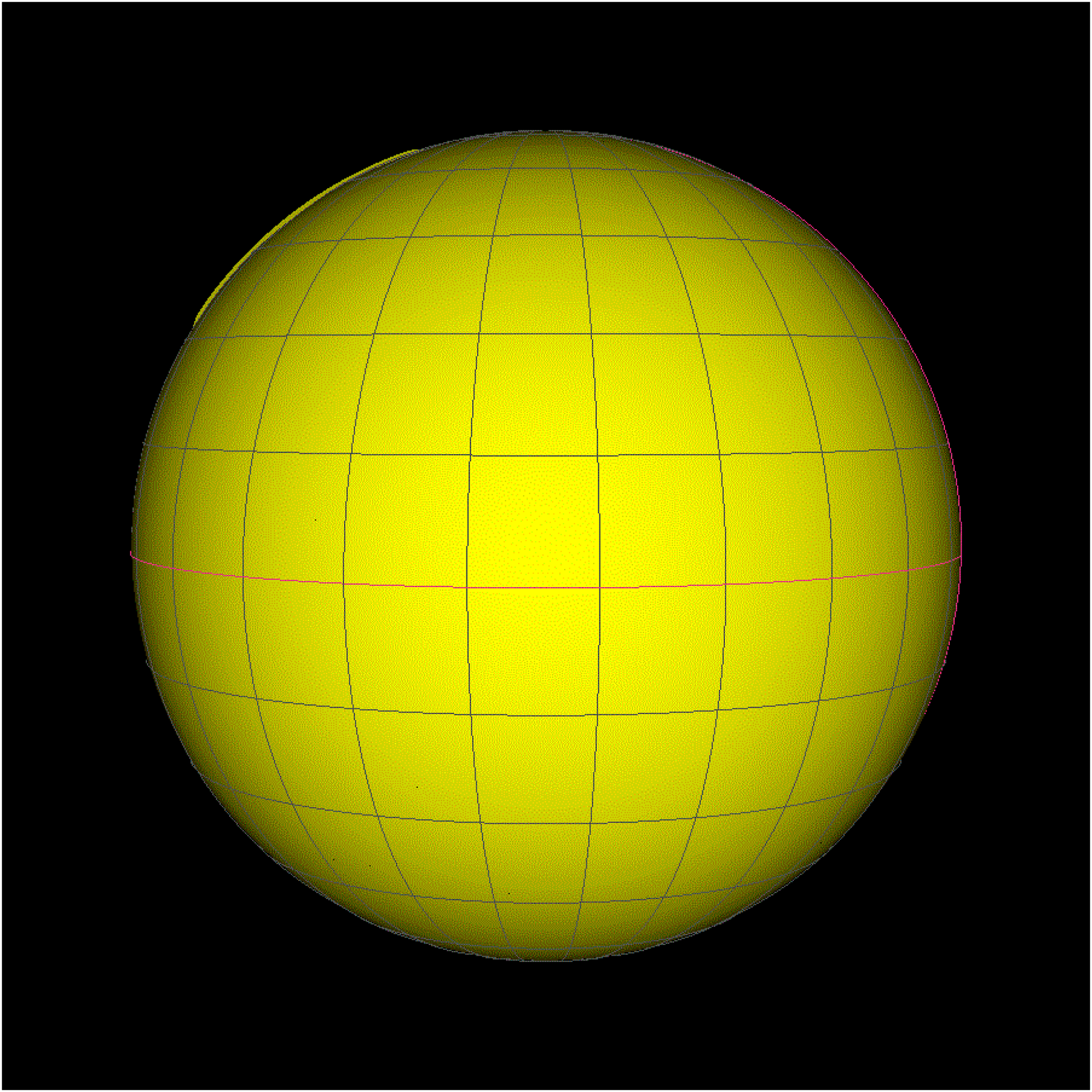}
\includegraphics[scale = 0.46, angle = 270]{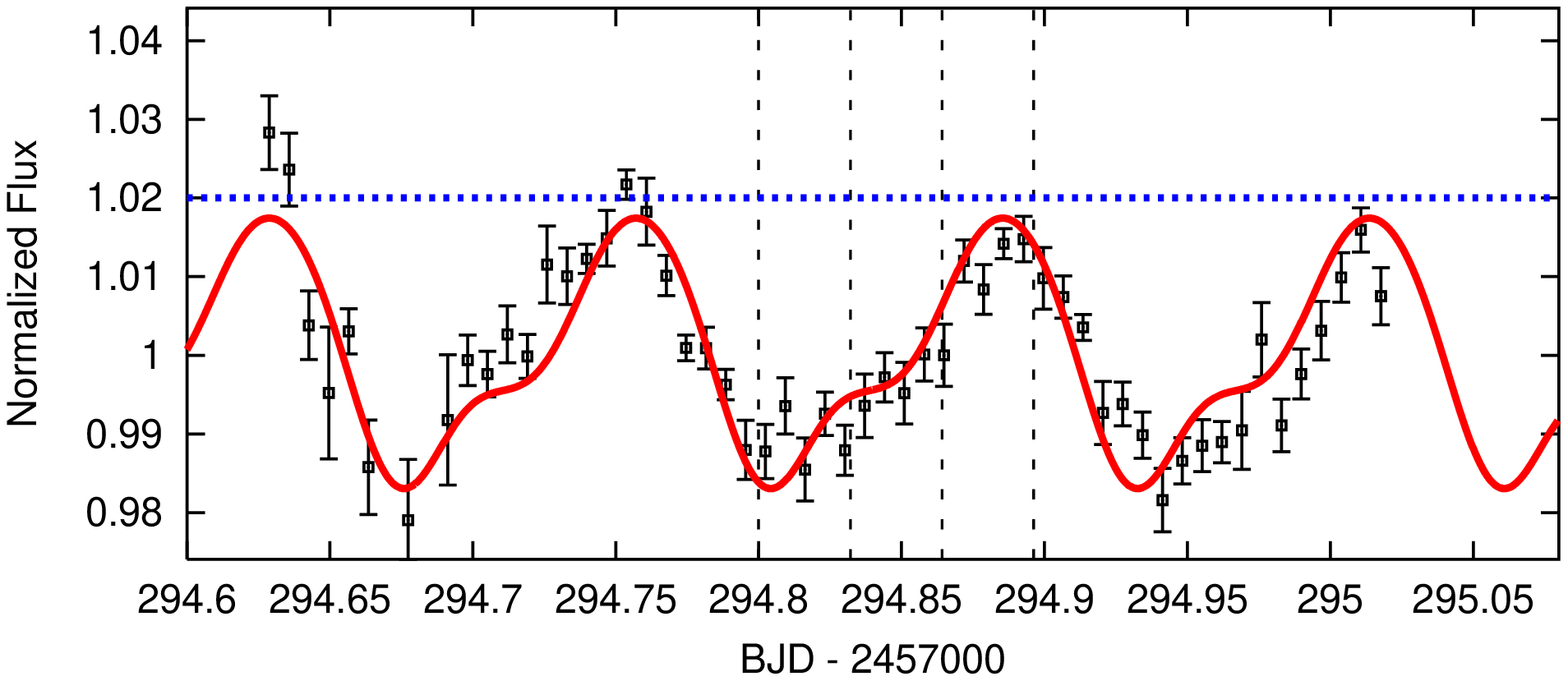}
\caption[]
	{			
	Top panels: possible spot model of 2MASS 0036+18 with cold spots ($\Delta$T = -100 $K$)
	as seen from the line of sight at stellar rotation phases increasing
	(from left).
	The bottom panel displays our {\it StarSpotz} cold spot fit
	(the red solid curve) to our z'-band photometry (black points
	with error bars) on UT 2015/09/29. The blue horizontal dotted line
	displays the unspotted photosphere of the star. The vertical dotted black
	lines indicate the phases corresponding to the spot models in
	the top panels (increasing from left).
	}
\label{FigSpotsCold}
\end{figure}

\begin{figure}
\centering
\includegraphics[scale = 0.045, angle = 0]{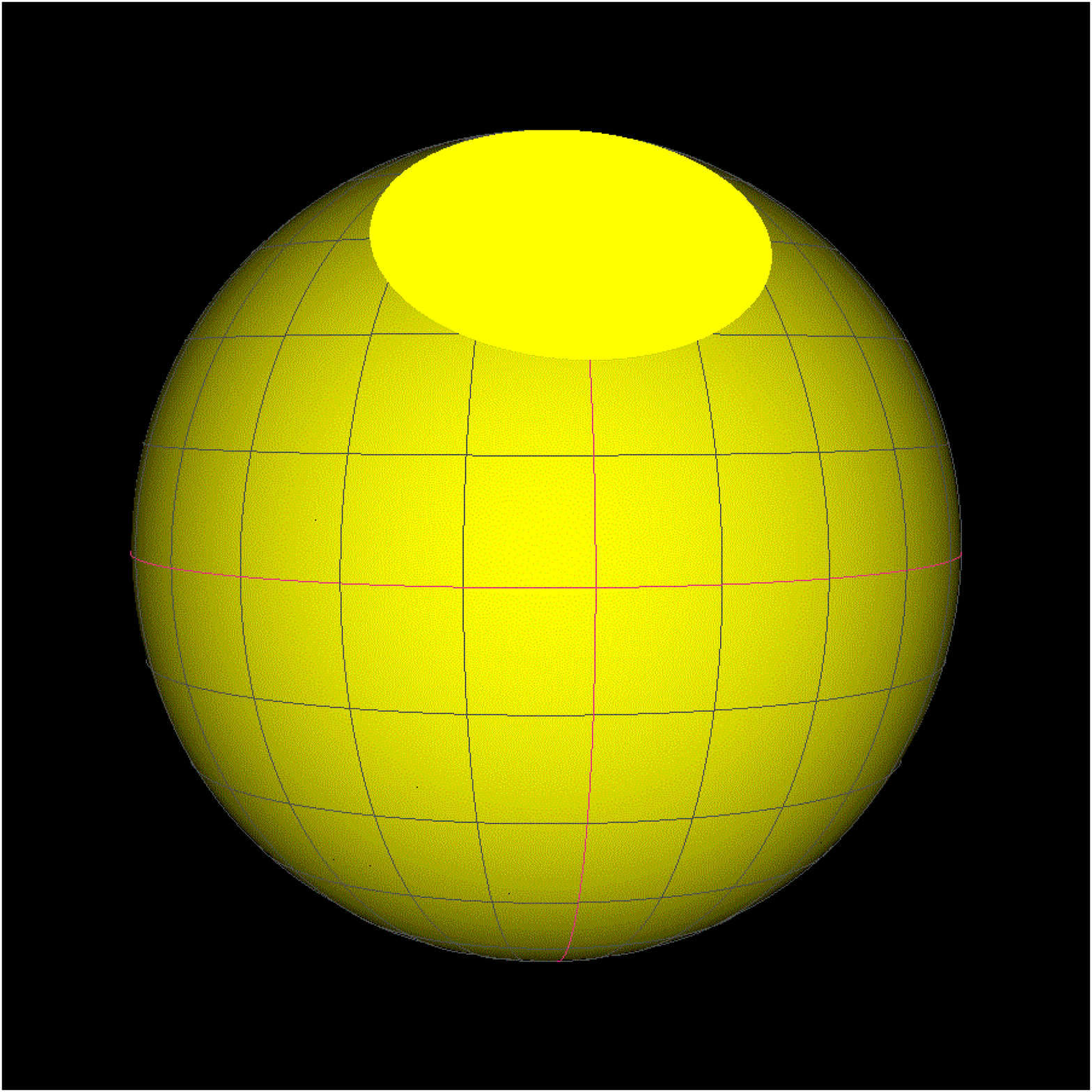}
\includegraphics[scale = 0.045, angle = 0]{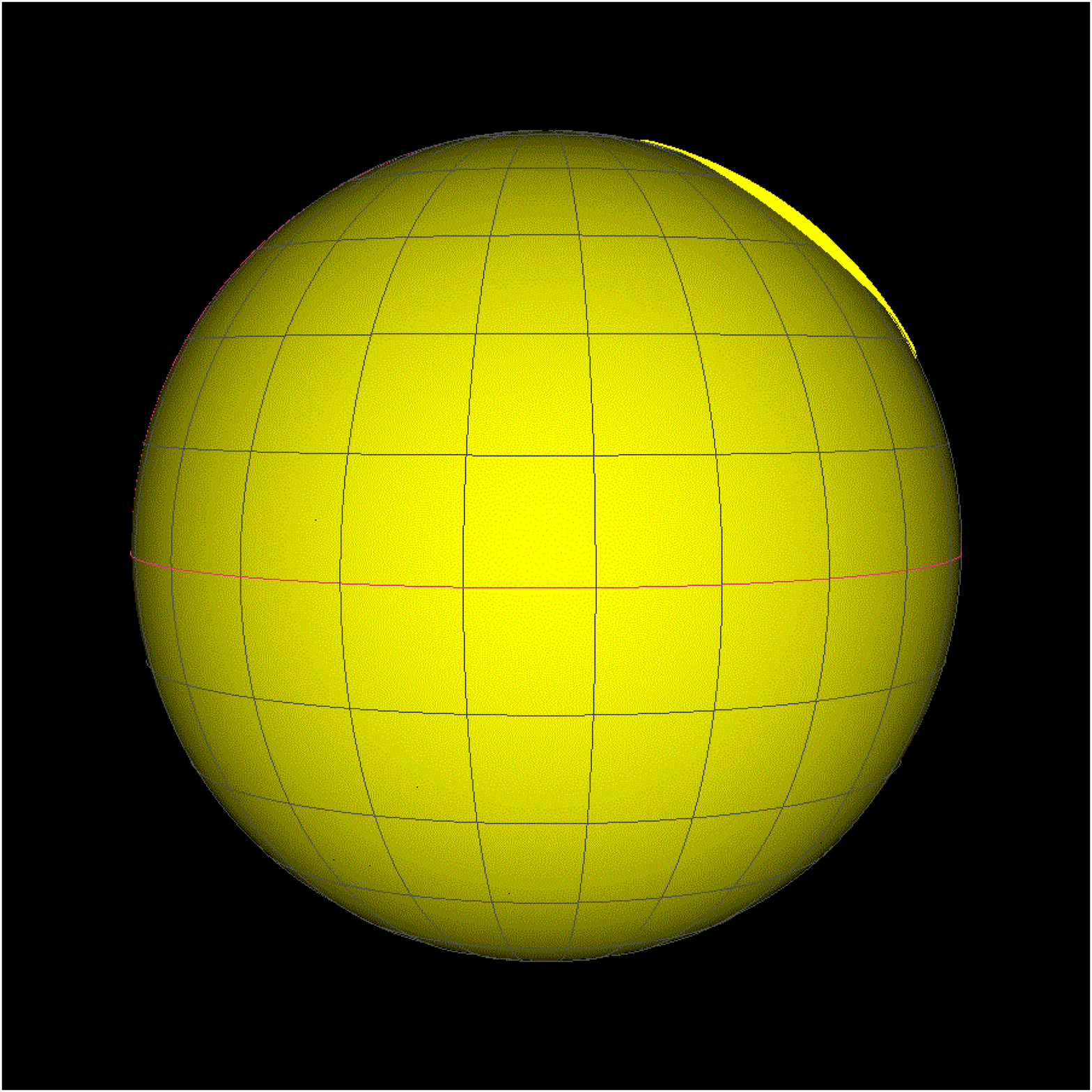}
\includegraphics[scale = 0.045, angle = 0]{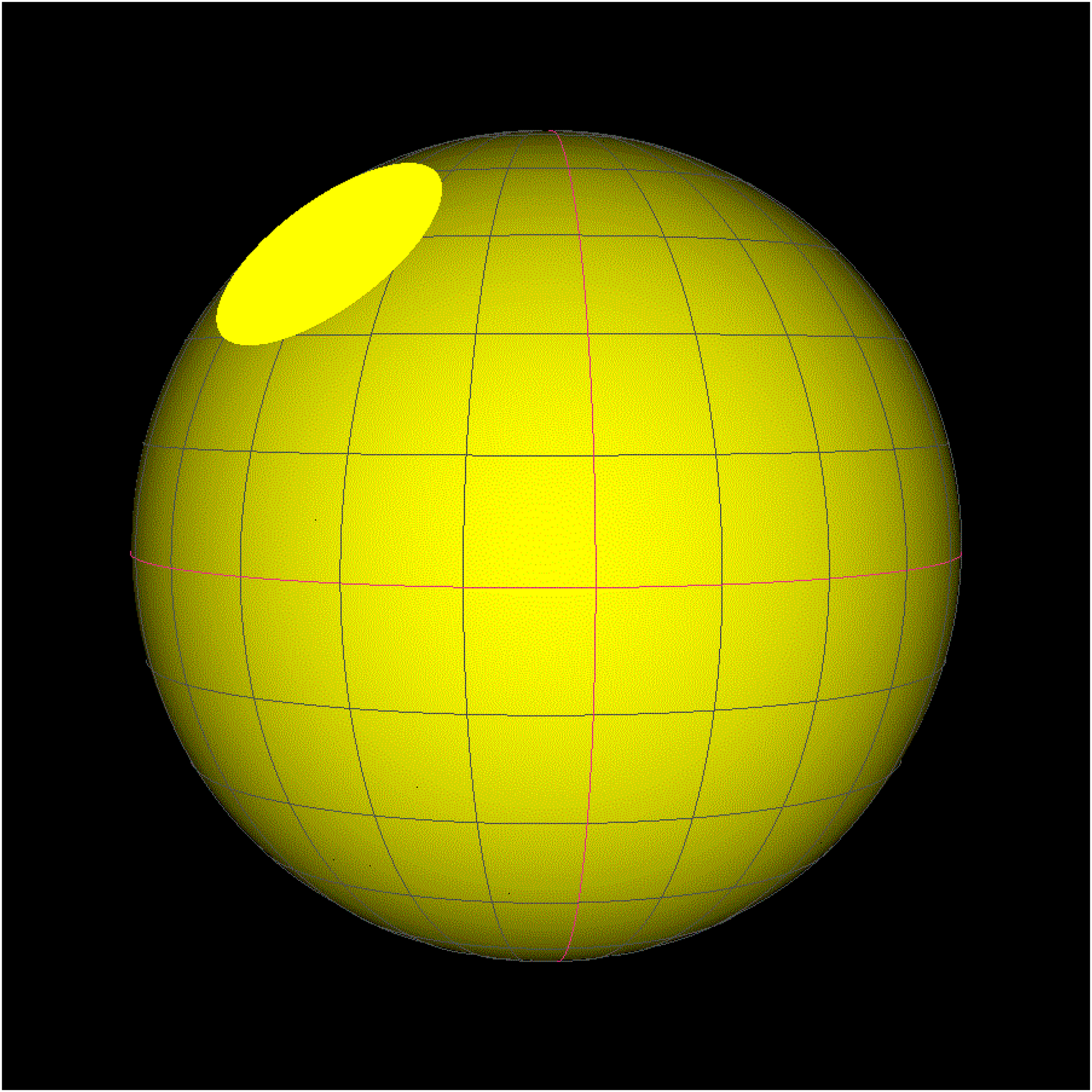}
\includegraphics[scale = 0.045, angle = 0]{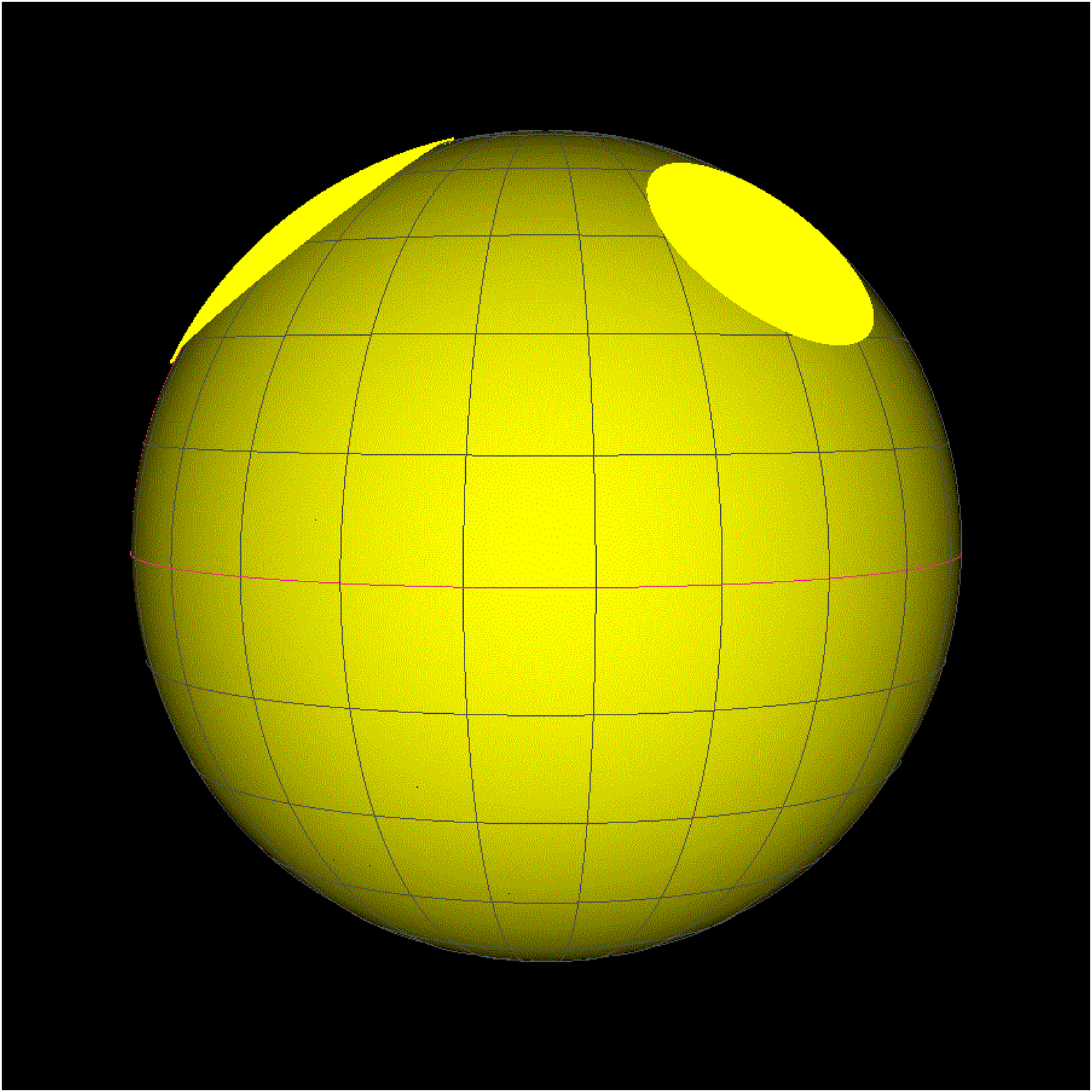}
\includegraphics[scale = 0.46, angle = 270]{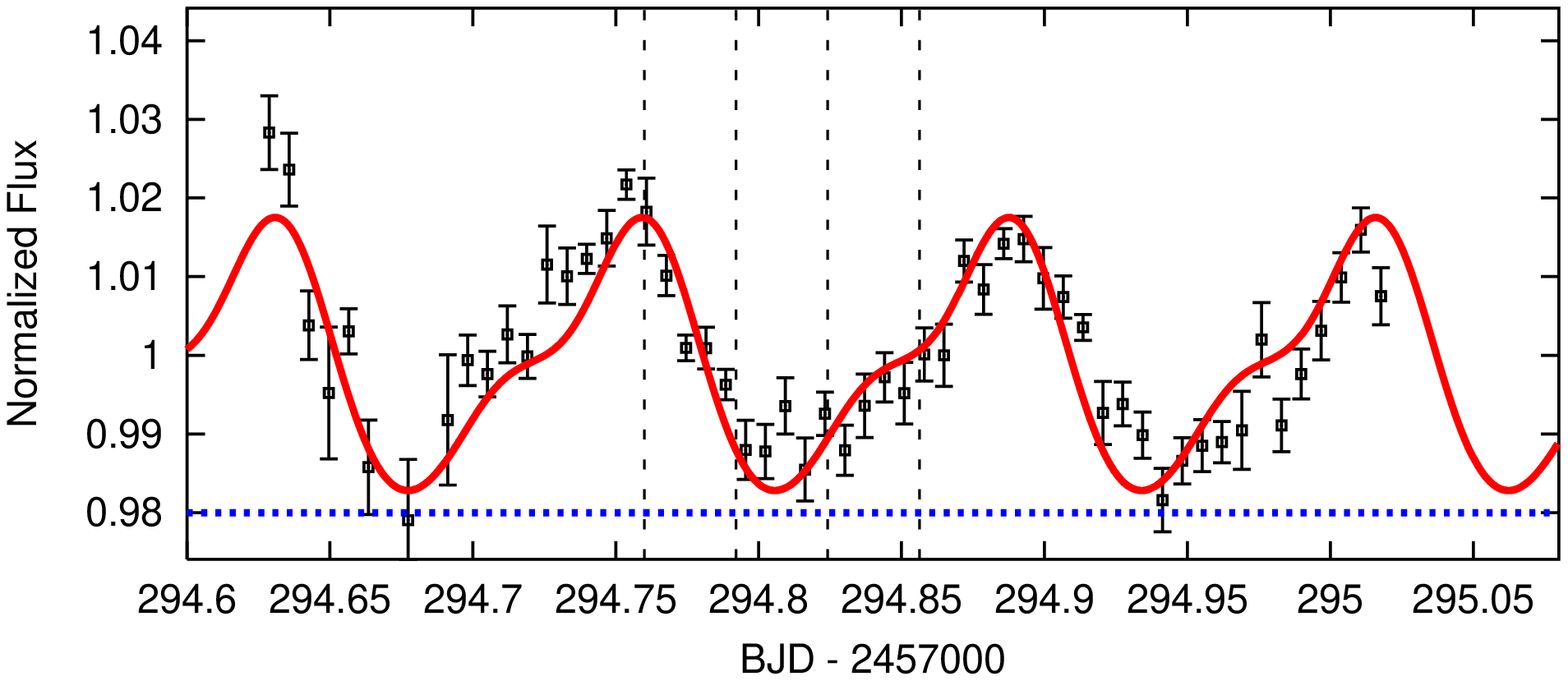}
\caption[]
	{			
	Top panels: possible spot model of 2MASS 0036+18 with hot spots ($\Delta$T = +100 $K$)
	as seen from the line of sight at stellar rotation phases increasing
	(from left).
	The bottom panel displays our {\it StarSpotz} hot spot fit
	(the red solid curve) to our z'-band photometry (black points
	with error bars) on UT 2015/09/29. The figure is otherwise identical
	to Figure \ref{FigSpotsCold}.
	}
\label{FigSpotsHot}
\end{figure}



 We have attempted to reproduce the variability of 2MASS 0036+18 using a photometric starspot model.
Our various 2MASS 0036+18 photometric light curves are fit using a
\citet{Budding77} model utilizing the {\it StarSpotz}
methodology \citep{Croll06,CrollMCMC,Walker07}.
Given that the combination of our inferred rotation period (\PeriodHoursTwoMassZeroZeroThirtySixAllbands \ $\pm$ \PeriodHoursErrorTwoMassZeroZeroThirtySixAllbands \ hr),
and the $v$sin$i$ measurements for this ultra-cool dwarf suggest a near edge-on 
inclination angle (Section \ref{SecVSinI}), the observed sinusoidal variability that persists for a large fraction
of the rotation period is difficult to reproduce with a single spot,
whether at low or high latitudes. Generically single spots at low latitudes for near edge-on inclination angles produce sharper decrements
in flux that do not persist for a large fraction of the rotation period (e.g. the variability displayed in the 2003 MOST photometry
of $\kappa$1 Ceti; \citealt{Rucinski04,Walker07}).
Multiple spots appear to be required to reproduce the observed variability for 2MASS 0036+18 for near edge-on inclination angles.
For spots near the equator for edge-on inclination angles,
many spots will be required to reproduce the near sinusoidal variability; as few as two spots might
reproduce the near sinusoidal variability if the spots are near the poles for near edge-on inclination angles.
If we are viewing 2MASS 0036+18 at a low inclination angle,
fewer star spots are likely required to reproduce the variability, although a near pole-on
viewing geometry would be 
difficult to reconcile
with the rotation period and $v$sin$i$ constraints for this object (see Section \ref{SecVSinI}).

We present a possible two-spot model with cool spots for a near edge-on inclination angle
for our z'-band light curve on UT 2015/09/29
in Figure \ref{FigSpotsCold}.
Also, spots hotter than the 2MASS 0036+18 photosphere are also able to reproduce the observed variability.
We present a possible two-spot model with hot spots in Figure \ref{FigSpotsHot}.
These models are not intended to display unique cold or hot spot solutions for the observed variability,
as photometric starspot models are infamously non-unique (e.g. \citealt{CrollMCMC,Walker07,Aigrain15}) - 
instead these models are simply intended to prove that cold and hot spot models with an edge-on inclination angle
can reproduce the profile of variability generally displayed by 2MASS 0036+18.

\subsection{A limit on the Temperature of Starspots on 2MASS 0036+18}
\label{SecSpotsTemperature}

If 2MASS 0036+18's variability is due to starspots then the multiwavelength 
amplitudes we observe allow us to
place a limit on the temperature of these starspots compared to that of the photosphere.
We first do this in Section \ref{SecSpotsBlackbody} employing
the approximation that 2MASS 0036+18 and its spots are blackbodies,
to give a first-order approximation of the temperatures of spots on this L3.5 dwarf.
Subsequently, in Section \ref{SecSpotsModels}, we determine the temperature of starspots on
this ultra-cool dwarf by comparing to 
one-dimensional cloud models. 

For both the blackbody models and the one-dimensional cloud models,
to facilitate this comparison we employ the method presented in \citet{Radigan12}.
Equation 2 of \citet{Radigan12} showed that the expected peak-to-peak variability amplitudes, $A$,  driven
by two regions with surface fluxes $\mathcal{F}_1$ and $\mathcal{F}_2$ rotating in and out of view can be expressed as:
\begin{equation}
A = \frac{ (1 - a - \Delta a) \mathcal{F}_1 + (a+\Delta a) \mathcal{F}_2 - (1-a) \mathcal{F}_1 - a \mathcal{F}_2}
{0.5[(1-a - \Delta a) \mathcal{F}_1 + (a + \Delta a) \mathcal{F}_2 + (1-a) \mathcal{F}_1 + a \mathcal{F}_2]}
\label{EquationRadigan}
\end{equation}
where 
$a$ is the minimum filling factor of the $\mathcal{F}_2$ region on the ultra-cool dwarf,
while $\Delta a$ is the change in this filling factor.
We assume that the atmosphere of 2MASS 0036+18 is dominated by the $F_1$ region of temperature $T_{UCD}$ with 
small spots/clouds of different temperatures, $T_{spot}$, and associated surface fluxes $F_2$,
rotating in and our of view.
We assume $a$ = 0 and therefore there is a time during each rotation period that no spots are in view along the line
of sight.

For these comparisons, we assume that the variability of 2MASS 0036+18 is constant, and therefore we can utilize the amplitudes
of variability observed at different wavelengths from different epochs (Table \ref{TableSinusoidBands}).
This is likely a reasonable approximation given that we do not observe significant evolution in the variability amplitudes
during our multiwavelength photometry from rotation period to rotation period (Figure \ref{FigMassJ0036HistogramAmplitude}).
In Section \ref{SecSpotsBlackbody} we also investigate whether different results are obtained 
if we allow the
spot sizes to change with time
and therefore must compare solely to 
the amplitudes from our simultaneous, multiwavelength photometry (Table \ref{TableSinusoidIndividualLightCurves}).
The conclusions we reach are qualitatively similar in both cases, and 
therefore henceforth make the assumption that the spot sizes are constant with time for most of the following analysis.


\subsubsection{Blackbody Models}
\label{SecSpotsBlackbody}

\begin{figure}
\includegraphics[scale=0.76, angle = 270]{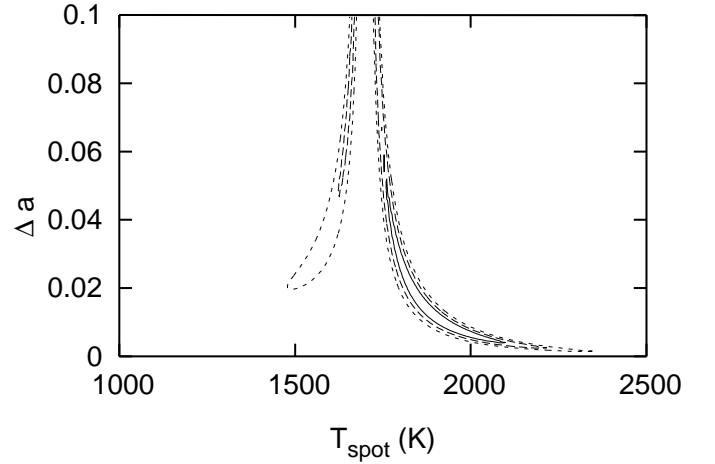}
\caption[]
	{	The change in filling factor, $\Delta a$ (a proxy for spot size),
		and temperature of spots, $T_{spot}$, that are consistent with the amplitudes of sinusoidal variability observed at various wavelengths
		assuming the spots and 2MASS 0036+18 radiate like blackbodies.
		We assume constant spots and therefore compare to the amplitudes of multiwavelength variability
		observed at different epochs (Table \ref{TableSinusoidBands}).
		The solid-black, dashed and dotted lines are the 1$\sigma$, 2$\sigma$ and 3$\sigma$ credible regions, respectively.
		Starspots on 2MASS 0036+18 can be either
		slightly hotter or slighter cooler than the photosphere to be consistent with the amplitudes of variability 
		that have been observed, if the spots and photosphere of this L3.5 dwarf radiate like blackbodies. 
	}
\label{FigTempSpots}
\end{figure}

We first determine the temperature of starspots on 2MASS 0036+18 assuming that the ultra-cool dwarf and its
starspots radiate like blackbodies. 
We use a $T_{UCD}$ = 1700 $K$ photosphere 
for 2MASS 0036+18 (\citealt{Cushing08} and see Section \ref{SecSpotsModels}),
and assume a single spot temperature.
We employ Planck functions to calculate the flux for the ultra-cool dwarf, $\mathcal{F}_1$,
and for the starspot, $\mathcal{F}_2$, with a temperature $T_{spot}$.
As we assume $a$ = 0 (and therefore there are times that no spots are visible), we compare
the predicted variability amplitudes for different values of the change in filling factor, $\Delta a$,
and the temperature of the spot, $T_{spot}$. 

For the assumption of constant spots, and therefore constant multiwavelength amplitudes,
we can compare to the
multiwavelength amplitudes that have been observed previously at different epochs (Table \ref{TableSinusoidBands}).
For blackbody emission, 
we present the best-fit spot temperatures in Figure \ref{FigTempSpots}. 
We scale the resulting reduced $\chi^2$ to $\sim$1, to allow for reasonable-sized error bars on $T_{spot}$.
The best-fit spot temperature can be slightly cooler than or slightly hotter than the photosphere of 
2MASS 0036+18. The temperature is highly correlated with the assumed change in the filling factor, $\Delta a$,
and therefore the size of the spots.
For blackbody emission, if the spots are slightly cooler than 2MASS 0036+18 
the allowed spot temperature ranges are approximately
$T_{spot}$ = \TSpotKelvinValue \ $\pm$ \TSpotKelvinValueError \ $K$,
while if the spots are slightly hotter than 2MASS 0036+18 the allowed temperature ranges are approximately
$T_{spot}$ = \TSpotHotKelvinValue \ $\pm$ \TSpotHotKelvinValueError \ $K$.

We also compare our blackbody models 
to the measured amplitudes solely from our simultaneous, multiwavlength photometry
(Table \ref{TableSinusoidIndividualLightCurves}).
We allow for different values of the change in filling factor ($\Delta a$) for each individual night of photometry,
and attempt to determine the spot temperature, $T_{spot}$, that provides the best-fit (lowest $\chi^2$)
to our simultaneous, multiwavelength amplitudes. 
Utilizing a $T_{UCD}$ = 1700 $K$ photospheric temperature for 2MASS 0036+18, the best-fit spot temperature
is approximately $T_{spot}$ $<$ 1500 $K$ 
for the Table \ref{TableSinusoidIndividualLightCurves} simultaneous, multiwavelength amplitudes;
this lower spot temperature is driven strongly by the 2015/10/14 DCT light curve, for which the measured
R-band peak-to-peak amplitude
(\SinusoidPeaktoPeakAmplitudePercentageTwoMassZeroZeroThirtySixFifteenOctoberFourteenRband \ $\pm$ \SinusoidPeaktoPeakAmplitudePercentageErrorTwoMassZeroZeroThirtySixFifteenOctoberFourteenRband \%)
is less than the z'-band amplitude
(\SinusoidPeaktoPeakAmplitudePercentageTwoMassZeroZeroThirtySixFifteenOctoberFourteenzband \ $\pm$ \SinusoidPeaktoPeakAmplitudePercentageErrorTwoMassZeroZeroThirtySixFifteenOctoberFourteenzband \%),
and has the smallest associated error-bars of the amplitudes listed in Table \ref{TableSinusoidIndividualLightCurves}.
However, the 2015/10/14 DCT light curve ends at high airmass ($\sim$2.31); as the airmass increases that night
the multiwavelength R-band and z'-band light curves diverge from their close agreement from earlier that evening. 
As we suspect this might be an airmass effect caused by the comparison of the redder target star to the bluer reference stars,
we also repeat our analysis utilizing the 2015/10/14 DCT light curve with data later than BJD $\sim$309.9 excluded.
The associated simultaneous, multiwavelength peak-to-peak amplitudes for our 2015/10/14 DCT light curve 
are then
\SinusoidPeaktoPeakAmplitudePercentageTwoMassZeroZeroThirtySixFifteenOctoberFourteenADJRband \ $\pm$ \SinusoidPeaktoPeakAmplitudePercentageErrorTwoMassZeroZeroThirtySixFifteenOctoberFourteenADJRband \%
in R-band
and
\SinusoidPeaktoPeakAmplitudePercentageTwoMassZeroZeroThirtySixFifteenOctoberFourteenADJzband \ $\pm$ \SinusoidPeaktoPeakAmplitudePercentageErrorTwoMassZeroZeroThirtySixFifteenOctoberFourteenADJzband \%
z'-band. Utilizing this simultaneous, multiwavelength amplitude,
and the others given in Table \ref{TableSinusoidIndividualLightCurves},
the best-fit spot 
temperatures are
$T_{spot}$ = \TSpotKelvinSimultaneousValue \ $\pm$ \TSpotKelvinSimultaneousValueError \ $K$
for spots slightly cooler than 2MASS 0036+18, and 
$T_{spot}$ = \TSpotHotKelvinSimultaneousValue \ $\pm$ \TSpotHotKelvinSimultaneousValueError \ $K$
for spots slightly hotter than 2MASS 0036+18.
Therefore, using either our simultaneous, multiwavelength photometry (Table \ref{TableSinusoidIndividualLightCurves}),
or the assumption of constant spots (Table \ref{TableSinusoidBands}), the conclusions are 
similar in that 
the spots on 2MASS 0036+18 can be up to a few hundred degrees cooler or hotter than the 2MASS 0036+18 photosphere, if
we assume the spot and ultra-cool dwarf radiate as blackbodies.

\subsubsection{Comparison to Models}
\label{SecSpotsModels}

\begin{figure}
\includegraphics[scale=0.55, angle = 270]{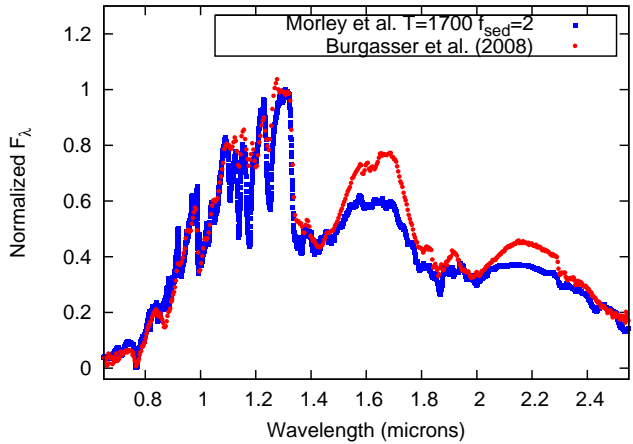}
\caption[]
	{	The \citet{Burgasser08} spectra (red points) of 2MASS 0036+18 compared to the best-fit
		\citet{MorleyPrep} 
		model spectra (blue points; utilizing $T_{UCD}$ = \TemperatureUCDUseSpectra \ $K$, $log \ g$ $\sim$ \loggUseSpectra, $f_{sed}$ = \fsedUseSpectra).
	}
\label{FigSpectra}
\end{figure}

\begin{figure}
\includegraphics[scale=0.45, angle = 270]{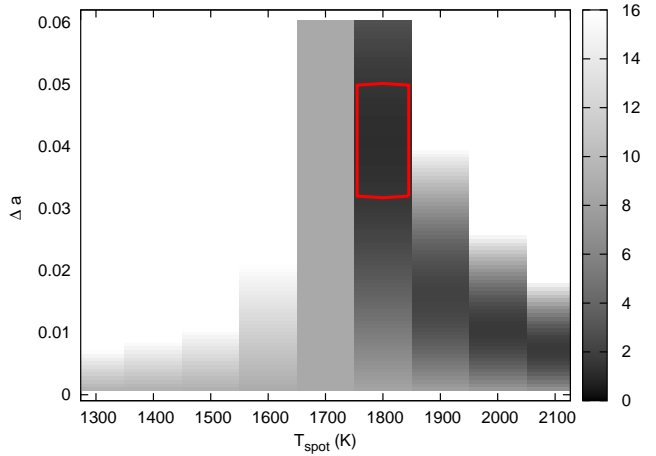}
\caption[]
	{	The change in filling factor, $\Delta a$ (a proxy for spot size),
		and temperature of spots, $T_{spot}$, that are consistent with the amplitudes of sinusoidal variability observed at various wavelengths
		assuming the spots and 2MASS 0036+18 radiate according to the 
		\citet{MorleyPrep} models.
		For 2MASS 0036+18 we utilize a 
		\citet{MorleyPrep} model spectra with
		$T_{UCD}$ = \TemperatureUCDUseSpectra \ $K$, $log \ g$ $\sim$ \loggUseSpectra, $f_{sed}$ = \fsedUseSpectra.
		We assume constant spots and therefore compare to the amplitudes of multiwavelength variability
		observed at different epochs (Table \ref{TableSinusoidBands}).		
		The intensity bar at right denotes the reduced $\chi^2$.
		The 1$\sigma$ credible regions are enclosed in the red solid line.
		Spots $\sim$100 $K$ hotter than our assumed 
		$T_{UCD}$ = \TemperatureUCDUseSpectra \ $K$ 2MASS 0036+18 photosphere 
		provide the best-fits to our observed multiwavelength amplitudes.
	}
\label{FigTempSpotsModel}
\end{figure}

Although blackbody models are instructive, the spectrum of 2MASS 0036+18 deviates significantly from that of a blackbody (see Figure \ref{FigSpectra}).
If starspots exist on the surface of 2MASS 0036+18,
then presumably the starspots on this ultra-cool dwarf also deviate significantly from blackbodies.
Therefore, to
perform a more realistic comparison, we next compare to 
one-dimensional cloud models of this L3.5 dwarf - these models will be presented in detail 
in \citet{MorleyPrep}. 
These models are new, updated versions of the \citet{SaumonMarley08} cloud models,
which utilize the \citet{AckermanMarley01} cloud models.
The models are coupled
to one-dimensional pressure-temperature profiles of atmospheres
in radiative-convective equilibrium.
These models have varying effective temperatures, $T$,
gravities, $log \ g$,
and $f_{sed}$ values. $f_{sed}$ is a parameter that denotes the efficiency of sedimentation:
high $f_{sed}$ values (e.g. $f_{sed}$=5) result in vertically thinner clouds, while a low
$f_{sed}$ value (e.g. $f_{sed}$=1) results in thicker clouds. 

To determine the characteristics of the unspotted photosphere of 2MASS 0036+18, we first
compare the 
\citet{MorleyPrep} generated models to the \citet{Burgasser08} spectra of 2MASS 0036+18 obtained
with the SpeX prism instrument \citep{Rayner03} on the NASA Infrared Telescope Facility (IRTF).
We test a limited grid of models from 
$T_{UCD}$ = 1500 to 2100 $K$, $f_{sed}$ = 1 - 2, and $log \ g$ $\sim$ 4.5 - 5.5.
For these range of parameters
the best-fit 
\citet{MorleyPrep} model is presented in Figure \ref{FigSpectra}, and features
the following parameters:
$log \ g$ $\sim$ \loggUseSpectra, $T_{eff}$ = \TemperatureUCDUseSpectra \ $K$, and $f_{sed}$ = \fsedUseSpectra.
Our best-fit model underpredicts the H and K-band flux of 2MASS 0036+18, as these types of models often do for L-dwarfs
(e.g. Figure 1 of \citealt{Morley12}). Further comparisons of the 2MASS 0036+18 
spectra to an expanded range of models that allow the 
metallicity, cloud patchiness, temperature variations, etc. to vary are warranted, but beyond
the scope of this work.

As we are interested in the temperature of starspots that might be driving the variability of this
ultra-cool dwarf, 
we first employ differences
in temperature only, and not in cloud thickness,
in our comparison of the 
\citet{MorleyPrep} models to
Equation \ref{EquationRadigan}. Thus, we
assume the $\mathcal{F}_1$ and $\mathcal{F}_2$ regions are identically cloudy ($f_{sed1}$ = $f_{sed2}$ = \fsedUseSpectra),
although with different temperatures ($T_{UCD}$ $\ne$ $T_{spot}$).

We increment the $T_{spot}$ values for the 
\citet{MorleyPrep} models in 100 $K$ steps to calculate $\mathcal{F}_2$,
and compare to the assumed best-fit photosphere of 2MASS 0036+18, $\mathcal{F}_1$ 
($log \ g$ $\sim$ \loggUseSpectra, $T_{UCD}$ = \TemperatureUCDUseSpectra \ $K$, and $f_{sed}$ = \fsedUseSpectra.).
We integrate the 
\citet{MorleyPrep} models over the associated bandpasses of our observations 
(Table \ref{TableSinusoidBands}).
We plot the best-fit models
compared to our measured multiwavelength amplitudes (Table \ref{TableSinusoidBands})
for a range of $\Delta a$ and $T_{spot}$ values in Figure \ref{FigTempSpotsModel}.
We again scale the resulting reduced $\chi^2$ to $\sim$1, to ensure reasonably-sized error bars on $T_{spot}$.
The best-fit spot temperatures from the 
\citet{MorleyPrep} models are approximately 
$\sim$100 $K$ hotter than the assumed $T_{UCD}$$\sim$1700 $K$ 2MASS 0036+18 photosphere.
As in Section \ref{SecSpotsBlackbody} there is an obvious correlation between the size of spots, $\Delta a$,
and the spot temperatures, $T_{spot}$, and small spots up to $\sim$300 $K$ hotter than the 
assumed $T_{UCD}$$\sim$1700 $K$ photosphere also provide reasonable fits to the multiwavelength variability. 

As our fit to the 2MASS 0036+18 spectra suggests the presence of a considerable cloud layer enveloping
2MASS 0036+18, 
if starspots hotter than the photosphere of this ultra-cool dwarf are responsible for the observed variability
there must exist some interplay between these two mechanisms.
We explore the impact of clouds in Section \ref{SecClouds}, and discuss the interplay between these two mechanisms 
in Section \ref{SecConclusions}.

\subsection{Clouds on 2MASS 0036+18}
\label{SecClouds}

\begin{figure}
\includegraphics[scale=0.45, angle = 270]{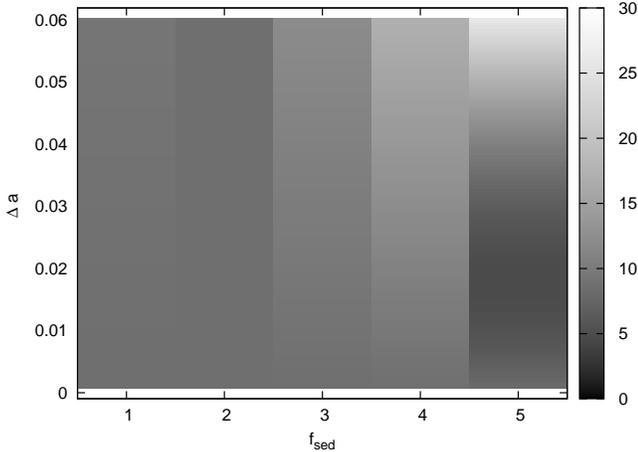}
\caption[]
	{	The change in filling factor, $\Delta a$ (a proxy for spot size),
		and cloudiness parameter, $f_{sed}$, that are consistent with the amplitudes of sinusoidal variability observed at various wavelengths
		assuming the spots and 2MASS 0036+18 radiate according to the 
		\citet{MorleyPrep} models.
		For 2MASS 0036+18 we utilize a 
		\citet{MorleyPrep} model spectra with 
		$T$ = \TemperatureUCDUseSpectra \ $K$, $log \ g$ $\sim$ \loggUseSpectra, $f_{sed}$ = \fsedUseSpectra.
		The intensity bar at right denotes the reduced $\chi^2$.
		We assume constant spots and therefore compare to the amplitudes of multiwavelength variability
		observed at different epochs (Table \ref{TableSinusoidBands}).
		None of these cloud models without temperature variations
		provide especially impressive fits to the observed amplitudes of variability.
	}
\label{FigCloudsSpotsModel}
\end{figure}

\begin{figure}
\includegraphics[scale=0.45, angle = 270]{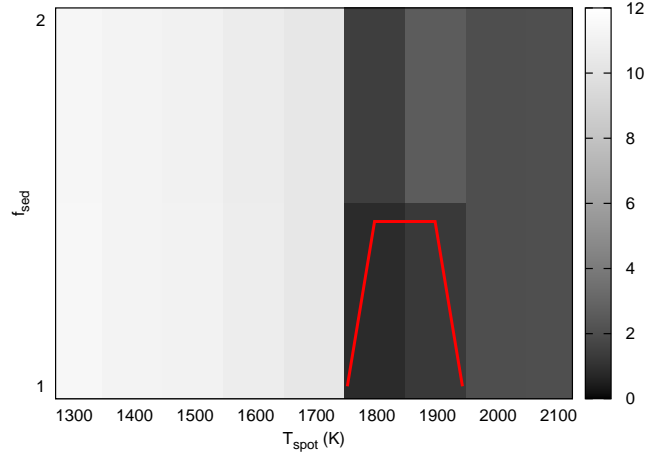}
\caption[]
	{	The change in spot temperature, $T_{spot}$,
		and cloudiness parameter, $f_{sed}$, that are consistent with the amplitudes of sinusoidal variability observed at various wavelengths
		assuming the spots and 2MASS 0036+18 radiate according to the 
		\citet{MorleyPrep} models.
		For 2MASS 0036+18 we utilize a 
		\citet{MorleyPrep} model spectra with 
		$T$ = \TemperatureUCDUseSpectra \ $K$, $log \ g$ $\sim$ \loggUseSpectra, $f_{sed}$ = \fsedUseSpectra.
		The intensity bar at right denotes the reduced $\chi^2$.
		The 1$\sigma$ credible region is enclosed in the red solid
		line (to the upper-right).
		We assume constant spots and therefore compare to the amplitudes of multiwavelength variability
		observed at different epochs (Table \ref{TableSinusoidBands}).
		Spotted regions slightly hotter than the 2MASS 0036+18 photosphere and with different cloud thicknesses can
		provide reasonable fits to the observed amplitudes of variability.
	}
\label{FigCloudsTemperatureSpotsModel}
\end{figure}

We also explore whether cloud opacity variations could be driving the variability of 2MASS 0036+18.
We note that if clouds, or aurorae, or another mechanism, are causing the variability on this ultra-cool
dwarf then this mechanism must result in multiwavelength phase shifts smaller than we have observed in Table \ref{TableSinusoidIndividualLightCurves}.
If clouds exist on 2MASS 0036+18 and are causing the variability that we observe than the lack of significant multiwavelength phase shifts
suggest that the cloud structures span multiple layers in the atmosphere of this L3.5 dwarf --
specifically the cloud gaps must span the pressure layers probed by our simultaneous, multiwavelength photometry.

We first explore the possibility that varying cloud thicknesses on this L3.5 dwarf could be leading to the observed variability.
We therefore employ a 
\citet{MorleyPrep} model where the spotted region 
has an identical temperature to that of 2MASS 0036+18, but different cloud thickness (i.e.
$f_{sed1}$ $\ne$ $f_{sed2}$, $T_{UCD}$ = $T_{spot}$).
For 2MASS 0036+18 we again assume that the 
$\mathcal{F}_1$ region is composed of the 
values that provide the best-fit to the \citet{Burgasser08} spectrum (Figure \ref{FigSpectra}):
$log \ g$ $\sim$ \loggUseSpectra, $T_{eff}$ = \TemperatureUCDUseSpectra \ $K$, and $f_{sed}$ = \fsedUseSpectra.
We increment the $f_{sed}$ values by 
1.0 and present the results in Figure \ref{FigCloudsSpotsModel}.
Of these varying 
$f_{sed}$ models without temperature variations
the best-fit to the observed amplitudes is provided by spots that are relatively cloud-free ($f_{sed}$=5).
However, 
none of these cloud models without temperature variations provide impressive fits to the observed amplitudes;
the varying temperature 
models (Section \ref{SecSpotsModels}; Figure \ref{FigTempSpotsModel})
provide considerably better fits to the observed amplitudes,
than the models that allow the cloud thicknesses but not the temperatures to vary.

Therefore, if the 2MASS 0036+18 photosphere and spots follow the 
\citet{MorleyPrep} models, and if this ultra-cool dwarf displays constant amplitude variability,
then clouds without temperature variations ($f_{sed1}$ $\ne$ $f_{sed2}$; $T_{UCD}$ = $T_{spot}$) 
cannot be responsible
for the variability we observe.

We next assume the spotted regions of 2MASS 0036+18
consist of different cloud thicknesses ($f_{sed1}$ $\ne$ $f_{sed2}$) 
and/or different temperatures ($T_{UCD}$ $\ne$ $T_{spot}$). We plot the 
best-fit comparisons to the data for a range of 
\citet{MorleyPrep} models with varying $f_{sed2}$ and $T_{spot}$ parameters
in Figure \ref{FigCloudsTemperatureSpotsModel}.
For this figure we only plot the optimal $\Delta a$ parameter (lowest $\chi^2$),
for the associated $f_{sed2}$ and $T_{spot}$ parameters.
Spotted regions consisting of either different temperatures, or different cloud thicknesses and different temperatures
provide reasonable fits to the observed amplitudes; the best-fit value is 
provided by a model
with a slightly different cloud thickness ($f_{sed}$=\fsedMinModel), and a different
temperature ($T_{spot}$ = \TemperatureMinModel \ $K$)
than our 2MASS 0036+18 photosphere.
If the 
\citet{MorleyPrep} models provide reasonable approximations to the
2MASS 0036+18 photosphere and spotted regions, then
spotted regions with different cloud thicknesses and temperatures
($f_{sed1}$ $\ne$ $f_{sed2}$ and/or $T_{UCD}$ $\ne$ $T_{spot}$) 
than
this ultra-cool dwarf can explain the observed, multiwavelength amplitudes.

In general, according to our varying cloud thickness and varying spot temperature models,
the best-fits to the observed amplitudes (Table \ref{TableSinusoidBands}) are provided by spots, or gaps in clouds,
that are
slightly hotter than the 2MASS 0036+18 photosphere.
However, this conclusion depends on our assumption
of constant spots and therefore that we can compare to
our multiwavelength amplitudes observed at different epochs (Table \ref{TableSinusoidBands});
if the spots vary in size, a better comparison would solely be to our 
multiwavelength amplitudes observed simultaneously (Table \ref{TableSinusoidIndividualLightCurves}).
For the latter, simultaneous multiwavelength amplitudes, cooler spots than the 2MASS 0036+18
photosphere that radiate as blackbodies provide reasonable fits to the observed amplitudes (as discussed in Section \ref{SecSpotsBlackbody}).
Therefore it seems prudent to conclude that spots up to a few hundred degrees 
cooler or hotter than the 2MASS 0036+18 photosphere can explain the
observed multiwavelength variability of this ultra-cool dwarf.

\section{Conclusion}
\label{SecConclusions}

 Our long-term, multiwavelength, ground-based
light curves of the L3.5 dwarf 2MASS 0036+18 allow us to address a number of unique science
cases.
First, our many nights of photometry allow us to demonstrate that the light curves of 2MASS 0036+18 do not
significantly evolve from rotation period to rotation period, or night-to-night.
This lack of evolution allows us to precisely determine the rotation period of this L3.5 dwarf to be
\PeriodHoursTwoMassZeroZeroThirtySixAllbands \ $\pm$ \PeriodHoursErrorTwoMassZeroZeroThirtySixAllbands \ hours.
This rotation period is recovered from the R-band to the J-band, and is similar to the rotation period
that has previously been recovered from radio \citep{Hallinan08} and infrared \citep{Metchev15} observations for this ultra-cool dwarf;
therefore there is no strong
evidence for differential rotation with latitude, or at different pressure layers as
probed by shorter wavelength observations peering deeper into
the atmosphere of this L3.5 dwarf.

These light curves also allow us to constrain the rate of flares exhibited by 2MASS 0036+18;
this ultra-cool dwarf must display significantly fewer optical/near-infrared flares
than radio flares \citep{Berger02,Berger05}. Also, we are able to rule out transiting super-Earth, and even some Earth-sized planets
in the habitable zone of this ultra-cool dwarf.

Our 
\INSERTNIGHTSSIMULTANEOUS \ nights of simultaneous, multiwavelength photometry do not display discernible
phase shifts, and therefore suggest that the variability of 2MASS 0036+18 is driven by starspots, or another
mechanism (clouds, aurorae, etc.) that results in similarly modest phase shifts.
The amplitude of variability generally decreases with increasing wavelength, a result consistent
with starspots slightly warmer or cooler than 2MASS 0036+18 being responsible for the observed variability, and
with the H$\alpha$ detection \citep{Pineda16} from this L3.5 dwarf. 
Our fit to the spectra of 2MASS 0036+18 suggests that considerable clouds envelope this ultra-cool dwarf.
Clouds, or another mechanism, resulting in temperature differences could also be responsible for driving the observed variability
of this ultra-cool dwarf, but this other mechanism would have to result in phase shifts as small or smaller than displayed in
our multiwavelength photometry; therefore, if clouds are causing the variability of 2MASS 0036+18 then the gaps in clouds 
would likely have to 
span the multiple pressure layers
probed by our observations
and expose hotter layers of the atmosphere beneath the cloud layer.
If a single mechanism is causing the variability of 2MASS 0036+18 then starspots 
or clouds that span multiple pressure layers are the leading explanations;
however, other mechanism(s) that result in temperature differences, but a lack of significant multiwavelength phase shifts 
are also viable.

The lack of significant evolution
of the 2MASS 0036+18 light curve
from rotation period to rotation period, or night-to-night, is in stark contrast
to what has been observed for L/T transition brown dwarfs (e.g. \citealt{Artigau09, Radigan12, Gillon13,CrollUCDII})
and their variability that has generally been attributed to clouds.
It is not clear if our detection of some, but not significant, evolution 
of the light curve of 2MASS 0036+18 favours an interpretation of clouds or spots
driving the variability of this ultra-cool dwarf.
Starspots on an active M-dwarf have been inferred to be quasi-stable for years
(e.g. \citealt{Davenport15}).
Naively, one might expect cloud structures to evolve rapidly, but 
multiwavelength photometric monitoring of an L1 dwarf \citep{Gizis13,Gizis15} have indicated the possibility
of a cloud feature that is stable on that ultra-cool dwarf for up to two years; in addition, 
Jupiter's Great Red Spot has likely been present since,
at least, soon after the invention of the telescope \citep{Hook1665,Cassini1666,Marcus93},
suggesting that one cannot rule out a priori the presence of long-lived cloud structures on 2MASS 0036+18.

The most likely explanation for 2MASS 0036+18's variability is arguably a mixture of mechanisms.
As our fit to the spectra of 2MASS 0036+18 suggests that significant clouds envelope this L3.5 dwarf, a mixture of clouds 
and starspots seems likely.
The variability of 2MASS 0036+18 is likely driven predominantly by starspots, with some complex
interplay between the hot/cool starspots and the clouds on this L3.5 dwarf that envelope this ultra-cool dwarf.

We encourage further long-term, multiwavelength
monitoring of this intriguing L-dwarf, as well as other L and ultra-cool dwarfs,
to determine the long-term variability amplitudes, phases, and evolution, or lack thereof, of these objects.
With the recent detection that nearly all L-dwarfs are variable \citep{Metchev15},
and that H$\alpha$ detections are common for many early L-dwarfs \citep{Schmidt15}, 
such monitoring may indicate whether the observed characteristics of 2MASS 0036+18 are common
for other early L-dwarfs.


\acknowledgements

We thank Dan Clemens for helpful discussions about the Mimir non-linearity correction.
We thank Francois-Rene Lachapelle, Etienne Artigau, \& Sandie Bouchard for attempting to observe
2MASS 0036+18 simultaneously with our observations. 
We also thank Saul Rappaport for helpful discussions that improved this manuscript.

These results made use of Lowell Observatory's Discovery Channel Telescope.
Lowell operates the DCT in partnership with Boston University, Northern Arizona University, the University
of Maryland, and the University of Toledo. Partial support of the DCT was provided by Discovery
Communications. LMI was built by Lowell Observatory using funds from the National Science Foundation
(AST-1005313).

\end{document}